\documentclass[aps,prd,showpacs,twocolumn,superscriptaddress,nofootinbib,preprintnumbers]{revtex4-1} 

\makeatletter
\def\p@subsection{}
\makeatother

\usepackage{graphicx,amsmath,amssymb,lipsum,bm}

\usepackage[usenames,dvipsnames]{xcolor}
\definecolor{xlinkcolor}{rgb}{0.7752941176470588, 0.22078431372549023, 0.2262745098039215}

\usepackage[colorlinks=true,citecolor=xlinkcolor,linkcolor=xlinkcolor,urlcolor=xlinkcolor, backref=false,pdfborder={0 0 0}]{hyperref}

\usepackage[normalem]{ulem}
\usepackage[utf8]{inputenc} 
\usepackage{mathtools}
\usepackage{aas_macros}
\usepackage{float}

\usepackage{booktabs}

\def\beqn#1{\begin{equation}\label{#1}}
\def\eeqn{\end{equation}}

\def\beqa#1{\begin{eqnarray}\label{#1}}
\def\eeqa{\end{eqnarray}}
\def\beq{\begin{eqnarray}}
\def\eeq{\end{eqnarray}}
\let\vec\mathbf

\newcommand{\hq}{\hat{\bm q}}

\newcommand{\vq}{\bm q}

\usepackage{ulem}
\usepackage{diagbox}

\usepackage[dvipsnames]{xcolor}
\usepackage{mathrsfs}

\newcommand{\be}{\begin{equation}}
\newcommand{\ee}{\end{equation}}

\renewcommand\k{{\bm k}}
\newcommand\q{\bm{q}}

\newcommand\G{\mathcal{G}_2}

\newcommand{\bseq}{\begin{subequations}}
\newcommand{\eseq}{\end{subequations}}

\renewcommand{\ln}{\mathop{\rm ln}\nolimits}

\newcommand{\kmax}{k_{\rm max}}
\newcommand{\ld}{\Lambda{\rm CDM}}

\newcommand{\wa}{w_0w_a{\rm CDM}}
\newcommand{\bao}{{\rm BAO}}
\newcommand{\cmb}{{\rm CMB}}

\def\gsim{\raise0.3ex\hbox{$\;>$\kern-0.75em\raise-1.1ex\hbox{$\sim\;$}}}
\def\lsim{\raise0.3ex\hbox{$\;<$\kern-0.75em\raise-1.1ex\hbox{$\sim\;$}}}

\def\beqn#1{\begin{equation}\label{#1}}
\def\eeqn{\end{equation}}

\def\beqa#1{\begin{eqnarray}\label{#1}}
\def\eeqa{\end{eqnarray}}

\def\kmax{{k_\text{max}}}
\def\hMpc{h{\text{Mpc}}^{-1}}
\def\Mpch{h^{-1}{\text{Mpc}}}

\def\Z2{$\mathcal{Z_2}$}

\defcitealias{desi1}{Paper 1}
\defcitealias{desi2}{Paper 2}
\defcitealias{desi3}{Paper 3}

\usepackage{xspace}
\newcommand{\paperone}{\citetalias{desi1}\xspace}
\newcommand{\papertwo}{\citetalias{desi2}\xspace}
\newcommand{\paperthree}{\citetalias{desi3}\xspace}

\newcommand {\ignore}[1]{}

\begin{document}

\preprint{MIT-CTP/5993}

\title{{\Large Reanalyzing DESI DR1:}\\
4. Percent-Level Cosmological Constraints from Combined Probes\\ and Robust Evidence for the Normal Neutrino Mass Hierarchy
}

\author{Mikhail M.~Ivanov}
\email{ivanov99@mit.edu}
\affiliation{Center for Theoretical Physics -- a Leinweber Institute, Massachusetts Institute of Technology, 
Cambridge, MA 02139, USA} 
 \affiliation{The NSF AI Institute for Artificial Intelligence and Fundamental Interactions, Cambridge, MA 02139, USA}

\author{James M. Sullivan}
\thanks{Brinson Prize Fellow} 
\email{jms3@mit.edu}
\affiliation{Center for Theoretical Physics -- a Leinweber Institute, Massachusetts Institute of Technology, 
Cambridge, MA 02139, USA}

\author{Roger de Belsunce}
\email{belsunce@mit.edu}
\affiliation{Center for Theoretical Physics -- a Leinweber Institute, Massachusetts Institute of Technology, 
Cambridge, MA 02139, USA} 
 \affiliation{The NSF AI Institute for Artificial Intelligence and Fundamental Interactions, Cambridge, MA 02139, USA}

\author{Shi-Fan Chen}
\email{sc5888@columbia.edu}
\affiliation{Department of Physics, Columbia University, New York, NY 10027, USA}
\affiliation{NASA Hubble Fellowship Program, Einstein Fellow}

\author{Anton~Chudaykin}
\email{anton.chudaykin@unige.ch}
\affiliation{D\'epartement de Physique Th\'eorique and Center for Astroparticle Physics,\\
Universit\'e de Gen\`eve, 24 quai Ernest  Ansermet, 1211 Gen\`eve 4, Switzerland}

\author{Mark Maus}
\email{mark.maus@berkeley.edu}
\affiliation{Department of Physics, University of California, Berkeley, 366 Physics North MC 7300, Berkeley, CA 94720-7300, USA}
\affiliation{Lawrence Berkeley National Laboratory, 1 Cyclotron Road, Berkeley, CA 94720, USA}

\author{Oliver~H.\,E.~Philcox}
\email{ohep2@cantab.ac.uk}
\affiliation{Leinweber Institute for Theoretical Physics at Stanford, 382 Via Pueblo, Stanford, CA 94305, USA}
\affiliation{Kavli Institute for Particle Astrophysics and Cosmology, 382 Via Pueblo, Stanford, CA 94305, USA}

\begin{abstract} 
\noindent 
We present cosmological parameter measurements from the full combination of DESI DR1 galaxy clustering data, described with large-scale structure effective field theory. By incorporating photometric galaxies and CMB lensing cross-correlations, and extending the bispectrum likelihood to smaller scales with a consistent one-loop computation, we achieve substantial gains in constraining power. Combined with the latest DESI baryon acoustic oscillation (BAO) data and cosmic microwave background (CMB) priors on the spectral tilt and baryon density, we find, in $\Lambda$CDM, $H_0=69.08\pm 0.37~\mathrm{km}\,\mathrm{s}^{-1}\mathrm{Mpc}^{-1}$, $\Omega_m=0.2974\pm 0.0050$, and {$\sigma_8 = 0.838\pm 0.017$} ({$S_8 = \sigma_8\sqrt{\Omega_m/0.3}
=0.834\pm 0.018$}).
Adding the Pantheon+ supernovae (SNe), we find a {$2.2\sigma$} preference for the $w_0w_a$ dynamical dark energy model from low-redshift data alone, rising to {$2.7\sigma$} when exchanging the SNe for \textit{Planck} CMB data.
Combining the full-shape, BAO, CMB, and SNe likelihoods improves the dark energy figure-of-merit by {$15\%$} and bounds the neutrino mass sum to {$M_\nu<0.049$ eV} ($\Lambda$CDM) and {$M_\nu<0.077$ eV} ($w_0w_a$CDM) at 95\% CL. {This is the strongest $w_0w_a$CDM bound to date, $37\%$ tighter than from the background expansion data alone}. 
The preference for the normal neutrino mass ordering thus holds regardless of the background model: the
inverted hierarchy is disfavored at ${\approx}\,3.5\sigma$ in $\Lambda$CDM and
${\approx}\,2.4\sigma$ in $w_0w_a$CDM, with the latter constraint free of
the geometric tension between CMB and BAO that is
known to sharpen the $\Lambda$CDM bound.
\end{abstract}

\maketitle

\section{Introduction}\label{sec:intro}

\noindent Neutrino masses are the only known non-trivial parameters
in the conventional model of particle physics
that have yet to be measured experimentally. 
These can provide crucial
insights into the high-energy completion of the Standard Model of Particle Physics and uncover a new mechanism of lepton mass generation~\cite{Donoghue_Golowich_Holstein_2023}. From neutrino oscillation experiments, we know that at least two mass states should have non-zero masses; moreover, there are two mass gaps with widths around $0.05~$eV
and $0.01~$eV~\cite{Esteban:2024eli}.\footnote{\url{http://www.nu-fit.org/}} This demarcates two scenarios: the ``inverted hierarchy'' (IH), where the two most massive states have a small gap between them, and the ``normal hierarchy'' (NH), with a small gap between the most massive state and the other two states. The neutrino oscillation data suggests a lower bound on the sum of neutrino masses 
$M_\nu\equiv \sum_{i} m_{\nu_i} $
in each scenario~\cite{Esteban:2024eli}:
\be 
\begin{split}
\label{eq:floor}
   &  M_\nu\geq 0.098~\text{eV}\quad [\text{IH}] \\
   &  M_\nu\geq 0.058~\text{eV}\quad [\text{NH}]\,.
\end{split}
\ee 
Neutrino beta-decay experiments
such as KATRIN provide 
an upper bound on the electron antineutrino of $0.45$~eV (at $90\%$ CL)~\cite{KATRIN:2024cdt}, which is significantly weaker
than the neutrino mass floor
dictated by oscillation
experiments~\eqref{eq:floor}. 

Cosmology provides a promising way to measure the sum of the neutrino masses~\cite{Lesgourgues:2006nd,Hannestad:2010kz,TopicalConvenersKNAbazajianJECarlstromATLee:2013bxd,Allison:2015qca,Lattanzi:2017ubx,Brinckmann:2018owf,Racco:2024lbu}.\footnote{Note that cosmological datasets do not have enough sensitivity to distinguish different mass states~\cite{Archidiacono:2020dvx}, thus cosmology can only probe mass hierarchies if the total neutrino mass happens to be below the inverted hierarchy floor of \eqref{eq:floor}.} In general, observations are sensitive to massive neutrinos through 
three main channels: modifications to the expansion history before recombination (when neutrinos behave as radiation), changes to the expansion history at low redshift (when they act as a dark matter sub-component), and the suppression of large-scale structure growth, due to neutrino free-streaming. Thanks to these effects, cosmology is currently a leading probe of 
neutrino masses~\cite{Palanque-Delabrouille:2019iyz,Ivanov:2019hqk,Garny:2020rom,Kumar:2022vee,Wang:2024hen,ACT:2023kun,Ivanov:2024jtl,DESI:2025zgx,DESI:2025ejh,Chebat:2025kes,desi2,Jiang:2024viw,Chudaykin:2019ock,Ivanov:2019hqk,Allali:2024aiv,Chebat:2025kes,RoyChoudhury:2024wri,Craig:2024tky,2022JHEAp..36....1T,2025PhRvD.111h3518H}. 
However, pushing the neutrino mass
constraints close to the oscillation floor ($M_\nu\sim 0.1$~eV) is a difficult endeavor, since in this regime the neutrinos have energy density less than $1\%$ of that of matter, which necessitates careful control of systematic uncertainties. In addition, the neutrino mass inference is significantly affected by tensions between
cosmological datasets, which may be interpreted as a dependence on the underlying cosmological model. 

Under the (optimistic) assumption that there are no unknown systematics nor unmodeled physical phenomena, combinations of cosmological datasets nominally provide extremely strong constraints on the total neutrino mass within the minimal cosmological model
$\Lambda$CDM. To date, the most stringent bound is~\cite{Wang:2024hen}
\be 
M_\nu < 0.043~\text{eV}~\quad \text{at 95\% CL,}
\ee 
from a combination of galaxy clustering, cosmic microwave background anisotropies, and various background expansion measurements. Since this value rules out the oscillation floor (albeit weakly), it must imply
either a breakdown of the $\Lambda$CDM model or the presence of unknown systematic effects.
Under the first option, the neutrino mass constraints 
should be derived under a more general cosmological model; however, it is well known that the limits depend strongly on the model assumed, with, for example, significant degradation seen in time-evolving dark energy models \citep[e.g.,][]{DESI:2025ejh}.

A less exciting interpretation of the tension between cosmological neutrino masses and oscillation experiments is the presence 
of systematics. One such source has been known for about a decade: the so-called $A_L$-anomaly seen in the cosmic microwave background 
(CMB) dataset. This refers to an excess of lensing power, which disfavors any suppression of large-scale structure predicted in the presence of massive neutrinos~\cite{Calabrese:2008rt,Ade:2013zuv,Planck:2018nkj}.\footnote{Here we use the definition of $A_L$ as a lensing power affecting both the smoothing of the acoustic peaks in the CMB primaries
and the four-point function used to reconstruct the CMB lensing auto spectrum. The latter drives the massive neutrino information in combined analyses~\cite{Green:2024xbb}.} Due to the lensing anomaly, CMB-based constraints on neutrino masses are significantly tighter than those suggested by sensitivity forecasts \cite[e.g.,][]{Chudaykin:2019ock}. 
In fact, the neutrino mass posterior has been found to peak in the nonphysical negative regimes ($M_\nu<0$), implying more 
large-scale power than in the fiducial model~\cite{Planck:2013nga,RoyChoudhury:2019hls}. 

The systematics and modeling choices in massive neutrino analyses have come under a new wave of scrutiny with the release of 
large-scale structure clustering data from the Dark Energy Spectroscopic Instrument (DESI) collaboration~\citep{DESI:2025zpo,DESI:2025zgx}. 
At face value, the expansion history implied by DESI observations is inconsistent with the CMB prediction under the minimal
$\Lambda$CDM model, featuring a preference for systematically smaller cosmological distances. As above, this discrepancy can be alleviated by formally allowing neutrino masses to be negative, which reduces the late-time matter density, $\Omega_m$ \cite{Craig:2024tky,Loverde:2024nfi,Green:2024xbb,Elbers:2024sha,Graham:2025fdt}. As for the $A_L$-anomaly, this does not imply that neutrinos have negative mass, but instead signals a mismodeling of the expansion history within $\Lambda$CDM. A corollary of the preference for negative neutrino masses is that the nominal constraints on $M_\nu$ are very strong when one restricts to the physical region $M_\nu>0$, disfavoring the oscillation floor~\cite{DESI:2025ejh}. Additional complications with the inference of neutrino masses
within $\Lambda$CDM arise due to potential discrepancies
in the measurements of the reionization optical depth, $\tau_{\rm reio}$, which biases the neutrino mass to negative values through its degeneracy with other cosmological parameters~\cite{Loverde:2024nfi,Craig:2024tky,Sailer:2025lxj,Sullivan:2026tas}. 

Concordance between cosmological datasets and neutrino oscillation experiments can be resolved by introducing a dynamical dark energy (DDE) component, \textit{i.e.}\ replacing $\Lambda$CDM
with the so-called $w_0w_a$CDM model, where the dark energy equation of state is allowed to vary with redshift. Evidence for such a background has been reported
by DESI (in combination with the CMB and supernovae) at the level of about $(3-4)\sigma$~\cite{DESI:2025zgx}. As before, this may point to either a breakdown of the $\Lambda$CDM model, or to unknown systematics in the expansion history, which are effectively marginalized over when varying the DDE parameters. 
In either case, inferring $M_\nu$ in a DDE-background should yield less biased and more model-independent constraints on the neutrino mass, though this comes at a cost of significantly weaker constraints \cite{Planck:2018nkj,RoyChoudhury:2019hls,DESI:2025zgx}.
In particular, previous $w_0w_a$CDM+$M_\nu$ analyses were
consistent with both normal 
and inverted neutrino mass hierarchies~\cite{DESI:2025ejh}.

In this work, we carry out a new precision analysis of the public DESI data, combining spectroscopic statistics with photometric clustering, CMB lensing, primary CMB anisotropies, BAO, and supernovae. For the first time, we find some evidence for the normal hierarchy of neutrino mass states within the DDE model (as well as the $\Lambda$CDM model, as found in previous works).
{As we discuss below, the $w_0w_a$CDM constraint is particularly meaningful because it is less affected by the parameter tensions that affect the $\Lambda$CDM bound.}
To achieve this, we perform a full-shape analysis of galaxy clustering data~\cite{Ivanov:2019pdj,DAmico:2019fhj,Chen:2021wdi}, which leverages a precision theoretical description of galaxy clustering utilizing {effective field theory (EFT)} concepts drawn from particle physics~\cite{Baumann:2010tm,Carrasco:2012cv,Ivanov:2022mrd}. Our treatment goes beyond that of the official DESI collaboration~\cite{DESI:2024hhd} in a number of ways:
\begin{itemize}
    \item \textit{Three-Point Functions}: In all spectroscopic analyses, we analyze both the galaxy power spectrum and bispectrum (as in \citep{desi1,desi2}), novelly including the bispectrum quadrupole. This utilizes the quasi-optimal bispectrum estimators introduced in~\cite{desi1} (building on \citep{polybin3d}), which accurately account for the survey window and fiber collision effects via a novel stochastic scheme. We find that the bispectrum offers significant additional constraining power on cosmological parameters.
    \item \textit{One-Loop Predictions}: We extend the bispectrum likelihood to smaller scales by utilizing a consistent one-loop EFT theory prediction in combination with a robust analysis pipeline. Our computation has been recently presented in~\cite{Bakx:2025pop} (see~\cite{Sefusatti:2009qh,Angulo:2014tfa,Baldauf:2014qfa,Eggemeier:2018qae,Philcox:2022frc,DAmico:2022ukl,Spaar:2023his} for important results on the EFT one-loop bispectrum), and utilizes the \textsc{cobra} factorization scheme for rapid evaluation of EFT loop integrals~\cite{Bakx:2024zgu} (which has not previously been applied to data). This facilitates percent-level-accuracy one-loop bispectrum computations for different cosmological models.
    \item \textit{Combined Probes}: In addition to the two- and three-point functions of the DESI DR1 galaxy samples, we analyze the cross-correlation of both photometric and spectroscopic DESI samples with CMB lensing maps, as well as the photometric auto-correlations, {optionally combined with the CMB lensing power spectra with the appropriate covariances}. This represents the most complete analysis of the first-year DESI data performed {so} far. Our analyses build on Refs.~\cite{Sailer:2024coh,Sailer:2025rks,Maus2025:joint_3d_lensing_dr1}, but focus primarily on perturbative scales with our EFT-based theory predictions, including all relevant nuisance parameters (in contrast to \citep{Maus2025:joint_3d_lensing_dr1}, which utilized a hybrid model, pushing to smaller scales) and combine with the spectroscopic dataset.
    \item \textit{Priors}: We use carefully chosen priors on EFT nuisance parameters to suppress the so-called prior volume effects (building on \citep{Ivanov:2019pdj,desi2,Maus:2024sbb,Tsedrik:2025hmj}). These allow for straightforward interpretation of the full-shape results, particularly for $w_0w_a$CDM+$M_\nu$ analyses without external distance information (e.g., from supernovae).\footnote{For prior volume effects in the EFT-based analysis, see the discussions in~\cite{Ivanov:2019pdj,Chudaykin:2020ghx,Philcox:2021kcw,Philcox:2022frc} as well as recent developments~\cite{Maus:2024dzi,Chudaykin:2024wlw,Maus2025:joint_3d_lensing_dr1,Tsedrik:2025hmj,desi2,Simon:2022lde}. We caution that dropping higher-order statistics and using uninformative priors on EFT parameters can pull the EFT coefficients into unphysical regions, which violate the perturbativity of the EFT expansion; see \cite{desi1} for further discussion.} If untreated, the prior volume effects may induce large shifts between the Bayesian and frequentist analyses \citep[e.g.,][]{DESI:2025hao}; for this reason, the official DESI papers do not include full-shape analyses of the $w_0w_a$CDM+$M_\nu$ model without relying on supernova distance measurements.
\end{itemize}
This work is the fourth in a series of works reanalyzing the DESI DR1 catalogs: \paperone~\cite{desi1} presented our custom estimation and analysis pipeline for the spectroscopic DESI dataset; \papertwo~\cite{desi2} placed constraints on various $\Lambda$CDM extensions in combination with CMB and supernovae datasets; \paperthree~\cite{desi3} used the DESI galaxies and quasars to constrain primordial non-Gaussianity, achieving the current tightest constraint on $f_{\rm NL}^{\rm local}$. Here, we throw in the kitchen sink, consistently analyzing photometric, spectroscopic, and CMB data, and integrating the aforementioned developments in bispectrum modeling, to place one of the tightest limits on the amplitude of dark matter clustering to date, leading to tight constraints on the neutrino sector and beyond.

\vskip 4pt
The remainder of our paper is structured as follows. Section~\ref{sec:data} outlines the details of our analysis, including the datasets and estimators, before our theoretical model and priors are discussed in Section~\ref{sec:theory}. Our main results are presented in Section~\ref{sec:results} and we draw conclusions in Section~\ref{sec:disc}. We present further details of the perturbation theory model in Appendix \ref{sec:one-loopEFT}.

\section{Data and Analysis}
\label{sec:data}

\subsection{Datasets}

\begin{table}[]
    \centering
    \begin{tabular}{|l|c|c|c|c|}
    \hline
      & $z_{\rm eff}$ & $\bar{n}^{-1}$ [$h^{-3}\mathrm{Mpc}^{3}$] & SN$_{\mathrm{2D}}$  & $s_\mu$\\\hline
     \textbf{BGS}   & $0.2953$ & $5.2\times 10^{3}$ & $1.3\times 10^{-5}$  & $1.113$\\
      \textbf{LRG1}   & $0.5096$ & $4.2\times 10^{3}$ & $6.2\times 10^{-6}$  & 1.016\\
      \textbf{LRG2}  & $0.7059$ & $4.7\times 10^{3}$ & $4.1\times 10^{-6}$ & 0.996\\
      \textbf{LRG3}   & $0.9184$ & $8.8\times 10^{3}$ & $3.8\times 10^{-6}$ & 1.032\\
      \textbf{ELG2}   & $1.3170$ & $9.6\times 10^{3}$ & $-$  & $-$\\
      \textbf{QSO}   & $1.4901$ & $4.5\times 10^{4}$ & $-$  & $-$\\
    \hline
      \textbf{pBGS1}   &$0.211$ & $90$ & $4.7\times 10^{-7}$  & 0.81\\
\textbf{pBGS2}   & $0.352$ & $430$ & $9.9\times 10^{-7}$  & 0.8\\
      \textbf{pLRG1}   & $0.470$ & $2.8\times 10^{3}$ & $3.9\times 10^{-6}$  & 0.972\\
\textbf{pLRG2}   & $0.625$ & $2.6\times 10^{3}$ & $2.1\times 10^{-6}$  & 1.044\\
      \textbf{pLRG3}  & $0.785$ & $3.4\times 10^{3}$ & $2.0\times 10^{-6}$  & 0.974\\
      \textbf{pLRG4} & $0.914$ & $5.3\times 10^{3}$ & $2.2\times 10^{-6}$  & 0.988\\
    \hline
    \end{tabular}
    \caption{\textbf{Galaxy Catalogs}: Key parameters of the six spectroscopic and six photometric galaxy catalogs used in this work, matching \citep{DESI:2024aax,DESI:2024jis,Maus2025:joint_3d_lensing_dr1}.
    $s_\mu$ is the galaxy magnification bias parameter and $\mathrm{SN}_{\rm 2D}$ is the Poissonian angular shot noise. The photometric samples are denoted by the prefix ``p'', and span a smaller range of redshifts than the spectroscopic samples. To minimize cross-correlations between catalogs, the photometric samples mask out galaxies inside the DESI DR1 spectroscopic footprint. Our fiducial full-shape analysis comprises: three-dimensional power spectra and bispectra from each spectroscopic chunk; CMB lensing cross-correlations with the BGS and LRG spectroscopic samples; two-dimensional power spectra from each photometric chunk; CMB lensing cross-correlations with each photometric sample.}\label{tab: desi-chunks}
\end{table}

\begin{figure*}[!t]
\includegraphics[width=1\columnwidth]{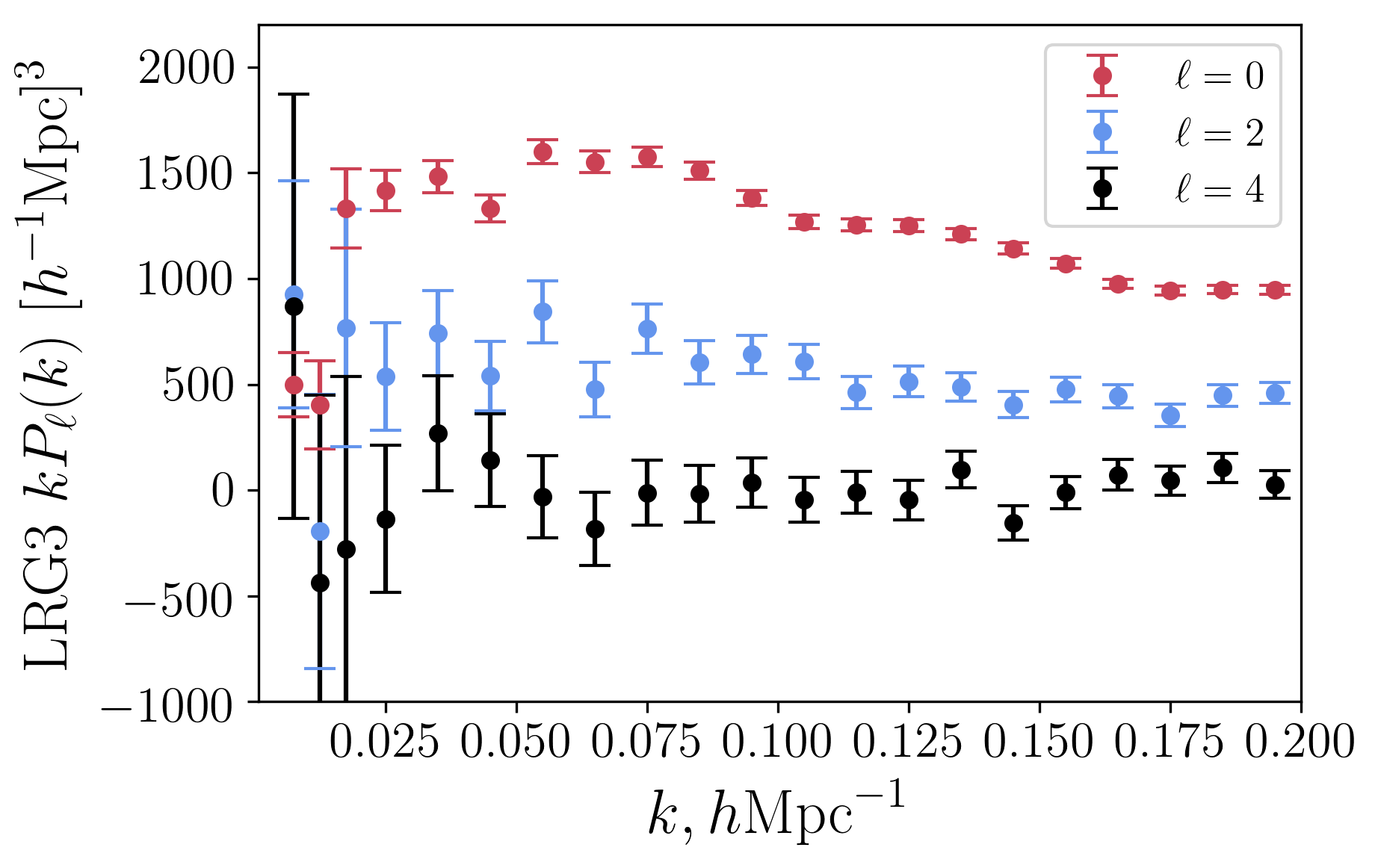}
\includegraphics[width=1\columnwidth]{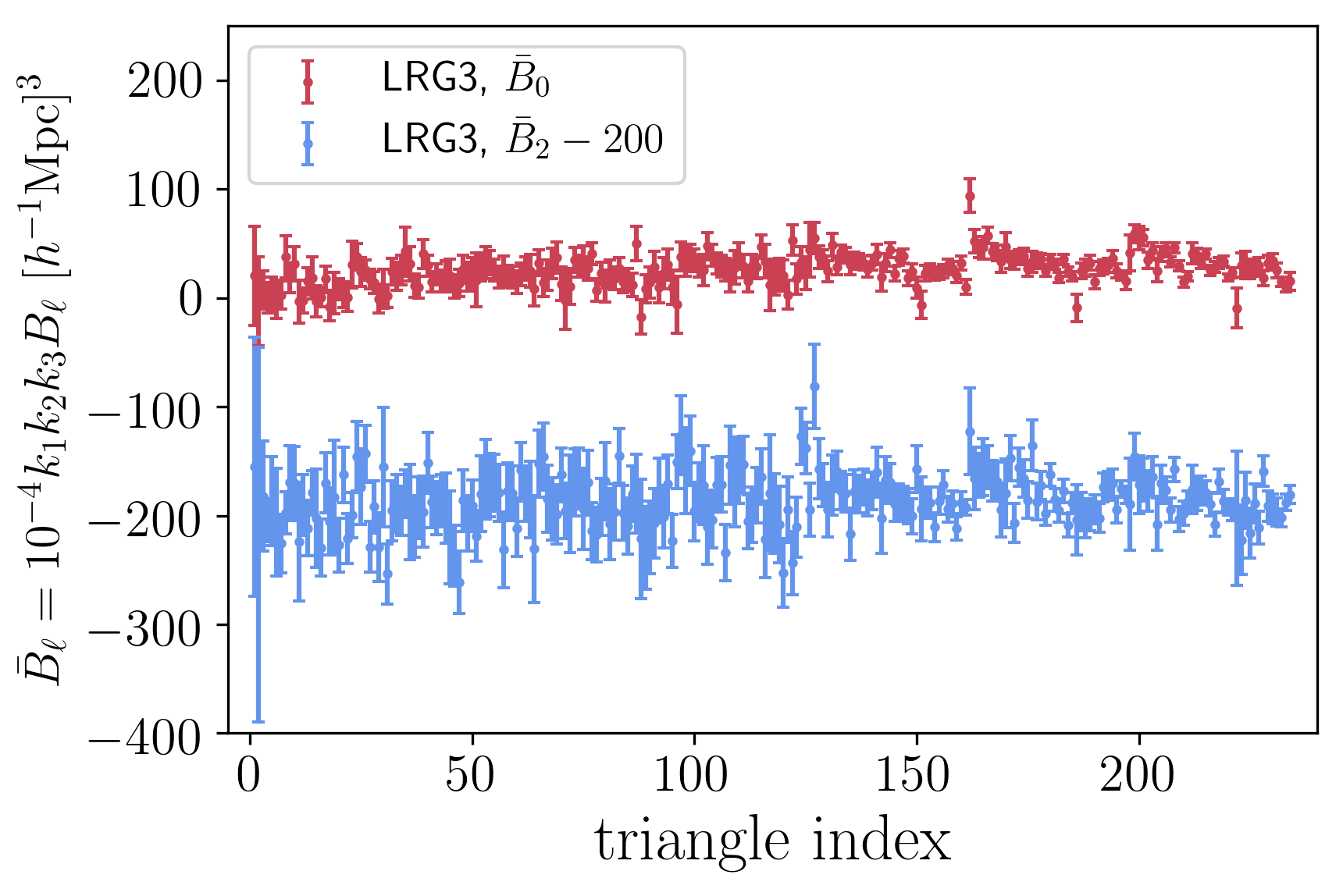}
\includegraphics[width=1\columnwidth]{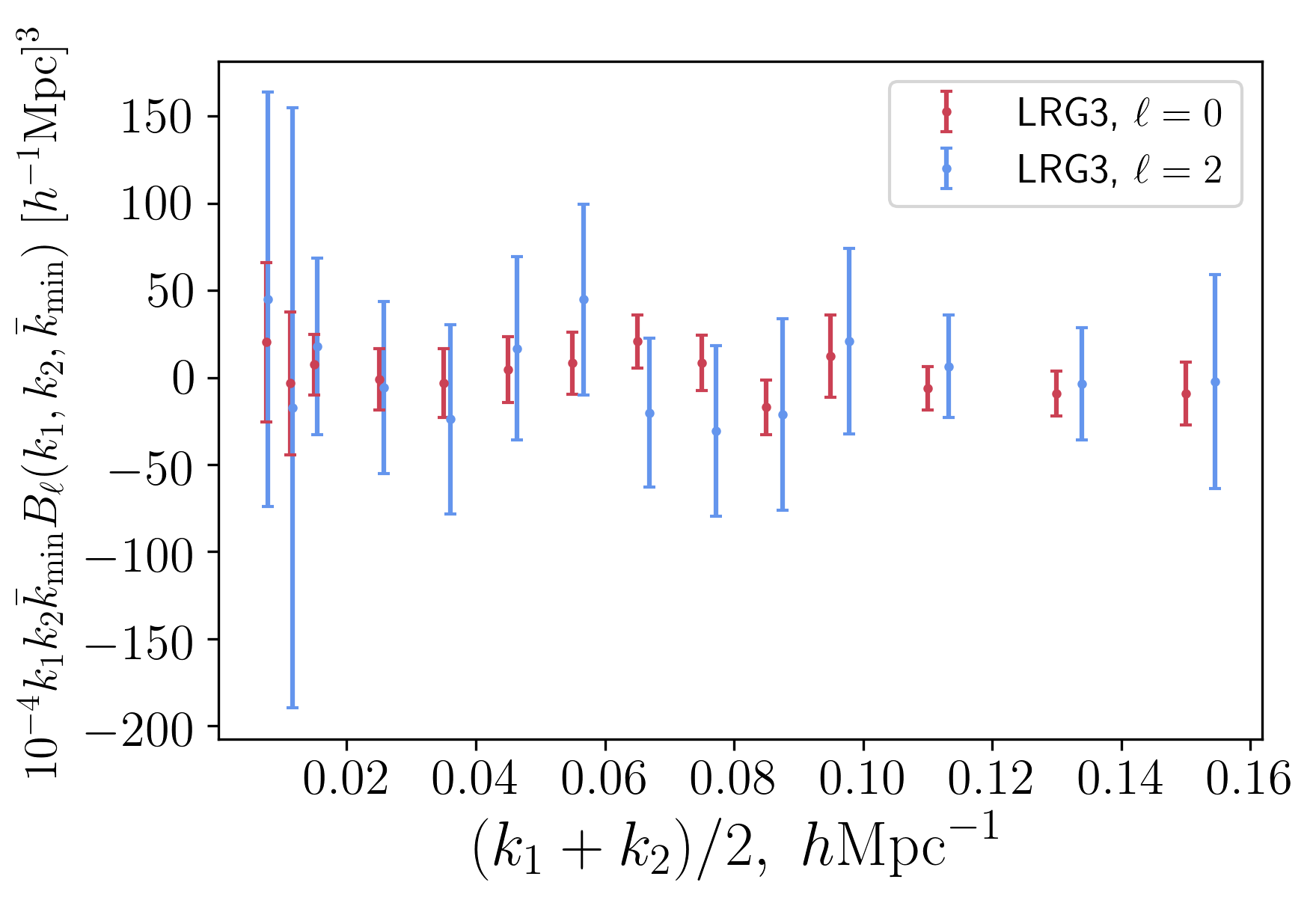}
\includegraphics[width=1\columnwidth]{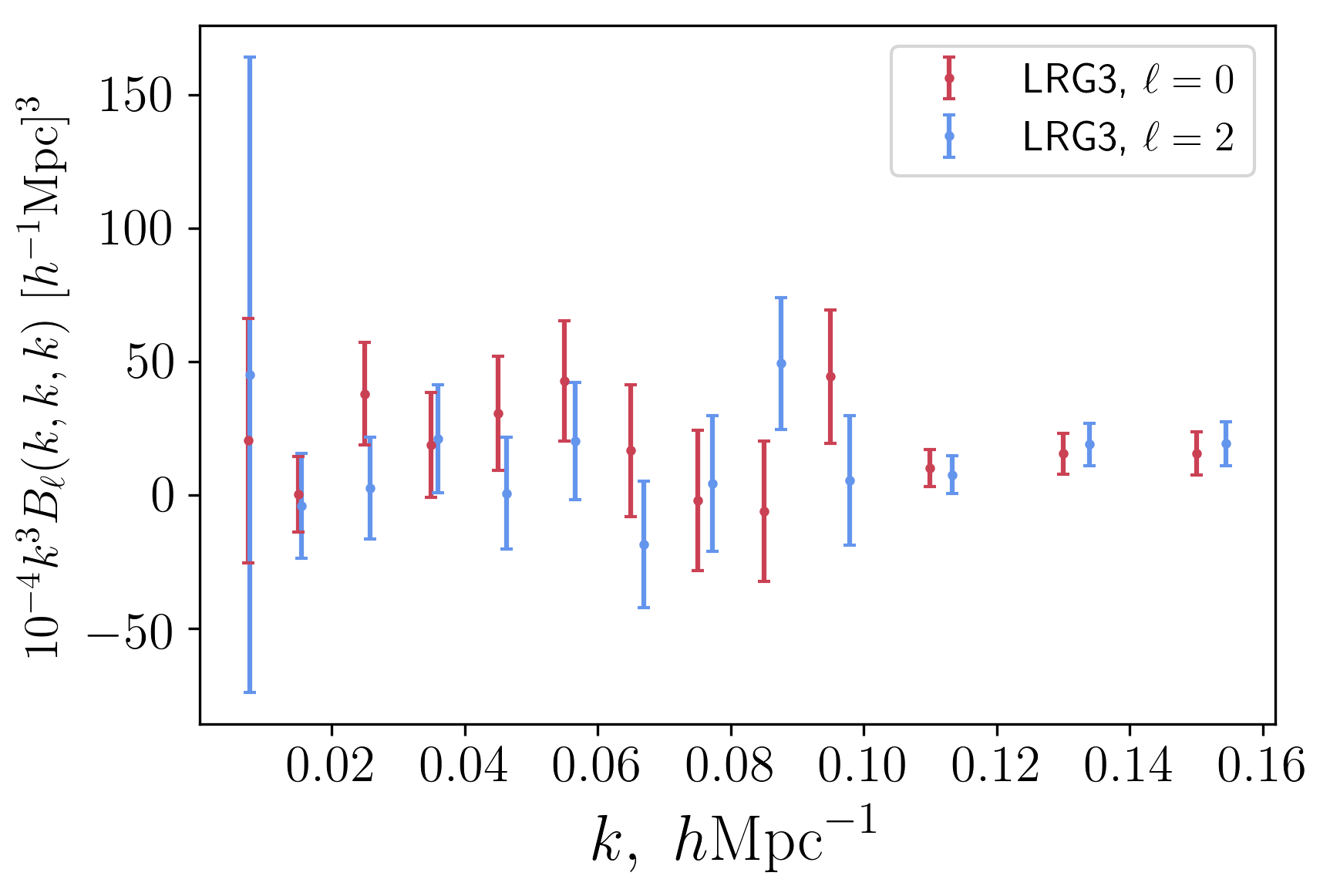}
\includegraphics[width=1\columnwidth]{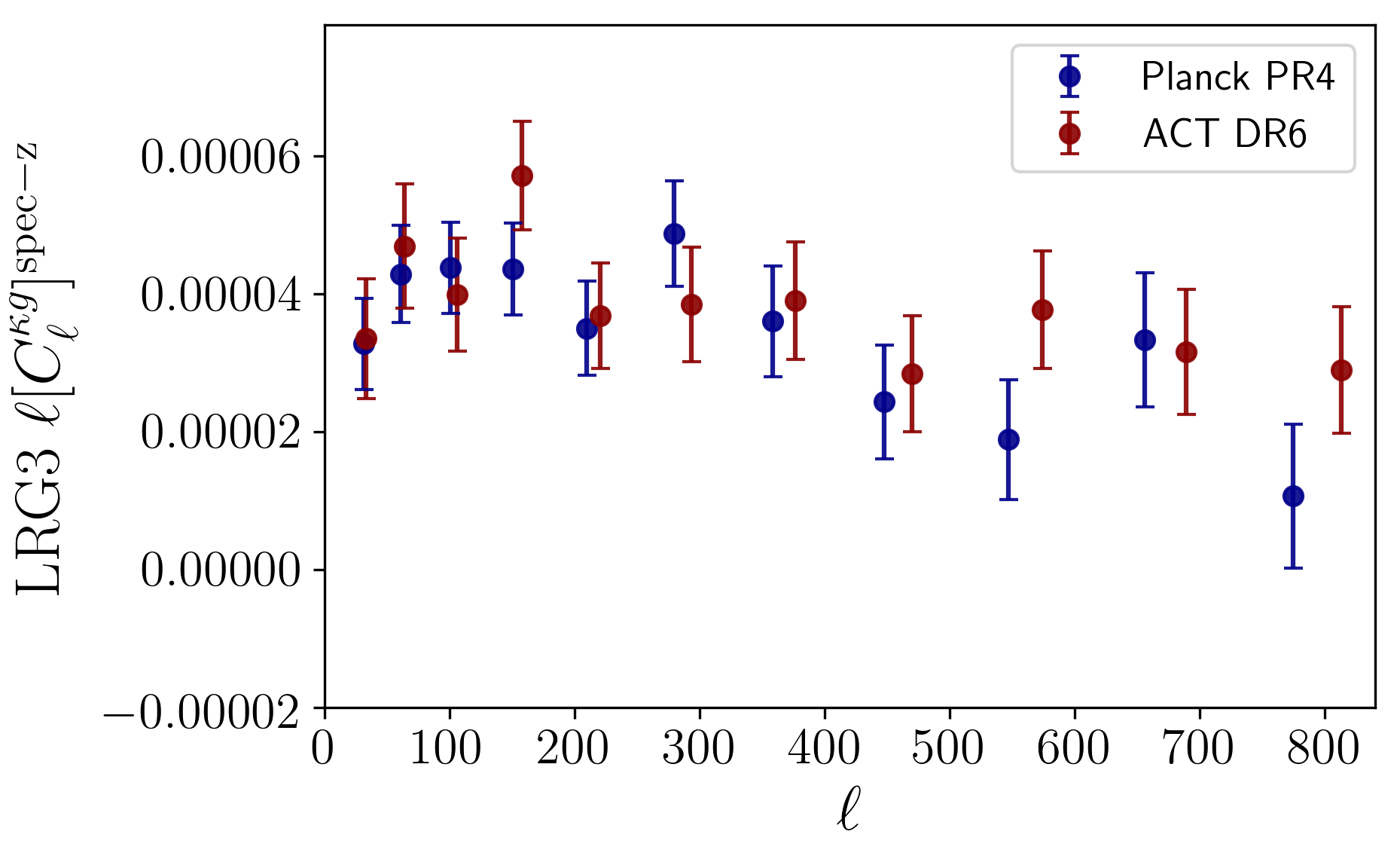}
\includegraphics[width=0.93\columnwidth]{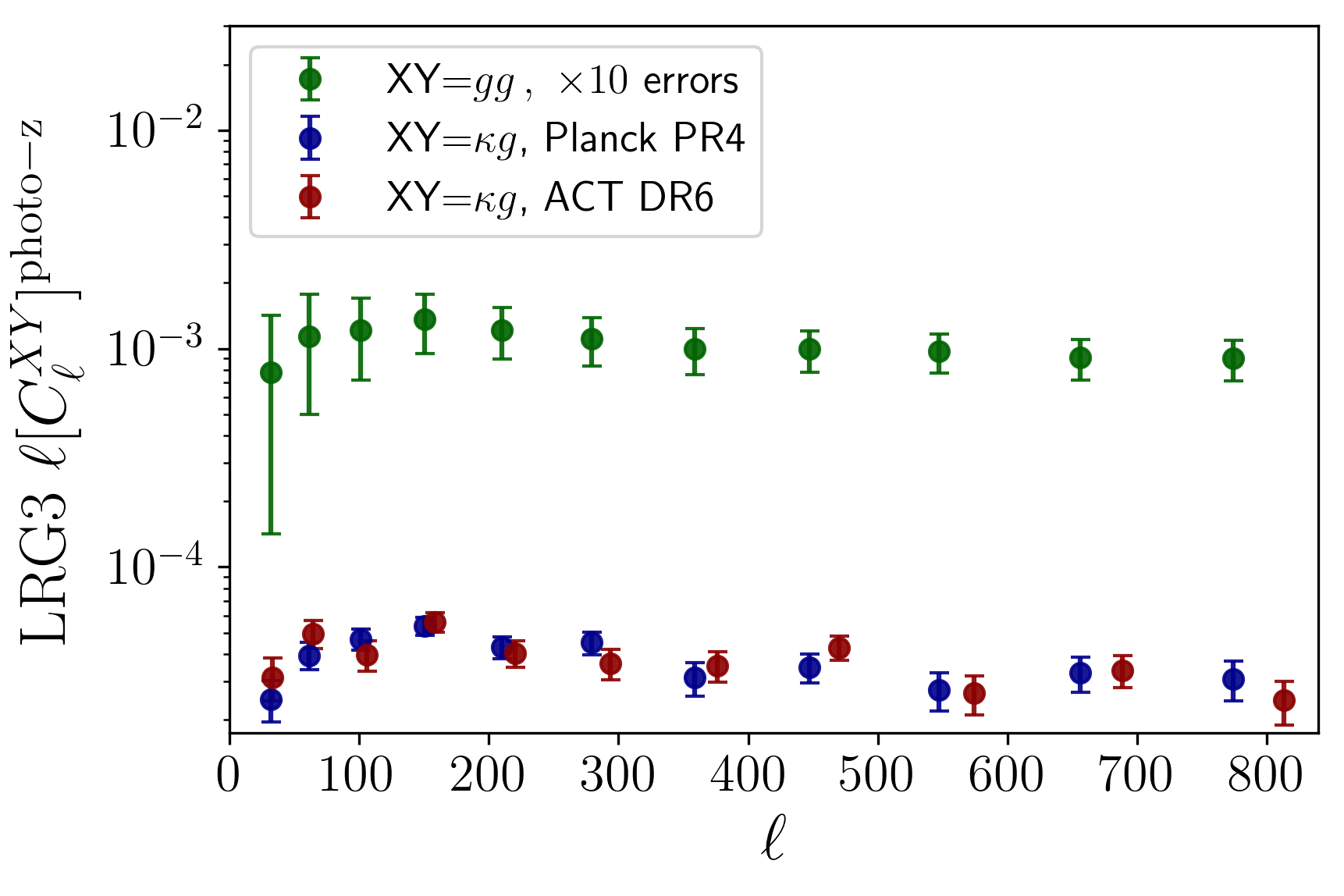}
\caption{\textbf{DESI Measurements}: \textit{Upper panel:} Three-dimensional power spectrum and bispectrum measurements for the spectroscopic DESI DR1 LRG3 sample at $z_{\rm eff}=0.92$.
The bispectrum dataset includes bins with centers ranging from $\bar{k}_{\rm min}=0.0075\,\hMpc$ to $\bar{k}_{\rm max}=0.15\,\hMpc$ (as described in the text), and the quadrupole moment, $B_2$, is offset along the $y$-axis for better visibility. 
\textit{Middle panel:} Bispectrum multipoles for squeezed ($\bar{k}_{\rm min} =0.0075\,\hMpc$) and equilateral configurations. The measurements are offset horizontally for better visibility.
\textit{Lower panel:} Cross-spectrum between the spectroscopic DESI DR1 LRG3 sample and the \textit{Planck} PR4 and ACT DR6 CMB lensing maps (left panel), as well as the auto-spectrum of photometric DESI pLRG4 galaxies at $z_{\rm eff}=0.91$ and their cross-correlation with the lensing maps (right panel). We increase the galaxy auto-spectrum errors by a factor of ten for better visibility. 
Poisson shot noise terms have been subtracted from all relevant observables. 
}
\label{fig:data_lrg3}
\end{figure*}

\textbf{Spectroscopic galaxies:} We use four sets of tracers: bright low-redshift galaxies (BGS), luminous red galaxies (LRG), emission line galaxies (ELG), and quasars (QSO), as in Table~\ref{tab: desi-chunks}. In all cases, our selection follows the main DESI DR1 full-shape analyses \citep{DESI:2024aax,DESI:2024jis,DESI:2025fxa}; notably, this does not include the low-redshift ELG sample (ELG1) nor the high-redshift $z>2.1$ quasars (whose impact is limited except for primordial non-Gaussianity studies \citep{Chaussidon:2024qni,desi3}). We do not use the full-shape Lyman-$\alpha$ data in this work. When computing the cross-correlations of spectroscopic galaxies with CMB lensing, we use only the BGS, LRG1, LRG2, and LRG3 samples, omitting the ELG2 and QSO samples following \cite{Maus2025:joint_3d_lensing_dr1}. In principle, these can be used to sharpen cosmological constraints even further (see \citep{deBelsunce:2025qku} for a joint analysis of the CMB lensing and QSO cross-correlations), though we expect only a marginal improvement over our baseline results, due to the high shot-noise in the QSO sample.
The magnification biases used for cross-correlations with the spectroscopic galaxies were measured in \cite{Heydenreich:2025}.

\textbf{Photometric galaxies and CMB lensing:} In areas outside the DESI DR1 footprint we use photometric samples derived from DESI targeting \cite{Hahn23,Zhou23a}, split into two and four redshift bins respectively for BGS and LRGs as in \cite{Chen:2024vvk,Zhou23b}. These photometric samples have well-characterized redshift distributions calibrated from DESI spectra \cite{Zhou23b}, and their magnification biases are measured directly at the catalog level \cite{Zhou23b,Wenzl24} (see Table~\ref{tab: desi-chunks}). We use CMB lensing maps derived from \textit{Planck} PR4 \cite{Carron22} and ACT DR6 observations, following \cite{Qu24,MacCrann24,Madhavacheril24}.

{In contrast to \cite{Maus2025:joint_3d_lensing_dr1}, which varied the magnification
bias parameters within Gaussian priors of width $0.1$ centered
on the measured values, we fix $s_\mu$ to these values
in our baseline analysis for both the photometric and spectroscopic samples. We have verified that marginalizing over all
$s_\mu$ with Gaussian priors of width $0.1$ leaves the posteriors on all cosmological parameters unchanged.}

\textbf{DR2 BAO:}
As in \paperone and \papertwo, we supplement our full-shape measurements with public baryon acoustic oscillation measurements from DESI DR2 \citep{DESI:2025zgx,DESI:2025zpo}. These comprise distance measurements from the six spectroscopic samples given in Table~\ref{tab: desi-chunks}, as well as BAO from the Lyman-$\alpha$ forest data and the LRG3 $\times$ ELG1 galaxy sample, which was omitted from the DR1 analysis due to concerns of systematic contamination. As argued in \paperone (and demonstrated in Appendices C \& D of the former work), we can neglect the covariance between DR2 BAO and DR1 full-shape due to the limited overlap in the galaxy sample and the weak correlation between the 
pre- and post-reconstructed statistics on small scales~\cite{Philcox:2021kcw} (which dominates the BAO measurements). {The conservative estimate of \paperone, based on mocks calibrated to SDSS-BOSS, bounded the impact of the neglected cross-covariance at
$\lesssim 0.4\sigma$; a subsequent dedicated study based on DESI-specific
simulations~\cite{Forero-Sanchez:2026bff} has since demonstrated that the effect on
cosmological parameters, including $\Omega_m$, is in fact negligible for DESI (see Fig.~3 there), additionally justifying our treatment of the full-shape and BAO
likelihoods as independent.}

\textbf{CMB primaries:}
We employ the high-$\ell$ \textsc{plik} TT, TE, and EE spectra, low-$\ell$ \textsc{SimAll} EE and low-$\ell$ \textsc{Commander} TT likelihoods from the official \textit{Planck} 2018 release~\cite{Planck:2018nkj}. This combination will be
called ``CMB'' in what follows. Note that this nomenclature
differs from that of \papertwo, where ``CMB'' 
included lensing auto-correlations. 

\textbf{CMB lensing auto-correlations:}
In addition, we include measurements of the lensing convergence power spectrum from the \textit{Planck} \textsc{npipe} PR4 maps~\cite{Carron:2022eyg} and the Atacama Cosmology Telescope (ACT) Data Release 6 (DR6)~\cite{ACT:2023kun,ACT:2023dou}. We will refer to these datasets simply as ``lens.''
Notably, we model non-linear corrections entering the lensing observables and the lensing-driven smoothing of the peaks of CMB primaries with the one-loop EFT matter power spectrum calculation as implemented in \texttt{CLASS-PT}~\cite{Chudaykin:2020aoj} (as used in several previous analyses of \textit{Planck} data \cite{Ivanov:2019hqk,Ivanov:2020ril,He:2023dbn,He:2023oke,McDonough:2023qcu}).\footnote{LSS perturbation theory was shown to give accurate results for both the lensed CMB power spectrum and the CMB lensing potential power spectra when compared to 
popular non-linear codes, such as \texttt{halofit}~\cite{Chudaykin:2020aoj}.}

\textbf{Supernovae:}
We also utilize Type Ia supernova data from the Pantheon+ sample \citep{Brout:2022vxf}, which consists of 1550 spectroscopically confirmed SNe in the redshift range $0.001 < z < 2.26$. Here, we use the public likelihood from~\cite{Chudaykin:2025gdn} and analytically marginalize over the supernova absolute magnitude. Hereafter, we will denote this supernova dataset as ``SNe.''

\subsection{Statistics}

\textbf{Two-dimensional correlators}: We use the angular auto- and cross-correlations $C_\ell^{gg}, C_\ell^{\kappa g}$ in harmonic space between spectroscopic and photometric galaxies and CMB lensing presented in \cite{Maus2025:joint_3d_lensing_dr1}. Specifically, we use the measurements presented in that work from \texttt{HEALPix} maps with $N_{\rm side} = 2048$ using \texttt{NaMaster} \cite{Alonso2019}, which also generates analytic covariances given a model for the angular clustering and the noise curves of the \textit{Planck} and ACT CMB lensing maps, as well as window matrices linking the measured bandpowers to theoretical $C_\ell$ predictions.\footnote{This is a slightly different choice than in \cite{Maus2025:joint_3d_lensing_dr1}, which used a more recently developed pixel-free harmonic-space measurement \cite{Baleato24}, though in both cases covariances are computed using the standard method.} We apply a normalization correction computed from mocks to the galaxy-CMB cross-correlation measurements due to mode coupling from masking and filtering in the latter \cite{Farren24}. {In extended analyses, we additionally include the CMB lensing auto-spectra $C_\ell^{\kappa\kappa}$ from the official \textit{Planck} PR4 and ACT DR6 lensing likelihoods, with a covariance that includes all cross-correlations with the projected galaxy statistics (discussed below).}

\textbf{Three-dimensional correlators}: 
We measure the power spectrum and bispectrum of each spectroscopic DESI chunk using the quasi-optimal estimators contained in the \textsc{PolyBin3D} code \citep{polybin3d}. Our power spectrum measurements are obtained similarly to those of \paperone, and include the monopole, quadrupole, and hexadecapole, using the binning described in Section~\ref{subsec: binning}.
We compute the bispectrum across a wider range of scales than in previous works, due to our use of a one-loop theoretical model, and additionally 
compute both the monopole and quadrupole moments ($B_{0,2}$), following \citep{Ivanov:2023qzb,Bakx:2025pop}. In all cases, we combine measurements from the Northern and Southern Galactic caps, weighting by the survey areas. To compute the statistics, we require a fiducial cosmology -- as in \cite{desi1}, we use the \textit{Planck} 2018 best-fit parameters $\{h = 0.6736, \omega_b = 0.02237, \omega_{\rm cdm} = 0.1200, N_{\rm ur} = 2.0328, \omega_\nu = 0.00064420\}$ assuming a single massive neutrino with $M_\nu=0.06$~eV~\citep{Aghanim:2018eyx}.

As discussed in \paperone, our power spectrum and bispectrum estimators correct for a number of known systematic effects. In particular, we account for the mask utilizing quasi-unwindowed estimators (plus a theory convolution matrix for the power spectrum, akin to the pseudo-$C_\ell$ approach of \cite{Alonso2019}), imaging systematics via particle weights (following \citep{DESI:2024aax}), radial integral constraints (based on \citep{deMattia:2019vdg}), and wide-angle effects in the power spectrum (as in \citep{Beutler:2021eqq}). Finally, we account for fiber collisions by removing pairs of points with angular separations below $0.05$ degrees, working deterministically for the power spectrum (following \citep{Pinon:2024wzd}), and stochastically for the bispectrum, using the efficient algorithm introduced in \paperone. When computing the bispectrum, we reweight the input data such that the statistic has the same effective redshift as the two-point function following \citep{Chen:2024vuf}, which simplifies the theoretical model. 

{Averaging over expectations, our stochastic scheme is equivalent to the exact removal of all pairs and triplets of galaxies affected by fiber collisions, which is computationally infeasible to perform directly. Moreover, the associated
modification of the effective selection function \citep{Pinon:2024wzd} is propagated self-consistently into the estimator normalization and window
treatment; this approach was validated for the two-point analog in \paperone. The radial integral constraint is treated identically for the
power spectrum and bispectrum, since it enters only through the response
matrix, which is linear in the data. As an explicit test of residual
fiber-collision systematics at our extended bispectrum scale cut, we have
repeated our baseline analysis without applying the fiber-collision
correction to the bispectrum, finding shifts of at most $0.12\sigma$
(in $\omega_{\rm cdm}$), with all other cosmological parameters shifting by
less than $0.1\sigma$ (cf.\ Section~\ref{sec:results} and
Table~\ref{tab:main2}); since this brackets the total impact of fiber
collisions, any residual effect after mitigation is negligible.}

Finally, we compute theoretical covariance matrices for the power spectrum and bispectrum using the FFT-based algorithms in \textsc{PolyBin3D}. These account for the survey geometry as well as inhomogeneous noise properties, imaging systematics marginalization (via linear marginalization), and integral constraints, though work in the Gaussian limit (e.g., ignoring trispectrum contributions to the power spectrum covariance). The code requires a fiducial power spectrum, which is obtained via a fit to a power spectrum obtained from an initial analysis run. All spectra, normalizations, fiber collision corrections, and covariances are computed on a high-performance computing platform, with the total computation requiring $\mathcal{O}(10^4)$ CPU-hours.

\subsection{Likelihood}
\noindent Our full DESI likelihood is a product of four contributions:
\be 
L_{\rm EFT-FS}=L_{P_\ell} 
\times L_{B_\ell} 
\times L_{\kappa g}^{\rm spec-z}
\times L_{\kappa g,gg}^{\rm photo-z}\,.
\ee 
The likelihoods for the three-dimensional power spectrum and bispectrum multipoles are given by
\be 
\begin{split}
& -2\ln L_{P_\ell}=\sum_{X}\sum_{\ell,\ell'=0,2,4}\sum_{i,j=1}^{N_{\rm bins}}\mathbb{C}_{ij,\ell\ell',XX}^{-1}\\
&(P^{{\rm th},X}_\ell(k_i)-P^{{\rm d},X}_\ell(k_i))(P^{{\rm th},X}_{\ell'}(k_j)-P^{{\rm d},X}_{\ell'}(k_j))\,,\\
& -2\ln L_{B_\ell}=\sum_{X}\sum_{\ell,\ell'=0,2}\sum_{I,J=1}^{N_{\rm triangles}}\mathbb{C}_{IJ,\ell\ell'}^{-1}\\
&\,\times\,(B^{{\rm th},X}_\ell(T_I)-B^{{\rm d},X}_\ell(T_I))(B^{{\rm th},X}_{\ell'}(T_J)-B^{{\rm d},X}_{\ell'}(T_J))\,,
\end{split}
\ee 
where `th' and `d' represent theory and data respectively, $\mathbb{C}$ is a covariance matrix, $T_{I}=(k_1,k_2,k_3)$ is a triplet of wavenumbers satisfying the triangle conditions, and $X$ indexes the spectroscopic sample with
\be 
X\in\{\rm{BGS},\rm{LRG1},\rm{LRG2},\rm{LRG3},\rm{ELG2},\rm{QSO}\}\,.
\ee 

We do not include the cross-covariance between the power spectra and bispectra. As argued in \paperone, this is justified as long as our analysis is restricted to large enough scales. In particular, \cite{Ivanov:2021kcd} explicitly checked that the cross-covariance can be ignored for the tree-level bispectrum likelihood restricted to $k_{\rm max}=0.08~\hMpc$. Similarly, simulation studies found that the use of a joint power and bispectrum likelihood without the cross-covariance does not bias parameter recovery from the tree-level bispectrum~\cite{Beyond-2pt:2024mqz} for $k_{\rm max}=0.08~\hMpc$,
and for the one-loop bispectrum at $k_{\rm max}=0.15~\hMpc$~\cite{Bakx:2025pop}, which match our fiducial scale cuts (see Section~\ref{subsec: binning}). Furthermore, our bispectrum covariances ignore non-Gaussian contributions, which can be important in the squeezed limit \citep[e.g.,][]{Biagetti:2021tua,Salvalaggio:2024vmx}. However, the squeezed triangles have both a small signal (in the models probed herein) and a large noise and hence do not contribute significantly to the cosmological parameter constraints~\cite{Ivanov:2021kcd}. {More generally, the marginalization over the large set of
EFT nuisance parameters ($\sim 45$ per redshift bin for the one-loop
bispectrum model) generates an effective theoretical-error contribution to
the covariance. This smooth highly correlated component dominates over
the neglected non-Gaussian terms on mildly non-linear scales, making its omission an accurate approximation. 
For the galaxy
power spectrum, this was demonstrated explicitly in~\cite{Wadekar:2020hax},
where full-shape analyses with Gaussian and fully non-Gaussian analytic
covariances yielded virtually identical cosmological constraints. 
The same logic applies to the bispectrum likelihood used here, particularly given the larger number of nuisance parameters contained therein. 
Nonetheless, we caution that an explicit test of the non-Gaussian bispectrum covariance and
of the power spectrum--bispectrum cross-covariance at
$k_{\rm max}^{B_\ell}=0.16~\hMpc$
for the one-loop model
and 
for the DESI DR1 volume is not yet
available in the literature 
(\citep[cf.][]{Euclid:2026glh} for the case of the tree-level bispectrum), and we leave such a
study for future work.}

To analyze the projected angular statistics, we use the following spectroscopic galaxy-CMB lensing likelihood:
\be 
\begin{split}
& -2\ln L_{\kappa g}^{\rm spec-z}=
% \sum_{\rm pair~of~tracers~(A,B)}\sum_{\rm CMB-exp}\\
\sum_{X,X'}\sum_{\kappa,\kappa'}\sum_{L,L'=1}^{N_{\rm bins}}\mathbb{C}_{X\kappa,X'\kappa',LL'}^{-1}\\\nonumber
&\quad\,\times\,(C^{{\rm th},X\kappa}_{L}-C^{{\rm d},X\kappa}_{L})(C^{{\rm th},X'\kappa'}_{L'}-C^{{\rm d},X'\kappa'}_{L'})
\end{split}
\ee 
where $X$ and $\kappa$ refer to spectroscopic and CMB lensing samples, with
\be 
\label{eq:sampls}
\begin{split}
&X\in\{\rm{BGS},\rm{LRG1},\rm{LRG2},\rm{LRG3}\}\,\\
&\kappa\in\{\kappa_{\rm Planck\,PR4},\kappa_{\rm ACT\,DR6}\}\,.
\end{split}
\ee 
where $L$ encodes the harmonic bins. Here, the covariance matrix $\mathbb{C}$ accounts for the correlations between the different CMB experiments and galaxy samples. {When the CMB lensing auto-spectrum is included in the analysis, the spectroscopic two-dimensional likelihood is extended to a joint $[C_\ell^{\kappa g}, C_\ell^{\kappa\kappa}]$ analysis, with the covariance matrix additionally incorporating $\mathrm{Cov}[C^{\kappa_a\kappa_b},C^{\kappa_c\kappa_d}]$, taken from the official lensing likelihoods, and the cross-covariance $\mathrm{Cov}[C^{\kappa_a \kappa_b},C^{\kappa_c g_i}]$, computed analytically.} We ignore the cross-covariance between these two-dimensional cross-spectra and our three-dimensional measurements due to the negligible overlap between the sampled modes, following \cite{Taylor22}. As in~\cite{Maus2025:joint_3d_lensing_dr1},
we do not use the spectroscopic $C_\ell^{gg}$ data
in our fiducial analysis due to its non-negligible correlation with the three-dimensional power spectra, $P_\ell(k)$.

Finally, our photometric likelihood is given by 
\be 
\begin{split}
& -2\ln L_{gg,\kappa g}^{\rm photo-z}=\sum_{XK,X'K'}\sum_{L,L'=1}^{N_{\rm bins}}\mathbb{C}_{XK,X'K',LL'}^{-1}\\
&\quad\,\times\,(C^{{\rm th},XK}_{L}-C^{{\rm d},XK}_{L})(C^{{\rm th},X'K'}_{L'}-C^{{\rm d},X'K'}_{L'})\,,
\end{split}
\ee 
where $XK$ runs over the six galaxy auto-spectra constructed from the photometric {samples} of Table~\ref{tab: desi-chunks}, and the twelve cross-spectra formed with the \textit{Planck} PR4 and ACT DR6 lensing maps. 
Note that our photometric likelihood includes both auto- and cross-correlation statistics, unlike the two-dimensional spectroscopic likelihood.
The covariance matrix $\mathbb{C}$ includes all relevant Gaussian cross-covariances, \textit{i.e.}\ $\mathrm{Cov}[C^{\kappa_{a} g_i},C^{\kappa_{b} g_j}]$,
$\mathrm{Cov}[C^{g_i g_i},C^{\kappa_{a} g_j}]$, and $\mathrm{Cov}[C^{g_i g_i},C^{g_j g_j}]$, where $i,j$ and $a,b$ indicate the photometric and lensing samples respectively. {When the CMB lensing auto-spectrum is included, the covariance is extended to the complete set of lensing cross-terms, $\mathrm{Cov}[C^{\kappa_a\kappa_b},C^{g_i g_i}]$, $\mathrm{Cov}[C^{\kappa_a\kappa_b},C^{\kappa_c g_i}]$, and $\mathrm{Cov}[C^{\kappa_a\kappa_b},C^{\kappa_c\kappa_d}]$, such that the photometric part constitutes a full `$3\times2$pt' analysis of $[C_\ell^{gg},C_\ell^{\kappa g},C_\ell^{\kappa\kappa}]$.}

{All covariance blocks entering the two likelihoods above are computed analytically in the Gaussian approximation \cite{GarciaGarcia19}, as implemented in \texttt{NaMaster}. Each covariance computation requires a fiducial spectrum with a signal and a noise component. For the lensing auto-spectra we combine a fiducial $C_\ell^{\kappa\kappa}$ with the reconstruction-noise curves $N_\ell^{\kappa\kappa}$ provided with the PR4 and ACT DR6 data releases.\footnote{Publicly available at \url{https://github.com/carronj/planck_PR4_lensing} and \url{https://github.com/ACTCollaboration/act_dr6_lenslike}.} The fiducial $gg$ and $g\kappa$ spectra are polynomial fits to the measured band powers (which include shot noise) in the spectroscopic case, while in the photometric case we follow the iterative approaches of Refs.~\citep{Sailer:2024coh, deBelsunce:2025qku}: we first fit cosmological and nuisance parameters using a covariance built from fits to the data, and adopt the resulting best-fit spectra, which include shot noise and magnification bias, as the fiducial spectra of the final covariance. Throughout these computations we assume ACT DR6 and $g_i$ correlations to equal those of \emph{Planck} PR4 and the same $g_i$ redshift bin.  To mitigate numerical artifacts, we compute the fiducial spectra up to $\ell_{\rm max} = 3\cdot N_{\rm side}=6144$ and extrapolate our fits, when needed, beyond $\ell\geq 1000$.}

We do not analyze the cross-correlations between different spectroscopic or photometric galaxy samples (\textit{i.e.}\ $C_\ell^{g_ig_j}$ for $i\neq j$). In principle, these were 
measured in~\cite{Sailer:2024coh,Sailer:2025rks}, and could be used in cosmological analyses. In practice, this would require additional assumptions on the time evolution of EFT parameters, and raise the issue of small-scale sensitivity, as these correlations receive significant contributions from magnification, which is challenging to model within EFT. Since the signal from cross-correlations is significantly smaller than that from auto-correlations, we do not expect that their excision will significantly degrade cosmological constraints~\cite{Sailer:2024coh}.

{Finally, we comment on the treatment of the lensing auto-spectrum ($C_\ell^{\kappa\kappa}$) in combination with the galaxy auto- and lensing cross-spectra ($C_\ell^{gg}$ and $C_\ell^{\kappa g}$). Since these observables probe the same underlying lensing and galaxy fields over overlapping sky areas, they are substantially correlated, and a consistent joint analysis requires the corresponding cross-covariances. 
This corresponds to performing a complete $3\times2$pt analysis for the photometric samples and a joint $[C_\ell^{\kappa g},C_\ell^{\kappa\kappa}]$ analysis for the spectroscopic samples. The results of this joint analysis are presented in Section~\ref{sec:results} (cf.\ the `+ lens' entry of Table~\ref{tab:main2} and the corresponding rows of Table~\ref{tab:mnu}).}

We sample our likelihoods using the Markov Chain Monte Carlo (MCMC) technique, implemented via the \texttt{Montepython} code~\cite{Audren:2012wb,Brinckmann:2018cvx}.
We consider the chains converged when they satisfy the Gelman-Rubin
convergence criterion~\cite{Gelman:1992zz} with $R-1<0.03$ for all sampled parameters. For frequentist tests, we compute best-fit parameters by minimizing the likelihood with a Jeffreys prior, which is exactly equivalent to minimizing the full unmarginalized likelihood. 

\subsection{Binning and Scale Cuts}\label{subsec: binning}

\textbf{Three-dimensional statistics:}
Our choice of binning and scale cuts broadly match those of \paperthree. For the power spectrum, we use a bin width of $\Delta k = 0.005\,\hMpc$ up to $k=0.02\,\hMpc$ and $\Delta k = 0.01\,\hMpc$ afterwards, whilst for the bispectrum multipoles, we use $\Delta k = 0.005\,\hMpc$ up to $k=0.01\,\hMpc$, $\Delta k = 0.01\,\hMpc$ up to $k=0.1\,\hMpc$ and $\Delta k = 0.02\,\hMpc$ beyond. This is chosen to balance retaining BAO information
with a relatively short data vector. As stated above, we use three power spectrum multipoles ($\ell\in\{0,2,4\}$), and two bispectrum multipoles ($\ell\in\{0,2\}$). 

We use the following scale cuts in our baseline analysis of the spectroscopic DESI samples: $k^{P_\ell}_{\rm min}=k^{B_\ell}_{\rm min}=0.005\,\hMpc$, $k^{P_\ell}_{\rm max}=0.2\,\hMpc$, $k^{B_\ell}_{\rm max}=0.16\,\hMpc$. 
When restricting to a tree-level theoretical model (see Section~\ref{sec:theory}), we use $k^{B_\ell}_{\rm max}=0.08\,\hMpc$, following \paperone. These scale cuts have been calibrated with various suites of simulations covering LRG, ELG and QSO-like tracers
~\cite{Ivanov:2019pdj,Chudaykin:2020ghx,Chudaykin:2020hbf,Ivanov:2021kcd,Ivanov:2021zmi,Chudaykin:2022nru,Ivanov:2023qzb,Chudaykin:2024wlw,Bakx:2025pop}; moreover our $k_{\rm max}$ choices for ELG and QSO are more conservative than those suggested by simulations~\cite{Ivanov:2021zmi,Chudaykin:2022nru,Ivanov:2024dgv}.

Our scale cuts ensure that systematic errors due to two-loop corrections to the power spectrum and bispectrum are subdominant. This is particularly important given that these corrections produce systematic biases that grow very quickly and can easily exceed statistical errors even for marginal increases in $\kmax$
\cite{Chudaykin:2024wlw,Bakx:2025pop}. 
This was previously demonstrated in analyses of BOSS-like LRG galaxies in synthetic PTChallenge and Nseries simulations, with
\cite{DAmico:2022osl,Chudaykin:2024wlw} finding that increasing $k_{\rm max}^{P_\ell}$ from $0.20~\hMpc$ to $0.23~\hMpc$ biased 
the $\sigma_8$ inference upwards by $(4-5)\%$ (which exceeds the current error bars). 
Moreover, increasing $k_{\rm max}^{B_\ell}$ from $0.16~\hMpc$ to $0.20~\hMpc$ biases the $\sigma_8$ inference by $3\%$ in the opposite direction \cite{DAmico:2022osl,Bakx:2025pop}. Whilst the ensuing cancellation sources a combined constraint that is not significantly biased, this conclusion is specific to the simulation set and does not imply that the result is reliable.

From a theoretical standpoint, the shifts in $\sigma_8$ induced by $k_{\rm max}^{P_\ell}> 0.20~\hMpc$ or $k_{\rm max}^{B_\ell}\geq 0.17~\hMpc$ imply that the
one-loop EFT prediction is already affected by
higher order corrections and cannot provide a robust description of data. Furthermore, these scale cuts approach the fingers-of-God scale $k_{\rm FoG}\approx 0.25~\hMpc$ \cite{Ivanov:2021fbu} wherein the EFT gradient expansion necessarily breaks down.
The above discussion suggests
that our scale cut choice
$k_{\rm max}^{B_\ell}=0.16~\hMpc$ and $k_{\rm max}^{P_\ell}=0.20~\hMpc$
is valid both from theoretical and data analysis considerations, avoiding systematic errors from higher-order corrections.

\textbf{Projected angular statistics:}
We use the same binning strategy for harmonic space observables as in \cite{Maus2025:joint_3d_lensing_dr1}. In particular, the public DESI projected spectroscopic and photometric data used in this work have a non-uniform $\ell$-space binning with edges
\be 
\begin{split}
& \{10, 20, 44, 79, 124, 178, 243, 317, \\
& \,\,\, 401, 495, 600, 713, 837, 971, 1132\}\,.
\end{split}
\ee 

Since it does not depend on the non-linear velocity field, the EFT in real space has an extended range of applicability compared to the redshift space model. This allows us to use larger $k_{\rm max}$; indeed, real-space power spectrum and bispectrum analyses \cite{Chudaykin:2020hbf} (including those with masked cosmological parameters \cite{Beyond-2pt:2024mqz}) suggest that EFT predictions are adequate up to $\kmax=0.4~\hMpc$. This result is further supported by applications of EFT to simulations at the field level~\cite{Schmittfull:2018yuk,Ivanov:2024dgv,Ivanov:2024hgq,Ivanov:2024xgb}.

Under the Limber-Kaiser approximation~\cite{1953ApJ...117..134L,1992ApJ...388..272K,Kaiser:1996tp}, we can translate a Fourier-space cut to one on harmonic multipoles via
\be 
\ell_{\rm max}=\chi(z_{\rm eff})\kmax - 1/2\,,
\ee 
where we convert comoving distances to redshift in the fiducial cosmological model described above.
For the spectroscopic samples this gives
\be 
\text{spec-z}:\quad 
\ell_{\rm max}=
[329, 533, 701, 863]\,,
\ee 
for BGS, LRG1, LRG2, and LRG3 respectively. 
Likewise, for the photometric samples we find
\be 
\text{photo-z}:\quad 
\ell_{\rm max}=
[240, 385, 498, 635, 763, 857]\,,
\ee 
for pBGS1, pBGS2, pLRG1, pLRG2, pLRG3, and pLRG4.
These are different from the scale cuts
adopted by the DESI collaboration \citep{Maus2025:joint_3d_lensing_dr1},
\be 
\label{eq:desilmax}
\begin{split}
& \text{spec-z}:\quad 
\ell^{\rm official}_{\rm max}=
[400,600,600,600]\,,\\
& \text{photo-z}:\quad 
\ell^{\rm official}_{\rm max}=
[243,401,600,600,600,600]\,,
\end{split}
\ee 
whose main analysis was based on a 
hybrid EFT modeling applied to scales smaller than those possible to model robustly using the perturbative EFT used in this work.
We set $\ell_{\rm min}=20$ for cross-correlations involving \textit{Planck} PR4 lensing, and 
$\ell_{\rm min}=44$ for ACT DR6 cross-correlations, and $\ell_{\rm min}=79$ for the galaxy auto-correlations following~\cite{Sailer:2024coh,Sailer:2025rks,Maus2025:joint_3d_lensing_dr1}.

{Finally, 
for the lensing auto spectrum we use $\ell^{\rm PR4}_{\rm max}=400$ and $\ell^{\rm DR6}_{\rm max}=750$ as
in the standard lensing auto spectrum likelihood~\cite{Carron:2022eyg,ACT:2023kun,ACT:2023dou}, and 
$\ell^{\rm PR4}_{\rm min}=20$ and $\ell^{\rm DR6}_{\rm min}=44$
as in our 
galaxy-lensing cross-correlation analyses.}

\section{Theoretical Model}\label{sec:theory}

\noindent In this section, we describe our theoretical model of the galaxy power spectrum and bispectrum, as well as the cross-correlation with lensing. Our treatment of the former follows \paperone, but is extended to consistently include the one-loop bispectrum, following~\cite{Bakx:2025pop}. Whilst the theoretical description of each observable has already been presented in the literature~\cite[e.g.,][]{Ivanov:2019pdj,Chudaykin:2020aoj,Ivanov:2021kcd,Philcox:2021kcw,Chen:2024vuf,Bakx:2025pop,DAmico:2025zui}, we find it useful to recapitulate the relevant predictions in a unified framework, since we are undertaking a joint analysis of multiple different large-scale structure statistics for the first time. This ensures a unified theoretical framework, which is particularly important for the treatment of nuisance parameters entering the various statistics. In addition, the inclusion of the one-loop bispectrum requires a 
new basis of nuisance parameters, which differs from that used in the previous analyses based on the one-loop power spectrum and tree-level bispectrum.

\subsection{Spectroscopic Power Spectra}
\noindent We utilize the one-loop EFT prediction for the anisotropic galaxy clustering in redshift space, as implemented in the public \texttt{CLASS-PT} code~\cite{Chudaykin:2020aoj}.\footnote{Various other perturbation theory codes exist, including \textsc{pybird}, \textsc{velocileptors}, \textsc{folps}, \textsc{pbj} and \textsc{class one-loop} \citep{DAmico:2020kxu,Chen:2020fxs,Noriega:2022nhf,Moretti:2023drg,Linde:2024uzr}. Several such codes are compared in \citep{Maus:2024sbb}.}
Our model for the power spectrum can be written~\cite{Scoccimarro:1995if,Scoccimarro:1996jy,Scoccimarro:1999kp,Assassi:2014fva,Perko:2016puo,Ivanov:2019pdj}
\be \label{eq:pgg-model}
\begin{split}
& P^{{\rm (s)}}_{\rm gg}(\k)= Z_1^2(\k)P_{\rm lin}(k)+P^{{\rm (s)}}_{\rm 1-loop}[b_1,b_2,b_{\G},b_{\Gamma_3}](\k)\\
& +P^{{\rm (s)}}_{\rm ctr}[b_{\nabla^2\delta},e_1,c_1,c_2,\tilde{c}](\k)+P^{{\rm (s)}}_{\rm stoch}[P_{\rm shot},a_0,a_2](\k)\,,
\end{split}
\ee 
where $P_{\rm lin}$ is the linear power spectrum, 
\be 
Z_1(\k)\equiv b_1+f\mu^2
\ee is the linear EFT kernel (\textit{i.e.}\ Kaiser factor~\cite{Kaiser:1987qv}), depending on the logarithmic growth rate 
\be
f=\frac{d\ln D_+}{d\ln a}\,,
\ee
and the line-of-sight angle $\mu\equiv (\k\cdot \hat{\bf z})/k$ (for line-of-sight direction $ \hat{\bf z}$). In \eqref{eq:pgg-model}, we explicitly write down the dependence on the relevant EFT Wilson coefficients,  which act as ``nuisance parameters'' for the theory. Further details on the one-loop power spectrum corrections are given in Appendix~\ref{sec:one-loopEFT}.

We model the effects of neutrinos on galaxy clustering via the Einstein-de-Sitter approximation,
supplemented with the so-called ``cb'' prescription~\cite{Chudaykin:2019ock,Ivanov:2019hqk,Villaescusa-Navarro:2013pva,VillaescusaNavarro:2012ag,Villaescusa-Navarro:2017mfx}. In particular, we evaluate the EFT predictions with the linear matter power spectrum 
of the baryons and cold dark matter fluid (which depends implicitly on $M_\nu$). In principle, both the non-linear kernels and the logarithmic 
growth factors appearing in the non-linear correction terms depend on the neutrino mass \citep[e.g.,][]{Aviles:2021que,Noriega:2022nhf}; given the current error-bars on $M_\nu$, it is sufficient to evaluate both in a massless neutrino cosmology however \cite{Bayer:2021kwg}. Moreover, for sufficiently small neutrino masses, the structure growth constraints on $M_\nu$ arise primarily from breaking the CMB degeneracy between
the primordial amplitude $A_s$ and $M_\nu$, given measurements of the clustering amplitude 
$\sigma_8$, which acts as a principal component of $A_s$ and $M_\nu$ in the LSS data~\cite{Ivanov:2019hqk}. 
To further test this assumption, we obtained measurements of  
$(\Omega_m,H_0,\sigma_8)$ in two analyses: (a) assuming a single 
massive neutrino with the mass fixed to the solar oscillation floor ($M_\nu=0.06$~eV); (b) assuming a 
perfectly massless neutrino sector. The measurements of $(\Omega_m,H_0,\sigma_8)$ parameters from these two analyses are consistent to within $\lesssim 0.1\sigma$. Whilst this test greatly overestimates the approximations made in our analysis (since we do include neutrinos at linear order), it clearly demonstrates the robustness of our constraints with respect to neutrino modeling in LSS contexts.
We leave
a more detailed
description
of massive neutrinos
on the matter and galaxy clustering observables,
e.g. as in~\cite{LoVerde:2013lta,Senatore:2017hyk,Chiang:2018laa},
for future work.

Compared to previous literature, we use a different basis for the redshift-space counterterms $e_1,c_1,c_2$, in order to match those appearing in the one-loop bispectrum model~\cite{Philcox:2022frc,DAmico:2022ukl,Bakx:2025pop}. Specifically, 
\be 
\begin{split}
    P^{{\rm (s)}}_{\rm ctr}(\k)=&2Z_1^{\rm ctr}(\k)Z_1(\k)P_{\rm lin}(k)\,,\\
Z_1^{\rm ctr}(\k) = & \frac{k^2}{k_{\rm NL}^2}\bigg(-b_{\nabla^2\delta}+\left(e_1 -\frac{1}{2}c_1 f\right)f\mu^2 -\frac{1}{2}c_2 f^2\mu^4\bigg),
\end{split}
\ee 
where $k_{\rm NL}$ is a non-linear scale, which we fix to $0.45~\hMpc$ as a fiducial choice following~\cite{Philcox:2021kcw}. In the absence of the one-loop bispectrum, $c_1$ and $e_1$ would be completely degenerate, thus they were replaced by a single parameter in our previous analyses \cite{desi1,desi2}. Since the one-loop bispectrum breaks this the degeneracy, it is important to keep both parameters to ensure a consistent theoretical model.
Finally, we include a higher-order counterterm $\tilde{c}$ as in previous works:
\be 
Z_1^{\rm ctr}(\k)\supset-\frac{\tilde{c}}{2}
\frac{k^4}{k_{\rm NL}^4}
f^4\mu^4  Z_1(\k)\,.
\ee
Though formally next-order in perturbation theory, this is introduced to account
for the large velocity dispersion (\textit{i.e.}\ ``fingers-of-God''~\cite{Jackson:2008yv}) of 
massive galaxies such as the LRG sample. The importance of this counterterm for
unbiased cosmological and EFT parameters measurements has been extensively discussed in the literature \cite[e.g,][]{Ivanov:2019pdj,Chudaykin:2020hbf,Ivanov:2024xgb,desi1} (see also~\cite{Lewandowski:2015ziq,Taule:2023izt,Chen:2025aqr} for the motivations for this counterterm  from 
other perspectives).

The stochasticity contribution includes both scale-independent and scale-dependent parts,
\be 
\label{eq:stochP}
P^{{\rm (s)}}_{\rm stoch}(\k) = \frac{1}{\bar n}\left(P_{\rm shot}+a_0\frac{k^2}{k^2_{\rm NL}} +a_2\mu^2 \frac{k^2}{k^2_{\rm NL}} \right)~\,,
\ee 
for mean number density $\bar{n}$, where we have subtracted the Poisson contribution, such that $P_{\rm shot}$ has a prior-mean of zero.

To account for non-perturbative long-wavelength displacements, we apply infrared (IR) resummation using the approach based on time-sliced perturbation theory~\cite{Blas:2015qsi,Blas:2016sfa,Ivanov:2018gjr,Vasudevan:2019ewf}
(see also~\cite{Senatore:2014via,Baldauf:2015xfa,Vlah:2015zda}), and account for coordinate distortions as described in~\cite{Chudaykin:2020aoj}.
Specifically, we take into account the following relationship between the true wavenumbers and angles 
$(k_{\rm true},\mu_{\rm true})$
and the observed ones ($k_{\rm obs},\mu_{\rm obs}$) ``reconstructed'' under a fiducial cosmology: 
\be
\begin{split}\label{eq:AP_k_mu}
   k_{\rm true}^2&=k_{\rm obs}^2\left[
   % \l
   q_\parallel^{-2}
   % \frac{H_\true}{H_\fid}
   % \r^2
   \mu_{\rm obs}^2+
   q_\perp^{-2}
   (1-\mu_{\rm obs}^2)\right]\,,\\
   \mu_{\rm true}^2&=
   q_\parallel^{-2}
   % \l\frac{H_\true}{H_\fid}\r^2
   \mu_{\rm obs}^2\left[
   q_\parallel^{-2}
   % \l\frac{H_\true}{H_\fid}\r^2
   \mu_{\rm obs}^2+
   q_\perp^{-2}
   % \l\frac{D_{A,\fid}}{D_{A,\true}}\r^2
   (1-\mu_{\rm obs}^2)\right]^{-1}\,,
\end{split}
\ee
where the Alcock-Paczynski parameters are defined as \citep{Alcock:1979mp}
\be
q_\parallel = \frac{H_{\rm fid}(z)}{H_{\rm true}(z)}
\frac{H_{0,\rm true}}{H_{0,\rm fid}}
\,,\quad q_\perp = \frac{D_{{\rm true}, A}(z)}{D_{{\rm fid}, A}(z)}
\frac{H_{0,\rm true}}{H_{0,\rm fid}}\,.
\ee
Here $D_A$ and $H$ denote the angular diameter distance 
and the Hubble parameter, and the factor $H_{0,\mathrm{true}}/H_{0,\mathrm{fid}}$ appears if one explicitly uses $\Mpch$ units.
To fit the observations,  we compute the power spectrum multipoles as
\begin{widetext}
    \be 
P_\ell(k_{\rm obs}) = \frac{(2\ell+1)}{2}\int_{-1}^1 d\mu_{\rm obs} ~P^{{\rm (s)}}_{\rm gg}(k_{\rm true}[k_{\rm obs},\mu_{\rm obs}],\mu_{\rm true}[k_{\rm obs},\mu_{\rm obs}])\mathcal{L}_\ell(\mu_{\rm obs})\,,
\ee 
\end{widetext}
where $\mathcal{L}_\ell$ is a Legendre polynomial of degree $\ell$. As detailed in \paperone, we compute the power spectrum on a fine $k$-grid, and then multiply by a rectangular theory-matrix, which includes corrections for discreteness effects, integral constraints, and wide-angle effects. 

\subsection{Spectroscopic Bispectra at Tree-Level}
\noindent Next, we discuss the three-dimensional bispectrum, starting from the tree-level model. The EFT computation used herein was presented in~\cite{Ivanov:2021kcd,Ivanov:2023qzb} (see also~\cite{Scoccimarro:1997st,Scoccimarro:1999ed,Sefusatti:2006pa,Sefusatti:2007ih,Sefusatti:2009qh,Oddo:2019run,Oddo:2019run,Rizzo:2022lmh}), and can be written
\be 
\begin{split}
&B_{\rm ggg}^{\rm tree~(s)}=
% 2Z_2(\k_1,\k_2)Z_1(\k_1)Z_1(\k_2)P_{\rm lin}(k_1)P_{\rm  lin}(k_2)\\
B^{\rm tree}_{\rm det}[b_1,b_2,b_{\G}]
+B^{\rm tree~(s)}_{\rm stoch}[A_{\rm shot}]
\\
&+B^{\rm tree~(s)}_{\rm mixed}[P_{\rm shot},B_{\rm shot}]+
B^{\rm tree~(s)}_{\rm ctr}[\tilde{c}_1,b_1,b_2,b_{\G}].
\end{split}
\ee 
Here, $B^{\rm tree~(s)}_{\rm det}$
is the deterministic 
contribution:
\be 
\begin{split}
B^{\rm tree~(s)}_{\rm det}=&2Z_2(\k_1,\k_2)Z_1(\k_1)Z_1(\k_2)P_{\rm lin}(k_1)P_{\rm  lin}(k_2)\\
&+\text{2 cyc.}\,,
\end{split}
\ee 
which is defined with the following quadratic kernel:
\begin{widetext}
\bseq 
\begin{align}
% &Z_1(\k)  = b_1+f\mu^2\,,\\
&Z_2(\k_1,\k_2)  =\frac{b_2}{2}+b_{\mathcal{G}_2}\left(\frac{(\k_1\cdot \k_2)^2}{k_1^2k_2^2}-1\right)
+b_1 F_2(\k_1,\k_2) +f\mu^2 G_2(\k_1,\k_2)\notag \\
& \qquad\qquad\quad~~
+\frac{f\mu k}{2}\left(\frac{\mu_1}{k_1}(b_1+f\mu_2^2)+
\frac{\mu_2}{k_2}(b_1+f\mu_1^2)
\right)
\,,\quad \text{where}\\
& F_2(\k_1,\k_2)=\frac{5}{7}
+\frac{1}{2}\left(
\frac{(\k_1\cdot \k_2)}{k_1^2}
+\frac{(\k_1\cdot \k_2)}{k_2^2}
\right)+\frac{2}{7}\frac{(\k_1\cdot \k_2)^2}{k_1^2 k_2^2}\,,\\
& G_2(\k_1,\k_2)=\frac{3}{7}
+\frac{1}{2}\left(
\frac{(\k_1\cdot \k_2)}{k_1^2}
+\frac{(\k_1\cdot \k_2)}{k_2^2}
\right)+\frac{4}{7}\frac{(\k_1\cdot \k_2)^2}{k_1^2 k_2^2}\,.
\end{align} 
\eseq
\end{widetext}
$B^{\rm tree~(s)}_{\rm stoch}$
captures the three-point stochasticity:
\be \label{eq:Bstochtree}
B^{\rm tree~(s)}_{\rm stoch}=\frac{A_{\rm shot}}{\bar n^2}\,,
\ee 
while $B_{\rm mixed}$ is the mixed deterministic-stochastic contribution, with a structure similar to the Poissonian prediction $\propto P/\bar n$~\cite{1980lssu.book.....P,Baldauf:2016sjb}: 
\be 
\label{eq:Bmixtree}
B^{\rm tree~(s)}_{\rm mixed}= \frac{
b_1 B_{\rm shot}+
f\mu^2 P_{\rm shot} }{\bar n}P_{\rm lin}(k_1)Z_1(\k_1)+\text{cyc.}
% \,.
\ee 
Finally, $B_{\rm ctr}$ is a phenomenological term that acts as a proxy
for the one-loop counterterms, whose contribution is important 
even on large scales. This is obtained by adding a fingers-of-God contribution to the 
linear theory kernel, \textit{i.e.} using 
\be 
Z^{\rm FoG}_1(\k)=(b_1+f\mu^2) -\tilde{c}_1\mu^2\frac{k^2}{k_{\rm NL}^2}
\ee 
instead of $Z_1(\k)$ in the deterministic contribution, with $k_{\rm NL}=0.45~\hMpc$, as before. This scale is somewhat larger than the scale of fingers-of-God $k_{\rm FoG}=0.25~\hMpc$~\cite{Ivanov:2021fbu}, which will be reflected in our priors on $\tilde{c}_1$. Note that we include the $\tilde{c}_1$ term only when performing tree-level bispectrum analyses, since our one-loop likelihood has EFT counterterms that consistently account for all fingers-of-God effects at order $\mathcal{O}(k^2/k_{\rm FoG}^2)$.

We account for tree-level IR resummation in the bispectrum using the formalism of time-sliced perturbation theory~\cite{Ivanov:2018gjr} (see also \cite{Chen:2024pyp}).
To compare theory and data, we compute the multipoles via the continuous definition~\citep{Scoccimarro:1999ed,Scoccimarro:2015bla,Ivanov:2023qzb},
\be
\begin{split}\label{eq:B-mult-no-AP}
    &B_\ell^{\rm ideal}(k_1,k_2,k_3) = (2\ell+1)\int_0^{2\pi}\frac{d\phi}{2\pi}\int_{-1}^{1}\frac{d\mu_1}{2}\\
    &\times B(k_1,k_2,k_3,\mu_1,\mu_2,\mu_3)\mathcal{L}_\ell(\mu_1)\,,
\end{split}
\ee
where we order the wavenumbers in triangle configurations as $k_1\geq k_2\geq k_3$, such that the the angular integration is performed with respect to the cosine between the line-of-sight and the largest wavevector.  The angles $\mu_2,\mu_3$ are defined by the polar and azimuthal angles $\mu$ and $\phi$ through $\mu_1 = \mu$,
\be\label{eq: mu-i-angles}
\mu_2 = \mu\cos\zeta-\chi\sin\zeta, \qquad \mu_3 = -\frac{k_1}{k_3}\mu_1-\frac{k_2}{k_3}\mu_2
\ee
where $\chi \equiv \sqrt{1-\mu^2}\cos\phi$, $\cos\zeta\equiv\hat{\k}_1\cdot\hat{\k}_2$ \citep[e.g.,][]{Scoccimarro:2015bla,Philcox:2022frc,Bakx:2025pop}.

As discussed in \citep{Philcox:2022frc}, it is important to account for the effects of finite bin widths in the bispectrum. This is achieved by computing the binned statistic following the approach of~\cite{Ivanov:2021kcd,Ivanov:2023qzb}, recently streamlined in~\cite{Bakx:2025pop}:
\begin{widetext}
\be
\begin{split}
\label{eq:B-mult-bin-w-AP}
    B_\ell(\bar{k}_1,\bar{k}_2,\bar{k}_3)&=\frac{1}{\mathcal{N}(\bar{k}_1,\bar{k}_2,\bar{k}_3)}\frac{2\ell+1}{\alpha_\parallel^2\alpha_\perp^4}\prod_{i=1}^3\left[\int_{\bar{k}_i-\Delta k_i/2}^{\bar{k}_i+\Delta k_i/2}k_i d k_i\right]\int_0^{2\pi}\frac{d \phi}{2\pi}\int_{-1}^{1}\frac{d\mu_1} {2} \\
    &  \mathcal{I}(k_1,k_2,k_3,\mu,\phi)\,B(k_1',k_2',k_3',\mu_1',\mu_2',\mu_3')\mathcal{L}_\ell(\mu_1)\,,\quad \text{where}\\
 \mathcal{N}(\bar{k}_1,\bar{k}_2,\bar{k}_3)   &     = \prod_{i=1}^3\left[\int_{\bar{k}_i-\Delta k/2}^{\bar{k}_i+\Delta k_i/2}k_i d k_i\right] \int_0^{2\pi}\frac{d\phi}{2\pi}\int_{-1}^{1}\frac{d\mu_1}{2}\mathcal{I}(k_1,k_2,k_3,\mu_1,\phi)\,.
    \end{split}
\ee
\end{widetext}
$k_i\in[\bar{k}_i-\Delta k_i/2,\bar{k}_i+\Delta k_i/2)$ are linearly spaced bins, whose centers satisfy the triangle conditions, \textit{i.e.}\ $|\bar{k}_1-\bar{k}_2|\leq \bar{k}_3\leq \bar{k}_1+\bar{k}_2$, and we restrict to $\bar{k}_1\geq \bar{k}_2\geq \bar{k}_3$ as before. Here, $\mathcal{I}(k_1,k_2,k_3,\mu,\phi)$ is a weighting function equal to one if $\k_1,\k_2,\k_3$ obey the triangle conditions and zero otherwise. In \cite{Ivanov:2021kcd,Ivanov:2023qzb}, the $\mathcal{I}$ function was not included, and weights were introduced to account for the differences between the continuous integral of \eqref{eq:B-mult-bin-w-AP} and the discrete computation on a grid of wavenumbers. When $\mathcal{I}$ is included, these weights are equal to one with negligibly small corrections (for $B_0$ and $B_2$), thus do not need to be included \cite{Bakx:2025pop}.

Finally, we account for coordinate distortions in the bispectrum by treating 
all observed momenta and angles in \eqref{eq:B-mult-bin-w-AP} as functions of the true
momenta and angles as in \eqref{eq:AP_k_mu} (see see~\cite{Ivanov:2021kcd,Ivanov:2023qzb,Bakx:2025pop}
for further details).

\subsection{Spectroscopic Bispectra at One-Loop}
\noindent At one-loop order, the redshift-space galaxy bispectrum is given by
\begin{widetext}
\be 
\begin{split}
B_{\rm 1-loop}&= B^{\rm tree~(s)}_{\rm det}[b_1,b_2,b_{\G}] +
B^{\rm 1-loop~(s)}_{\rm det}[b_1,b_2,b_{\G},b_{\Gamma_3},b_3,\gamma_2^\times,\gamma_3,\gamma_{21}^\times,\gamma_{211},\gamma_{22},\gamma_{31}] 
\\
&+ 
B^{\rm ctr~(s)~I}[b_1,b_{\nabla^2\delta},b_{\nabla^2\delta^2},b_{\nabla^2\G},b_{(\nabla \delta)^2},b_{(\nabla t)^2},e_1,e_4,e_5,c_1,...,c_4,c_6,c_7] +B^{\rm ctr~(s)~II}[b_1,b_2,b_{\G},b_{\nabla^2\delta},e_1,c_1,c_2] \\&
+B_{\rm mixed}^{\rm tree~(s)}[b_1,P_{\rm shot},B_{\rm shot}]+B_{\rm mixed}^{\rm 1-loop~(s)}[b_1,b_2,b_{\G},b_{\Gamma_3}P_{\rm shot},B_{\rm shot},d_2,d_{\G},d_{\Gamma_3}]\\ &
+B_{\rm mixed}^{\rm ctr~(s)~I}[b_1,a_0,a_3,...,a_5,a_7,...,a_{12}]
+B_{\rm mixed}^{\rm ctr~(s)~II}[b_1,P_{\rm shot},B_{\rm shot},b_{\nabla^2\delta},e_1,c_1,c_2,\tilde{c}] \\
&+B_{\rm stoch}[A_{\rm shot},a_1,a_6]\,.
\end{split} 
\ee 
\end{widetext}
\cite{Bakx:2025pop} (see also \cite{Eggemeier:2018qae,Eggemeier:2021cam,Philcox:2022frc,DAmico:2022ukl}). Here, $B^{\rm tree~(s)}_{\rm det}$ and $B^{\rm 1-loop~(s)}_{\rm det}$ refer to mode-coupling contributions, $B^{\rm ctr~(s)~I,II}$ are deterministic higher-derivative corrections, $B^{\rm tree~(s)}_{\rm mixed}$ and $B^{\rm 1-loop~(s)}_{\rm mixed}$ are the mixed deterministic-stochastic terms at leading- and next-to-leading order respectively, $B^{\rm ctr~(s)~I,II}_{\rm mixed}$ are mixed terms involving higher derivative deterministic and stochastic corrections, 
while $B_{\rm stoch}$ is a purely stochastic contribution, which now includes scale-dependent bispectrum stochasticity.
The explicit expressions for these terms are given in Appendix~\ref{sec:one-loopEFT}. 
{We caution that the mixed counterterm--stochastic
contribution $B^{\rm ctr\,(s)\,II}_{\rm mixed}$, being second order in the
EFT parameters, is treated via a linearized approximation around fiducial
parameter values (see Appendix~\ref{sec:one-loopEFT}); since our baseline
analysis sets the fiducial values equal to the prior means, this
contribution vanishes identically and is effectively excluded from the
theory prediction. This is justified as $B^{\rm ctr\,(s)\,II}_{\rm mixed}$
is numerically small at our baseline scale cut
$k^{B_\ell}_{\rm max}=0.16~\hMpc$, and our parameter constraints show no
appreciable sensitivity to it, as detailed in
Appendix~\ref{sec:one-loopEFT}.}

Together with the one-loop power spectrum prediction, this involves 45 free parameters, which parametrize all relevant baryonic and galaxy formation effects at one-loop order. These are treated as independent for each spectroscopic DESI redshift bin, and, upon marginalization, effectively induce a smooth correlated covariance in the bispectrum likelihood. To account for long-wavelength displacements, we use the efficient one-loop IR resummation technique outlined in~\cite{Ivanov:2018gjr} and implemented in \cite{Philcox:2022frc}. In particular, we apply the full anisotropic damping to the tree-level bispectrum expressions and use an isotropic damping approximation for the one-loop components. We treat the binning effects in the one-loop bispectrum via \eqref{eq:B-mult-bin-w-AP}, as in the tree-level analysis.

Due to the large number of convolution integrals contained within $B^{\rm 1-loop~(s)}_{\rm det}$ and the many triangle configurations analyzed, computing the one-loop corrections to the bispectrum is much more expensive than for the power spectrum, even with analytic methods such as \textsc{fftlog}~\cite{Simonovic:2017mhp}. This prohibits full cosmological studies, where one wishes to scan over non-linear cosmological parameters, which requires many evaluations of the bispectrum model. To ameliorate this, one can decompose the bispectrum into a suitable cosmology-independent basis, such that the expensive pieces of the calculation can be pre-computed and the cosmology-dependent result can be assembled as a matrix multiplication. Here, we use the \textsc{cobra} technique of \cite{Bakx:2024zgu} (see also \citep{DAmico:2022ukl,Anastasiou:2022udy}), which employs a singular-value decomposition of the input IR-resumed tree-level power spectrum~\cite{Blas:2016sfa} (see also \cite{Philcox:2020zyp,Chen:2024pyp}), allowing the full cosmological parameter prior to be expressed using {$N_{\rm cobra}=6$} basis functions. After pre-computing the $N_{\rm cobra}^3$ loop integrals using \textsc{fftlog} \citep{Simonovic:2017mhp} and applying the bin integration, we can compute the full one-loop bispectrum in less than a second. To implement the coordinate-distortion effect
on the one-loop corrections, we perform a linear-order expansion in $q_{\parallel}, q_{\perp}$, pre-computing all terms. We refer the reader to  \cite{Bakx:2025pop} for further details on this procedure.

\subsection{Projected Galaxy Clustering \& CMB Lensing Cross-Correlations}
\noindent To predict the angular power spectrum of galaxies and their cross-correlation with CMB lensing, we use the one-loop EFT models implemented in \texttt{CLASS-PT}~\cite{Chudaykin:2020aoj}. 
Specifically, the galaxy auto-spectrum depends on the real-space galaxy-galaxy power spectrum:
\be 
\begin{split}
& P^{\rm (r)}_{\rm gg}(k)= b_1^2P_{\rm lin}(k)+P^{\rm (r)}_{\rm 1-loop,gg}[b_1,b_2,b_{\G},b_{\Gamma_3}](k)\\
& +P^{\rm (r)}_{\rm ctr,gg}[b_{\nabla^2\delta}](k)+P^{\rm (r)}_{\rm stoch,gg}[P_{\rm shot},a_0](k)\,,
\end{split}
\ee 
where the superscript (r) indicates real space. The real-space model can be obtained from the redshift-space formulae defined above by setting all redshift-space quantities to zero, \textit{i.e.} with $f=a_2=0$. In this limit, the counterterm and stochastic contributions simplify significantly:
\be 
\begin{split}
& P^{\rm (r)}_{\rm ctr,gg}(k)=-2b_{\nabla^2\delta}\frac{k^2}{k_{\rm NL}^2} b_1 P_{\rm lin}(k)\,,\\
& P^{\rm (r)}_{\rm stoch,gg}(k)=\frac{1}{\bar n}\left(P_{\rm shot}+a_0\frac{k^2}{k_{\rm NL}^2}\right)~\,,
\end{split}
\ee 
where $b_{\nabla^2\delta}\equiv b'_{\nabla^2\delta}+b_1 c_s$ is an exact
reparametrization of the usual higher-derivative bias $b'_{\nabla^2\delta}$
and the dark matter sound speed $c_s/k^2_{\rm NL}$ (the latter containing
corrections due to baryonic feedback~\cite{Lewandowski:2014rca}).\footnote{
{We stress that this redefinition does not identify the
counterterms of different observables: $c_s$ remains an independent free
parameter, entering the galaxy--matter cross-spectrum below, so that the
pair $\{b_{\nabla^2\delta},c_s\}$ spans the full space of $k^2P_{\rm lin}$
counterterms allowed by the EFT for the auto- and cross-spectra.
Furthermore, the redshift-space and bispectrum counterterms, which
renormalize different loop integrals, are kept as independent parameters
throughout (cf.\ Sections~3.1 and~3.3).}}

% where $b_{\nabla^2\delta}$ is a higher-derivative bias, which absorbs the dark matter sound-speed term $c_s/k_{\rm NL}^2$ (which itself contains corrections due to baryonic 
% feedback~\cite{Lewandowski:2014rca}). 

To model the cross-correlation of galaxies and both CMB lensing and magnification bias, we require the cross-spectrum of galaxies and matter at one-loop order:
\be 
\begin{split}
& P^{\rm (r)}_{\rm gm}(k)= b_1P_{\rm lin}(k)+P^{\rm (r)}_{\rm 1-loop,gm}[b_1,b_2,b_{\G},b_{\Gamma_3}](k)\\
& +P^{\rm (r)}_{\rm ctr,gm}[b_{\nabla^2\delta},c_s](k)+P_{\rm stoch,gm}^{\rm (r)}[a^{\rm gm}_0](k)\,.
\end{split}
\ee 
The deterministic one-loop piece is given in~\cite{Chudaykin:2020aoj,Chudaykin:2020hbf}, and the counterterm contribution is 
\be 
P^{\rm (r)}_{\rm ctr,gm}(k)=-(b_{\nabla^2\delta}+c_s b_1)\frac{k^2}{k_{\rm NL}^2} P_{\rm lin}(k)\,.
\ee 
where we use a different convention to previous works in order to match the one-loop bispectrum counterterm parametrization~\cite{Bakx:2025pop}. Note that the degeneracy between $b_{\nabla^2\delta}$ and $c_s$ is broken when we analyze both the galaxy cross- and auto-spectrum.
The stochastic contribution is obtained by cross-correlating the matter and galaxy stochastic density components. Due to mass and momentum conservation, the latter scales like $k^2$ in the $k\to 0$ limit~\cite{Baumann:2010tm,Carrasco:2012cv,Ivanov:2022mrd}, while the former goes as $k^0$~\cite{Assassi:2014fva,Desjacques:2016bnm}, implying that the cross-spectrum 
in the $k\to 0$ limit
must be given by
\be 
P_{\rm stoch,gm}^{\rm (r)}= a^{\rm gm}_0\frac{k^2}{k_{\rm NL}^2}\frac{1}{\bar n}\,,
\ee 
where $a^{\rm gm}_0$ is an order-one dimensionless constant. 

An important contribution to the galaxy cross-spectrum is sourced by magnification and its cross-correlation with CMB lensing. To describe these effects, we require a
theoretical model for matter clustering, \textit{i.e.}\ the real-space matter power spectrum $P_{mm}$.
Due to the wide redshift-extent of the magnification and lensing kernels, this necessitates modeling $P_{mm}$ down to small scales, where the EFT description does not provide a well-controlled approximation. Even on large-scales, the contributions depend on the time-evolution of $P_{mm}$, which, in EFT models, is not known \textit{a priori}. 
To ameliorate this, we adopt a phenomenological model for the matter power spectrum,
which acts as a Pad\'e resummed version of the one-loop EFT prediction, supplemented with higher order counterterms~\cite{Chen:2024vuf} (see also \citep{Philcox:2020rpe}
for an alternative approach):
\be \label{eq: Pmm-pade}
\begin{split}
P^{\rm (r)}_{\rm mm}(k,z)= & P_{\rm lin}(k,z)+P^{\rm (r)}_{\rm 1-loop,mm}(k,z)\\
& +\frac{2\alpha(z) k^2 P_{\rm lin}(k,z)}{1+\beta(z) k^4}\,,
\end{split}
\ee 
where we have restored the explicit time-dependence of the counterterm coefficients. In the $k\to 0$ limit this reduces to the usual one-loop EFT prediction \cite[e.g.,][]{Carrasco:2012cv,Chudaykin:2020aoj}, 
\be 
P^{\rm (r)}_{\rm mm}(k,z)= P_{\rm lin}(k,z)+P^{\rm (r)}_{\rm 1-loop,mm}(k,z) + 2\alpha(z)  k^2 P_{\rm lin}(k)\,.
\ee 
When computing the one-loop term, we use the Einstein-de-Sitter approximation to compute the time-dependence (ignoring IR resummation effects, which violate this scaling, though are largely irrelevant for projected statistics): 
\be 
P^{\rm (r)}_{\rm 1-loop,mm}(k,z)=D^4_+(z)P^{\rm (r)}_{\rm 1-loop,mm}(k,z=0)\,
\ee 
where $D_+(z)$ is the growth factor. We fit the parameters $\alpha(z)$ and $\beta(z)$ appearing in \eqref{eq: Pmm-pade} to the \texttt{HMcode} output~\cite{Mead:2015yca}, using a redshift grid spanning $z\in [0,1]$. \cite{Chen:2024vuf} found that a simple two-parameter model for each fitting parameter: 
\be 
f=f_0\times [D_+(z)]^{N_{f}}\,,
\ee
for $f\in\{\alpha,\beta\}$, provides a $\lesssim 3\%$ accurate match to \texttt{HMcode} matter power spectrum on non-linear scales $0.4 \leq k/(\hMpc)\leq 1$ across this redshift range -- this will be adopted henceforth to describe the magnification bias contributions. {The fits for $\alpha(z)$ and $\beta(z)$ were obtained at
the fiducial  \textit{Planck} 2018 best-fit
$\Lambda$CDM cosmology. Whilst the accuracy of the
fitting functions cannot be directly validated in the $w_0w_a$CDM and
massive-neutrino extensions, for which no calibrated simulation suite is
available (a limitation shared by \texttt{HMcode} itself), the model is
expected to retain its accuracy over the relevant range of scales, since it
is dominated by the linear and one-loop contributions, which are recomputed
exactly for each sampled cosmology, with the calibrated terms providing
only a subdominant small-scale correction. Moreover, the small-scale
behavior of $P_{mm}$ affects our observables only through the magnification
contributions, which we demonstrate to be negligible for our constraints
(cf.\ Section~\ref{sec:data}, where we show that marginalizing over the magnification
bias parameters leaves the cosmological posteriors unchanged).}

Whilst our simple model clearly does not fully capture the complex physics of dark matter clustering on small scales, this is not a concern for our analysis, given that our main goal to have a rough model that does not bias the extraction of cosmological parameters from the angular clustering observables. 
We further note that the galaxy magnification is not used to model cross-spectra between different photometric samples since we do not include the galaxy cross-spectra
($C_{\ell}^{g_i g_j}$ for $i\neq j$) in our analysis. These were studied in detail
in~\cite{Sailer:2024coh,Sailer:2025rks}.

Under the Limber-Kaiser approximation~\cite{1953ApJ...117..134L,1992ApJ...388..272K,Kaiser:1996tp}, we can compute the projected clustering 
statistics from their three-dimensional real-space counterparts, by first forming the building blocks
\be
\label{eq:Cxy}
\hat{C}_{\ell}^{XY}=\int d\chi \frac{W^X(\chi)W^Y(\chi)}{\chi^2}P_{XY}\left(k=\frac{\ell+1/2}{\chi},z(\chi)\right)\,,
\ee
where $\chi$ is the comoving distance, $W^{X,Y}$ are the projection  
kernels,
\be 
\begin{split}
& W^g(\chi) =\frac{1}{\mathcal{N}} H(z)\frac{dN}{dz} ~\,,\\
& W^\kappa(\chi) = \frac{3}{2}\Omega_m H_0^2 (1+z)\frac{\chi(\chi_*-\chi)}{\chi_*}\,,\\
& W^\mu(\chi)= (5s_\mu -2)\frac{3}{2}\Omega_m H_0^2 (1+z) \\
& ~~~~~~~~~\times \int_{\chi}^{\infty}d\chi' \frac{\chi(\chi'-\chi)}{\chi'} W^g(\chi)\,,
\end{split}
\ee 
for galaxies, CMB lensing, and magnification respectively. Here, $\frac{dN}{dz}$ is the redshift
distribution of galaxies with normalization factor
\be 
\mathcal{N} = \int d\chi H(z(\chi)) \frac{dN}{dz} (\chi)\,,
\ee
$H(z)$ is the Hubble parameter at redshift $z$, $\chi_*$ is the comoving distance to the last scattering surface, and $s_\mu$
is the magnification bias amplitude. 
The full two-dimensional power spectra ${C}_\ell^{\kappa g},{C}_\ell^{g g}$ required for our analysis can be obtained using the building blocks of \eqref{eq:Cxy} as follows:
\be
\begin{split}
& C_{\ell}^{\kappa g}  = \hat{C}_{\ell}^{\kappa g} + \hat{C}_{\ell}^{\kappa \mu}\,,\\
& C_{\ell}^{g g}  = \hat{C}_{\ell}^{g g} + 2\hat{C}_{\ell}^{g \mu}+ \hat{C}_{\ell}^{\mu \mu}\,,
\end{split} 
\ee
with the identification 
\be\begin{split}
& P_{\kappa g}=P_{\mu g}=P_{\rm gm}\,,\quad P_{\kappa\mu} =  P_{\kappa\kappa} = P_{\mu\mu} \equiv  P_{\rm mm}\,.
 \end{split}
\ee
In practice, we can adopt the effective redshift approximation for the galaxy auto- and cross-correlations, due to a narrow width of the 
clustering kernel, which corresponds to the replacement
\be 
\begin{split}
& P_{\rm gg/gm}\left(k=\frac{\ell+1/2}{\chi},z(\chi)\right)\\
& ~~~~~~~~\approx P_{\rm gg/gm}\left(k=\frac{\ell+1/2}{\chi},z_{\rm eff}\right)\,,
\end{split}
\ee 
with $z_{\rm eff} = \int {\rm d}\chi\,z(\chi)\,(W^g(\chi)/\chi)^2 / \int {\rm d}\chi \,(W^g(\chi)/\chi)^2$.
This approximation is accurate to sub-percent 
precision for the DESI galaxy samples~\cite{Sailer:2024coh}. 

As discussed above, we do not include the cross-spectra between different galaxy samples in this work, whose modeling requires violation of the effective redshift assumption.\footnote{In essence, we assume that $C_\ell^{g_ig_j}(i\neq j)=0$. Note that we do not assume the different photometric spectra to be uncorrelated, \textit{i.e.} we allow for non-zero Cov$[C_\ell^{g_ig_i},C_\ell^{g_jg_j}]$.} 
This approximation is justified as long as the redshift overlap 
between the galaxy samples
in question is negligible. For the spectroscopic samples, this is true by construction, whilst the overlap for photometric samples occurs only at the tails  
of redshift distributions~\cite{Maus2025:joint_3d_lensing_dr1}, resulting in a 
suppressed amplitude of the cross-correlation signal~\cite{Sailer:2024coh}.

To model the effect of massive neutrinos on projected observables, we adopt the ``cb''+EdS approximation, as for the three-dimensional statistics, which is exact at linear order.
In principle, one should use the $P_{\rm cb,m}$ spectrum
to model the galaxy-CMB lensing cross-correlations since the lensing observables probe the total matter fluctuations including those of massive neutrinos, whilst the galaxies are sensitive only to dark-matter and baryon fluid. However, given the negligible difference with respect to $P_{\rm cb,cb}$ 
\be
P_{\rm cb,cb}/P_{\rm cb,m}-1\approx f_\nu\simeq 0.76\%\frac{M_\nu}{0.1~\text{eV}}\frac{0.14}{\omega_m}\,,
\ee
as well as the weak sensitivity of LSS neutrino mass constraints to the modeling of free-streaming, we can safely use $P_{\rm cb,cb}$
as an input in our computations of CMB-lensing cross-correlations. 

\subsection{Priors}

%\begin{widetext}
    \begin{table*}[!htb] 
    \centering
    \begin{tabular}{|llll|}
    \hline
    Type & Parameter & Default & Prior   \\  
    \hline
    \textbf{spectroscopic} 
    & $b_1 \sigma_8(z)$ &  & $\mathcal{U}[0, 3]$  \\  \textbf{nuisance} 
    & $b_2 \sigma_8^2(z)$ &  & $\mathcal{N}[0, 5^2]$  \\  
    (sampled) 
     & $b_{\mathcal{G}_2} \sigma_8^2(z)$ &  & $\mathcal{N}[0, 5^2]$  \\ 
    \hline
    \textbf{spec-z} & $b_{\Gamma_3}\,A_{\rm AP}\,A_{\rm amp}^2$ &  & $\mathcal{N}\left(\frac{23}{42}(b_1-1),1^2\right)$  \\
    \textbf{nuisance} & $b_{\nabla^2\delta}\,A_{\rm AP}\,A_{\rm amp}$ &  & $\mathcal{N}(0,6^2)$  \\
   (analytically  & $e_1\,A_{\rm AP}\,A_{\rm amp}$ &  & $\mathcal{N}(0,6^2)$  \\
   marginalized)
   & $c_1\,A_{\rm AP}\,A_{\rm amp}$ &  & $\mathcal{N}(0,6^2)$  \\
    & $c_2\,A_{\rm AP}\,A_{\rm amp}$ &  & $\mathcal{N}(0,6^2)$  \\
    & $\tilde{c}\,A_{\rm AP}\,A_{\rm amp}$ &  & $\mathcal{N}(16,16^2)$  \\
    & $\tilde{c}_1\,A_{\rm AP}\,A_{\rm amp}$ &  & $ \mathcal{N}(0,11^2)$  \\
    & $P_{\rm shot}A_{\rm AP}$ &  & $\mathcal{N}(0,1^2)$  \\
    & $(a_{0},a_2)A_{\rm AP}$ &  & $\mathcal{N}(0,1^2)$  \\
    & $a_0^{\rm gm}$ &  & $\mathcal{N}(0,1^2)$  \\
    & $B_{\rm shot}\,A^2_{\rm AP}\,A_{\rm amp}$, $A_{\rm shot}A_{\rm AP}^2$ &  & $\mathcal{N}(0,1^2)$  \\
    % & $A_{\rm shot}$ &  & $\mathcal{N}(0,1^2)$ &---\\
    & $(b_3,\gamma_{\mathcal{O}_i})A_{\rm amp}^3A^2_{\rm AP}$ &  & $\mathcal{N}(0,6^2)$  \\
    & $b_{\nabla^2\mathcal{O}_i}A_{\rm amp}^2A^2_{\rm AP}$ &  & $\mathcal{N}(0,6^2)$  \\
     & $(e_4,e_{5},c_{3},c_4,c_6,c_{7})A_{\rm amp}^2A^2_{\rm AP}$ &  & $\mathcal{N}(0,6^2)$  \\
  & $(d_2,d_{\G},d_{\Gamma_3})A_{\rm amp}^3A^2_{\rm AP}$ &  & $\mathcal{N}(0,1^2)$  \\
  & $(a_{3,4,5,7,...,12})A_{\rm amp}A^2_{\rm AP}$ &  & $\mathcal{N}(0,6^2)$  \\
   & $(a_1,a_6)A^2_{\rm AP}$ &  & $\mathcal{N}(0,6^2)$  \\
  & $c_s A_{\rm amp}$ &  & $\mathcal{N}(0,0.2^2)$  \\
    % & $ $ &  & $\mathcal{N}(0,1^2)$&---\\
   \hline 
    \textbf{photometric} 
    & $b_1 A^{1/2}_{\rm amp}$ &  & $\mathcal{U}[0, 4]$  \\  \textbf{nuisance} 
    & $b_2 A_{\rm amp}$ &  & $\mathcal{N}[0, 1^2]$  \\  
    (sampled) 
     & $b_{\mathcal{G}_2} A_{\rm amp}$ &  & $\mathcal{N}[0, 1^2]$  \\ 
    \hline
    \textbf{photometric} & $b_{\Gamma_3} \,A_{\rm amp}^2$ &  & $\mathcal{N}\left(\frac{23}{42}(b_1-1),1^2\right)$  \\
    \textbf{nuisance} & $b_{\nabla^2\delta} \,A_{\rm amp}$ &  & $\mathcal{N}(0,6^2)$ 
   \\
   % (analytically  & $c_2\,A_{\rm AP}\,A_{\rm amp}$ &  & $\mathcal{N}(30,30^2)$ & $[\Mpc/h]^2$\\
   % marginalized)
   %  & $c_4\,A_{\rm AP}\,A_{\rm amp}$ &  & $\mathcal{N}(0,30^2)$ & $[\Mpc/h]^2$\\
   %  & $\tilde{c}\,A_{\rm AP}\,A_{\rm amp}$ &  & $\mathcal{N}(400,400^2)$ & $[\Mpc/h]^4$\\
   %  & $c_1\,A_{\rm AP}\,A_{\rm amp}$ &  & $ \mathcal{N}(0,5^2)$ & $[\Mpc/h]^2$\\
   (analytically
    & $P_{\rm shot}$ &  & $\mathcal{N}(1,1^2)$  \\
    marginalized)
    & $a_{0}$ &  & $\mathcal{N}(0,1^2)$  \\
& $ a_0^{\rm gm}$ &  & $\mathcal{N}(0,1^2)$  \\
& $c_s A_{\rm amp}$ &  & $\mathcal{N}(0,0.2^2)$ \\
% & $b_1 A_{\rm m}^{1/2}$ &  & $\mathcal{U}[0,4]$ &---\\
% & $b_2 A_{\rm m} $, $b_{\G} A_{\rm m} $ &  & $\mathcal{N}(0,1^2)$ &---\\
    % &   &  & $\mathcal{N}(0,1^2)$ &---\\
    \hline
    \end{tabular}
    \caption{\textbf{Model parameters}: 
    Parameters and priors used in the galaxy clustering and CMB lensing cross-correlation analyses of this work. We split the parameters into those appearing in the spectroscopic and photometric theory models (which are treated independently). Here, $\mathcal{U}[a,b]$ stands for a uniform prior between $a$ and $b$, while $\mathcal{N}(\mu,\sigma^2)$ denotes a normal distribution with mean $\mu$ and variance $\sigma^2$.
    Furthermore, we use the shorthands $b_{\nabla^2\mathcal{O}_i}\in\{b_{\nabla^2\delta^2},b_{\nabla^2\G},b_{(\nabla \delta)^2},b_{(\nabla t)^2}\}$ and $\gamma_{\mathcal{O}_i}\in \{\gamma_2^\times,\gamma_3,\gamma_{21}^\times,\gamma_{211},\gamma_{22},\gamma_{31}\}$. The bias parameters $b_1\sigma_8$, $b_2\sigma_8^2$ and $b_{\mathcal{G}_2}\sigma_8^2$ enter the model non-linearly, thus must be directly sampled in the MCMC chains; all other nuisance parameters appear linearly in the theory prediction and are marginalized over analytically. To reduce prior-volume shifts, we apply rescalings to the bias parameters \citep{Ivanov:2019pdj,desi1,desi2,Tsedrik:2025hmj}, e.g., $b_1\to b_1\sigma_8(z)$, defining the Alcock-Paczynski parameter, $A_{\mathrm{AP}}\equiv 
\left( H^{\mathrm{fid}}_0/H_0 \right)^3 
\left(H(z)/H^{\mathrm{fid}}(z)\right)
\left(D^{\mathrm{fid}}_A(z)/D_A(z) \right)^2$, and the relative amplitude $A_{\rm amp}\equiv\sigma_8^2(z)/\sigma_{8,{\rm fid}}^2(z)$, where $\sigma_{8}^2(z)$ is the mass fluctuation amplitude and ``fid'' refers to quantities evaluated in the fiducial \textit{Planck} 2018 cosmology, 
    }
    \label{tab:priors}
\end{table*}
%\end{widetext}

% \newpage

\noindent
Table~\ref{tab:priors} presents a full summary of the EFT parameters used in this work and their priors. All in all, the EFT prediction for the three-dimensional spectroscopic galaxy observables depends on 45 Wilson coefficients per redshift bin, with the cross-correlation between spectroscopic samples and CMB lensing depending on a further two parameters: the dark matter sound speed $c_s$
and the matter-galaxy stochasticity coefficient $a_0^{\rm gm}$. When analyzing the photometric samples, we require only angular two-point functions describing clustering and galaxy-CMB lensing; these depends on nine free parameters per sample. 

For the EFT parameters appearing in the joint one-loop power spectrum and tree-level bispectrum model, we use the priors introduced and tested in \papertwo (see also \paperone).\footnote{To compare our Table~\ref{tab:priors} to the priors used in previous works, note that (a) $\tilde{c}_1^{\rm here}={c}_1^{\rm there}$, (b) the $A_{\rm AP}$ factor is not absorbed into the definition of the stochastic parameters in this work, and (c) we here use a different normalization for scale-dependent parameters.} To streamline the treatment of the bispectrum counterterms, we here normalize all such 
parameters by appropriate powers of $k_{\rm NL}=0.45~\hMpc$, e.g. $\tilde{c}=400~[\Mpch]^4=16.4~k_{\rm NL}^{-4}$, as reflected in the above table.\footnote{Note the normalization parameter $k_{\rm NL}$ is lower than the non-linear scale of dark matter at the redshifts of interest~\cite{Chudaykin:2020aoj} but larger than the physical cutoff of EFT in redshift space, $k_{\rm FoG}\simeq 0.25~\hMpc$. The numerical coefficient $16$ in our prior on $\tilde{c}$ reflects this hierarchy ($ (k_{\rm NL}/k_{\rm FoG})^4\sim 10$).}
These priors encode conservative assumptions about the physics of 
galaxy formation, and also insert non-linear combinations of cosmological parameters that appear in front of certain the one-loop contributions (with $b_{\Gamma_3}$ replaced by $b_{\Gamma_3}A_{\rm AP}A_{\rm amp}^2$ for example). This choice leads to reduced prior volume effects in EFT-based full shape analyses \citep[e.g.,][]{Ivanov:2019pdj,Chudaykin:2020ghx,Philcox:2021kcw,Maus:2024sbb,Tsedrik:2025hmj}.

For the one-loop bispectrum, we use priors whose width matches the 
conservative choices outlined in~\cite{Bakx:2025pop}, but add non-linear prefactors to mitigate prior volume effects (which are particularly important in dynamical dark energy analyses).
For each operator, these are obtained by looking for the contributions to the one-loop bispectrum involving the largest power of $\sigma_8(z)$, 
which should reduce the prior volume effects induced by strong non-Gaussian correlations between 
EFT parameters and the growth rate amplitude that are expected to dominate 
the prior volume effects. 

Finally, for the photometric statistics we rescale variables according to $\sigma_8(z_{\rm eff})/\sigma_{8,\mathrm{fid}}(z_{\rm eff})$, which produces physical priors
that respect the perturbative nature of the EFT expansion. This is 
particularly important in order to avoid sampling non-physical regions where the EFT parameters become large. Our prior on $c_s$ is also conservatively large in order to 
be agnostic about the strength of the baryonic feedback. Specifically, this allows for up to $100\%$ suppression of power at $k=1~\hMpc$, which is far stronger than that suggested by contemporary hydrodynamical simulations or Sunyaev-Zel'dovich observations~\cite[e.g.,][]{Salcido:2023etz,RiedGuachalla:2025byu,Siegel:2025ivd}. 

When performing analyses not including the CMB primaries, we sample the parameters $\{\omega_{\rm cdm},H_0,\ln(10^{10}A_s)\}$, using the flat uninformative priors given in \paperone. 
We also vary $n_s$ and $\omega_b$, subject to the \textit{Planck} 2018 priors~\cite{Aghanim:2018eyx}:
 \be 
 \begin{split}
 \label{eq:nsob_prior}
     & n_s\sim \mathcal{N}(0.9649,0.0042^2)\,,\\
     & \omega_b \sim \mathcal{N}(0.02237,0.00015^2)\,.
 \end{split}
 \ee 
Whilst one could use a less restrictive prior on $n_s$ or a BBN-derived prior on $\omega_b$ (as in \paperone), we adopt the tight priors in this work to aid comparison with the DESI+CMB analyses, which are the key focus of this work. In such joint studies (which use wide uninformative priors on all cosmological parameters), the information content on $\omega_b$ and $n_s$ is heavily dominated by the \textit{Planck} primary likelihood (in a nearly model-independent fashion~\cite{Planck:2018nkj}).
Notably, our DESI-only results remain essentially unchanged if one places a BBN prior on $\omega_b$ (from~\cite{Schoneberg:2024ifp}) instead of the CMB-informed constraint.

When varying the neutrino mass, we assume three degenerate massive states, which has been previously shown to produce accurate results~\cite{Lesgourgues:2006nd}. In contrast, when $M_\nu$ is fixed, we assume a single massive state whose mass is given by the oscillation floor $M_\nu^{\rm min}=0.06$ eV. When exploring extended cosmological models, we use the $w_0-w_a$ priors on cosmological parameters specified in \papertwo.
To ensure the existence of a  matter-dominated epoch, we impose the constraint $w_0+w_a<0$. 

Finally, we note that our choice of priors ensures that we are well within the sampling range for the 
\textsc{cobra} basis~\cite{Bakx:2025pop}. Since we do not explore the effect of clustering dark energy models
in our work, the effect of the non-trivial expansion
history due to dark energy is fully captured by changes of the distances and the linear growth factor.

\section{Results}
\label{sec:results}

\begin{figure*}[!t]
\includegraphics[width=1.4\columnwidth]{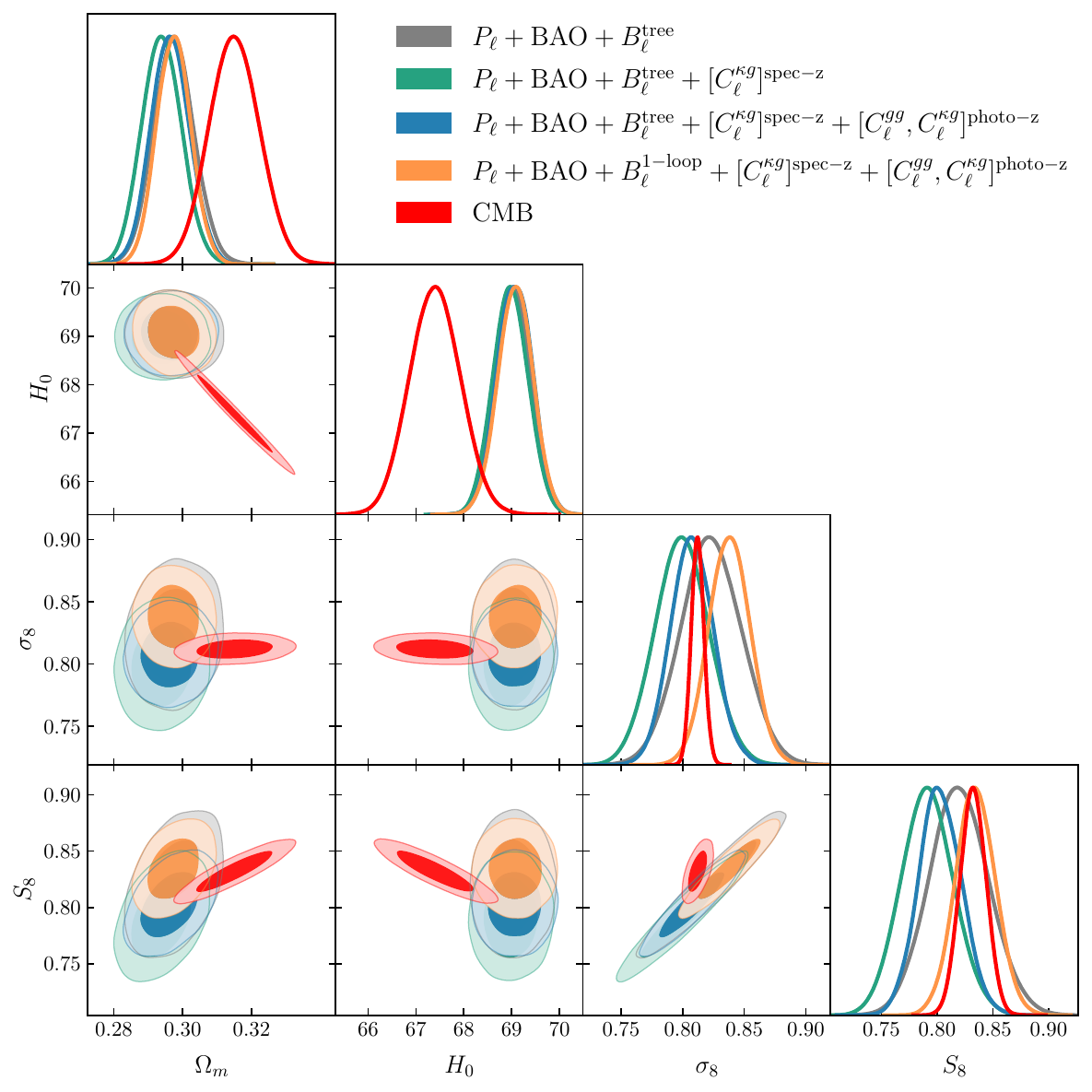}
\caption{\textbf{$\Lambda$CDM constraints:}
Posterior distributions for the $\Lambda$CDM parameters $\Omega_m$, $H_0$, $\sigma_8$ and $S_8$ obtained by combining the DESI DR2 BAO measurements
with the following combinations of data: (1) the DESI DR1 redshift-space power spectrum $P_\ell$ and tree-level 
bispectrum $B_\ell^{\rm tree}$ (gray); (2) adding the cross-correlation between the DESI spectroscopic samples and the CMB lensing reconstruction obtained from \textit{Planck} and ACT observations (green);
(3) adding also the (non-overlapping) DESI photometric galaxy samples, both in auto-correlation and through their cross-correlations with lensing (blue); (4) further adding small-scale spectroscopic bispectrum data-points, using a one-loop model for the galaxy three-point function (orange).
In all cases, we impose \textit{Planck} priors on the primordial power spectrum tilt and the physical baryon density. For comparison, we show constraints obtained from the \textit{Planck} 2018 CMB two-point function combined with the \textit{Planck} PR4 and ACT DR6  lensing reconstructions (red). Marginalized constraints are given in Table~\ref{tab:main2}. For $H_0$ and $\Omega_m$, our DESI constraints are comparable to those from the CMB. For the amplitude parameters, we find that the constraints sharpen significantly when we add lensing cross-correlations and the one-loop bispectrum.}
\label{fig:lcdm}
\end{figure*}

\begin{table*}[!htb]
    \centering
    %\rowcolors{2}{white}{vlightgray}
  \begin{tabular}{|c|cc|ccc|} \hline
    \textbf{Dataset} 
    & $\omega_{\rm cdm}$
    & $H_0$ 
    % & ${\ln(10^{10}A_s)}$ 
    & $\Omega_m$ 
    & $\sigma_8$
    & $S_8$
    \\
    \hline
    %\enspace
     $[C_{\ell}^{g g },C_{\ell}^{\kappa g }]^{\text{photo-z}}+{\rm BAO}$ 
    & $0.1204_{-0.0046}^{+0.0046}$ %%
    & $68.87_{-0.42}^{+0.42}$ %%
    % & $3.043_{-0.093}^{+0.093}$ %%
    & $0.3023_{-0.0083}^{+0.0083}$ %%
    & $0.816_{-0.029}^{+0.029}$ %%
    & $0.819_{-0.030}^{+0.030}$ %%
    \\
    \color{purple}
    $\quad$ + varied $s_\mu$
    & \color{purple}$0.1203_{-0.0046}^{+0.0046}$ %%
    & \color{purple}$68.87_{-0.42}^{+0.42}$ %%
    & \color{purple}$0.3021_{-0.0083}^{+0.0083}$ %%
    & \color{purple}$0.816_{-0.029}^{+0.029}$ %%
    & \color{purple}$0.819_{-0.030}^{+0.030}$ %%
    \\
    \color{purple}
    ${\rm lens}+{\rm BAO}$
    & \color{purple}$0.1176_{-0.0040}^{+0.0040}$ %%
    & \color{purple}$68.72_{-0.40}^{+0.40}$ %%
    & \color{purple}$0.2979_{-0.0074}^{+0.0074}$ %%
    & \color{purple}$0.8195_{-0.0095}^{+0.0095}$ %%
    & \color{purple}$0.817_{-0.016}^{+0.016}$ %%
    \\
    \color{purple}
    $[C_{\ell}^{g g },C_{\ell}^{\kappa g },C_{\ell}^{\kappa\kappa}]^{\text{photo-z}}+{\rm BAO}$
    & \color{purple}$0.1167_{-0.0039}^{+0.0035}$ %%
    & \color{purple}$68.69_{-0.40}^{+0.40}$ %%
    & \color{purple}$0.2962_{-0.0069}^{+0.0069}$ %%
    & \color{purple}$0.8184_{-0.0093}^{+0.0093}$ %%
    & \color{purple}$0.813_{-0.015}^{+0.015}$ %%
    \\
    $P_\ell+{\rm BAO}+B_\ell^{\rm tree} 
    % \q(\mathrm{PBB}_{\rm tree})
    $ 
    & $\enspace 0.1187_{-0.0032}^{+0.0032}\enspace$ %%
    & $\enspace 69.03_{-0.37}^{+0.37}\enspace$ %%
    % & $\enspace 3.046_{-0.065}^{+0.068}\enspace$ %%
    & $\enspace 0.2975_{-0.0059}^{+0.0059}\enspace$ %%
    & $\enspace 0.823_{-0.025}^{+0.025}\enspace$ %%
    & $\enspace 0.820_{-0.027}^{+0.027}\enspace$ %%
    \\
     $P_\ell+{\rm BAO}+B_\ell^{\rm tree}+[C_{\ell}^{\kappa g }]^{\text{spec-z}}$ 
    & $0.1169_{-0.0031}^{+0.0031}$ %%
    & $68.98_{-0.37}^{+0.37}$ %%
    % & $3.038_{-0.063}^{+0.063}$ %%
    & $0.2939_{-0.0057}^{+0.0057}$ %%
    & $0.800_{-0.022}^{+0.022}$ %%
    & $0.791_{-0.024}^{+0.024}$ %%
    \\
    $P_\ell+{\rm BAO}+B_\ell^{\rm tree}+[C_{\ell}^{\kappa g }]^{\text{spec-z}}+[C_{\ell}^{g g },C_{\ell}^{\kappa g }]^{\text{photo-z}}$ 
    & $0.1184_{-0.0030}^{+0.0030}$ %%
    & $69.06_{-0.37}^{+0.37}$ %%
    % & $3.041_{-0.053}^{+0.053}$ %%
    & $0.2964_{-0.0056}^{+0.0056}$ %%
    & $0.808_{-0.017}^{+0.017}$ %%
    & $0.803_{-0.019}^{+0.019}$ %%
    \\
    \color{blue}
   ${P_\ell+{\rm BAO}+B_\ell^{\rm 1-loop}+[C_{\ell}^{\kappa g }]^{\text{spec-z}}+[C_{\ell}^{g g },C_{\ell}^{\kappa g }]^{\text{photo-z}}}$ 
    & \color{blue}${0.1189_{-0.0026}^{+0.0026}}$ %%
    & \color{blue}${69.08_{-0.37}^{+0.37}}$ %%
    % & $3.109_{-0.049}^{+0.049}$ %%
    & \color{blue}${0.2974_{-0.0050}^{+0.0050}}$ %%
    & \color{blue}${0.838_{-0.017}^{+0.017}}$ %%
    & \color{blue}${0.834_{-0.018}^{+0.018}}$ %%
    \\
    \color{purple}
    $\quad$ + ${\rm lens}$
    & \color{purple}$0.1176_{-0.0024}^{+0.0024}$ %%
    & \color{purple}$69.00_{-0.36}^{+0.36}$ %%
    & \color{purple}$0.2954_{-0.0046}^{+0.0046}$ %%
    & \color{purple}$0.8235_{-0.0082}^{+0.0082}$ %%
    & \color{purple}$0.817_{-0.011}^{+0.011}$ %%
    \\
    \color{purple}
    $\quad$ no $B_\ell$ fiber correction
    & \color{purple}$0.1186_{-0.0028}^{+0.0025}$ %%
    & \color{purple}$69.06_{-0.36}^{+0.36}$ %%
    & \color{purple}$0.2970_{-0.0050}^{+0.0050}$ %%
    & \color{purple}$0.837_{-0.017}^{+0.017}$ %%
    & \color{purple}$0.833_{-0.018}^{+0.018}$ %%
    \\
    \color{purple}
    $B_\ell^{\rm 1-loop} \longleftrightarrow$ $B_{\ell}^{\rm 1-loop}(\kmax=0.08~\hMpc)$
    & \color{purple}$0.1179_{-0.0031}^{+0.0031}$ %%
    & \color{purple}$69.03_{-0.38}^{+0.38}$ %%
    & \color{purple}$0.2958_{-0.0057}^{+0.0057}$ %%
    & \color{purple}$0.816_{-0.018}^{+0.018}$ %%
    & \color{purple}$0.810_{-0.019}^{+0.019}$ %%
    \\\color{purple}
    $B_\ell^{\rm 1-loop} \longleftrightarrow$ $B_{\ell}^{\rm 1-loop}(\kmax=0.2~\hMpc)$
    % $P_\ell+{\rm BAO}+B_{\ell,\kmax=0.2}^{\rm 1-loop}+[C_{\ell}^{\kappa g }]^{\text{spec-z}}+[C_{\ell}^{g g },C_{\ell}^{\kappa g }]^{\text{photo-z}}$ 
    & \color{purple}$0.1181_{-0.0024}^{+0.0024}$ %%
    & \color{purple}$69.30_{-0.35}^{+0.35}$ %%
    & \color{purple}$0.2939_{-0.0045}^{+0.0045}$ %%
    & \color{purple}$0.825_{-0.016}^{+0.016}$ %%
    & \color{purple}$0.817_{-0.017}^{+0.017}$ %%
    \\
    %\hdashline
    $\,\, {\rm CMB+lens}\,\,$ 
    & $0.1200_{-0.0012}^{+0.0012}$
    & $67.40_{-0.53}^{+0.53}$
    % & $3.047_{-0.013}^{+0.013}$ %
    & $0.3150_{-0.0072}^{+0.0072}$
    & $0.812_{-0.005}^{+0.005}$
    & $0.832_{-0.012}^{+0.012}$
    \\
  \hline
    \end{tabular}
    \caption{
    \textbf{$\Lambda$CDM constraints}:  Mean and 68\% confidence intervals on $\Lambda$CDM parameters obtained from various combinations of DESI statistics and CMB lensing, following Fig.~\ref{fig:lcdm}. The row marked in blue shows the baseline analysis of this work.
    In the row denoted $B_{\ell}^{\rm 1-loop}(\kmax=0.2~\hMpc)$, we show constraints obtained from a combined analysis with the bispectrum scale-cut set to $k_{\rm max}^{B_\ell} = 0.20~\hMpc$; whilst this is outside the formal validity of our EFT model and is thus not systematically robust, it serves to illustrate the potential gains from pushing to smaller scales. In purple, we
   show variations of the main analyses: the photometric analysis with the magnification bias parameters $s_\mu$ marginalized over Gaussian priors of width $0.1$ centered on the measured values (`+ varied $s_\mu$'); the baseline analysis without correcting for fiber collisions in $B_\ell$ (`no $B_\ell$ fiber correction'); the baseline analysis supplemented by the CMB lensing auto-spectra with the full covariance of all lensing observables, corresponding to a $3\times2$pt analysis for the photometric samples (`+ lens'); and the baseline analysis with the one-loop bispectrum restricted to the tree-level scale cut, $B_{\ell}^{\rm 1-loop}(\kmax=0.08~\hMpc)$, which isolates the modeling difference between the tree-level and one-loop bispectrum likelihoods at fixed data (see Section~\ref{sec:results}). We further include the ${\rm lens}+{\rm BAO}$ and photometric $3\times2$pt$+{\rm BAO}$ dataset combinations, which isolate the information content of the CMB lensing auto-spectra. The last line gives CMB constraints derived from the the \textit{Planck} 2018 two-point likelihood~\cite{Aghanim:2018eyx} and the \textit{Planck} PR4 and ACT DR6 lensing likelihoods~\cite{Carron:2022eyg,ACT:2023kun,ACT:2023dou}. Constraints on the Hubble parameter are quoted in $\mathrm{km}\,\mathrm{s}^{-1}\mathrm{Mpc}^{-1}$ units here and throughout this work.
    }
\label{tab:main2}
\end{table*}

\subsection{DESI $\Lambda$CDM results}
\noindent We begin by presenting constraints on $\Lambda$CDM parameters from the various DESI datasets discussed in Section~\ref{sec:data}. Throughout this subsection, 
we assume a minimal $\Lambda$CDM model with $M_\nu=0.06~\text{eV}$, and impose CMB priors on $n_s$ and $\omega_b$ from \eqref{eq:nsob_prior}.
Our key results are shown in Fig.~\ref{fig:lcdm}, with marginalized constraints summarized in Table~\ref{tab:main2}.

First, we consider the joint analysis of the DESI photometric sample (including cross-correlations with lensing) in combination with the growth history extracted from DR2 BAO. Due to the limited geometric information provided by projected statistics, the $\Omega_m$ and $H_0$ measurements obtained from this combination are dominated by the BAO and the prior on $\omega_b$. In this case, the only novel information provided by the photometric full-shape analyses is a constraint on the growth of structure: $\sigma_8=0.808\pm 0.029$. This constraint is broadly consistent with the official collaboration result for a similar BAO+photometric analysis: $\sigma_8=0.791\pm 0.021$~\cite{Sailer:2025rks}
The larger errorbars and slight shifts found in our study are attributed to 
our more conservative priors on EFT parameters, as well as the additional EFT parameters $(a_0,b_{\Gamma_3},a_0^{\rm gm})$ which were fixed in \cite{Sailer:2025rks}. We have found that our photometric-plus-BAO results are insensitive to the scale-cuts, with very similar results found when adopting the $\ell_{\rm max}$ values of \eqref{eq:desilmax}, matching the official DESI analysis \citep{Maus2025:joint_3d_lensing_dr1}.

\begin{figure*}[!htb]
\includegraphics[width=1.5\columnwidth]{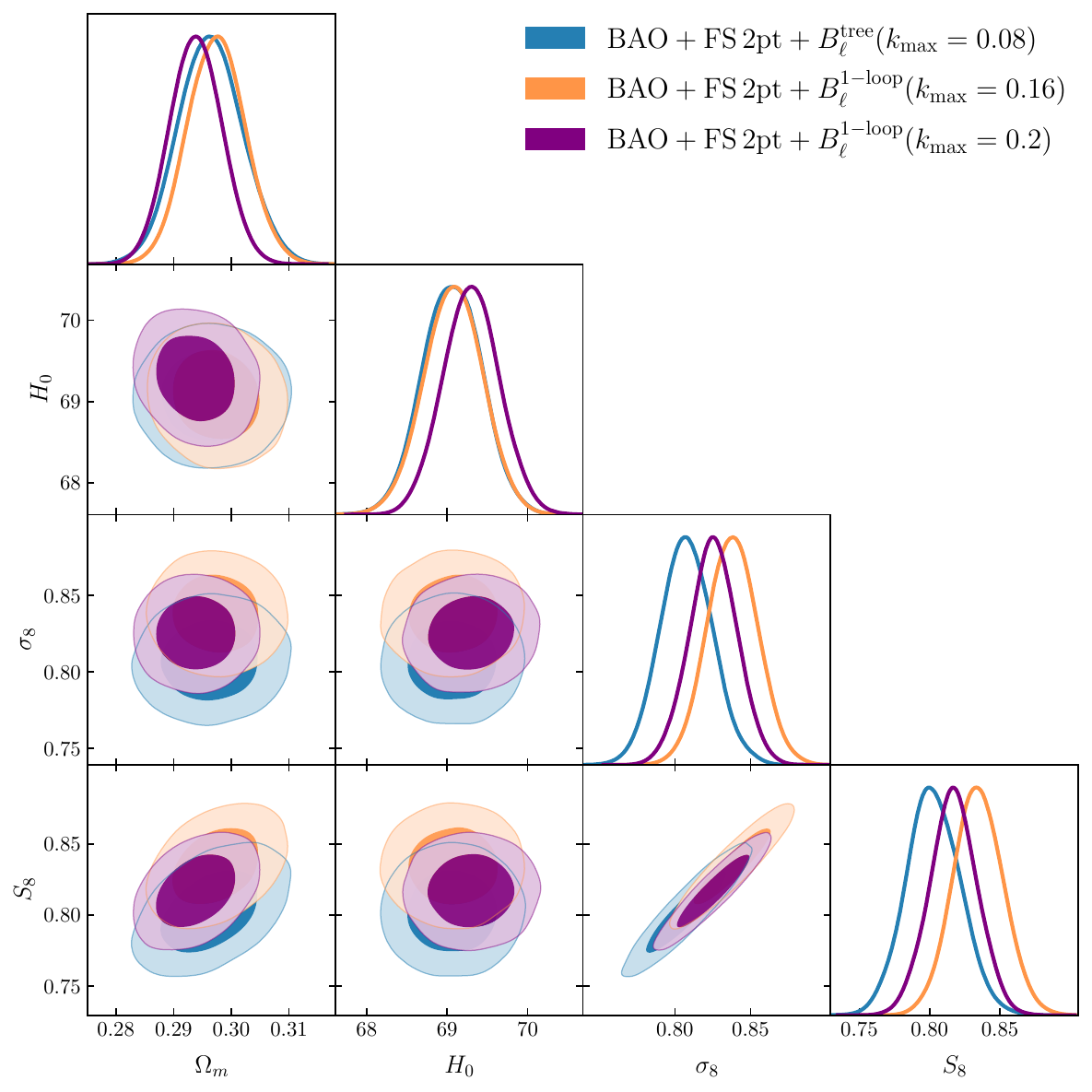}
\caption{\textbf{$\Lambda$CDM constraints from bispectrum analysis variants:} We show results from the full combination of 
spectroscopic and photometric two-point functions $P_\ell+[C_\ell^{\kappa g }]^{\text{spec-z}}+[C_{\ell}^{g g },C_{\ell}^{\kappa g }]^{\text{photo-z}}$ (denoted ``FS 2pt''), combined with the tree-level bispectrum likelihood at $k_{\rm max}^{B_\ell}=0.08~\hMpc$ (blue), and the one-loop likelihood at $k_{\rm max}^{B_\ell}=0.16~\hMpc$ (orange) and $k_{\rm max}^{B_\ell}=0.2~\hMpc$ (purple). The constraints improve slightly when upgrading the bispectrum likelihood from tree-level to one-loop, at the fiducial choice of $k_{\rm max}^{B_\ell}=0.16~\hMpc$. Increasing the maximum scale to $k_{\rm max}^{B_\ell}=0.2~\hMpc$ does not greatly reduce the error-bars, but produces significant shifts of the posterior contours. This behavior is similar to that observed in previous simulation-based studies \cite{Bakx:2025pop}, and implies that two-loop corrections become important in this regime.
}
\label{fig:bisp}
\end{figure*}

Results obtained using the three-dimensional $P_\ell+B^{\rm tree}_\ell$ likelihood in combination with that of the DESI DR2 BAO (assuming independence, as discussed above) are broadly consistent with those obtained in \paperone. Though the constraint on $H_0$ is dominated by the BAO dataset, we find slightly improved bounds on $\omega_{\rm cdm}$ and $\sigma_8$ with respect to the former work due to the tighter prior on $n_s$ and the inclusion of the tree-level bispectrum quadrupole.
These constraints are further improved 
upon adding the spectroscopic galaxy -- lensing 
cross-correlations, which tighten
the $\sigma_8$ constraint by $12\%$, giving $\sigma_8=0.800\pm 0.022$. Further adding the photometric samples, the growth of structure measurements tighten by $23\%$, yielding $\sigma_8=0.808\pm 0.017$. This result is nearly identical to the official DESI constraint obtained from the three-dimensional full-shape power spectra combined with the BAO and the same projected clustering observables~\cite{Maus2025:joint_3d_lensing_dr1}: $\sigma_8=0.803\pm 0.017$ (noting that we use more data, but less restrictive priors).
Finally, extending the bispectrum model to one-loop order yields $\sigma_8 = 0.838\pm 0.017$, with an error-bar comparable to that of the tree-level analysis; at the level of the lensing parameter $S_8\equiv\sigma_8\sqrt{\Omega_m/0.3}$, our fiducial error-bars are ${\approx}5\%$ narrower than those from the tree-level bispectrum analysis.

Notably, the $H_0$ and $\Omega_m$ constraints
do not tighten significantly when new statistics are added, with $\Omega_m$ improving by just $5\%$ when the full photometric and spectroscopic cross-correlation observables are added to the $\text{BAO}+P_\ell+B^{\rm tree}_\ell$ likelihood. Further including the one-loop bispectrum sharpens the $\Omega_m$ constraint by an additional $11\%$ (similar to \citep{Bakx:2025pop}), but does not change the $H_0$ result appreciably, due to the dominance of BAO information. 

{Whilst the error-bars are thus only mildly affected by the one-loop bispectrum, the inferred central value of $\sigma_8$ is $1.8\sigma$ higher than its tree-level counterpart (the corresponding shifts in $\Omega_m$ and $H_0$ are just $0.20\sigma$ and $0.05\sigma$). The one-loop analysis includes a large number of additional triangle configurations, each with at least one wavenumber in the range $0.08<k/(\hMpc)<0.16$, which pull the inferred amplitude upwards; at the same time, the associated gain in precision is largely offset by the simultaneous marginalization over the much larger one-loop EFT parameter space.
The comparison of the tree-level and one-loop results is, however, not apples-to-apples because the corresponding models
have very different complexities ($15$ vs $45$ parameters per tracer).
The na\"ive expectation that parameter shifts should be bounded by the reduction of the error-bars therefore does not apply here. This decomposition can be verified explicitly: repeating the one-loop analysis restricted to the tree-level scale cut, $k_{\rm max}^{B_\ell}=0.08~\hMpc$, yields $\sigma_8 = 0.816\pm 0.018$ (cf.\ Table~\ref{tab:main2}), within $0.5\sigma$ of the tree-level result, with $\Omega_m$ and $H_0$ agreeing to ${\approx}0.1\sigma$. This comparison, at fixed data, provides a direct estimate of the theory modeling error of the tree-level bispectrum likelihood: ${\approx}1\%$ ($0.5\sigma$) on $\sigma_8$ and ${\lesssim}0.1\sigma$ on the remaining parameters, consistent in sign and magnitude with the tree-level bias found in simulations,}\footnote{Simulations suggest a small systematic bias on $\sigma_8$
from the tree-level bispectrum at $k_{\rm max}^{B_\ell}=0.08~\hMpc$
to shift $\sigma_8$ by up to $-0.6\%$~\cite{Ivanov:2021kcd}. This bias (which is usually subdominant to statistical errors) is removed when the tree-level bispectrum computation is upgraded to one-loop order.}
{ whilst the remaining $1.3\sigma$ shift is driven by the newly included smaller-scale triangles. Shifts of a similar magnitude (${\sim}1\sigma$) between the tree-level and one-loop bispectrum analyses were observed in the simulation-based study of~\cite{Bakx:2025pop}, in which the one-loop pipeline used in this work was validated end-to-end at $k_{\rm max}^{B_\ell}=0.15~\hMpc$, recovering the input cosmology without significant biases. The observed shift is thus fully consistent with the behavior seen in mocks. We also note that the upwards shift brings our growth-of-structure measurement into excellent agreement with the amplitude inferred from the primary CMB.}

{We further consider supplementing our baseline dataset with the CMB lensing auto-spectra (``lens''), employing the full covariance of the lensing observables, including the cross-covariance of $C_\ell^{\kappa\kappa}$ with the projected galaxy statistics -- \textit{i.e.}, a complete $3\times2$pt analysis for the photometric samples and a joint $[C_\ell^{\kappa g},C_\ell^{\kappa\kappa}]$ analysis for the spectroscopic samples -- with results shown in Table~\ref{tab:main2}. This leads to a significant sharpening of the growth-of-structure constraints, yielding a $1\%$ measurement of the mass fluctuation amplitude, $\sigma_8 = 0.8235\pm 0.0082$, and $S_8 = 0.817\pm 0.011$, whilst the geometric parameters shift only mildly ($\Omega_m = 0.2954\pm 0.0046$, $H_0 = 69.00\pm 0.36~\mathrm{km}\,\mathrm{s}^{-1}\mathrm{Mpc}^{-1}$).}

{To isolate the information content of the lensing auto-spectrum, we additionally analyze the ${\rm lens}+{\rm BAO}$ and full photometric $(3\times2$pt$)+{\rm BAO}$ combinations, with results reported in Table~\ref{tab:main2}. The former yields $\sigma_8 = 0.8195\pm 0.0095$, a $1.2\%$ measurement of the mass fluctuation amplitude that is fully independent of the three-dimensional galaxy clustering, whilst the latter gives $\sigma_8 = 0.8184\pm 0.0093$ ($S_8 = 0.813\pm 0.015$). The $\sigma_8$ constraint in the $3\times2$pt analysis is thus dominated by $C_\ell^{\kappa\kappa}$, with the galaxy auto- and cross-correlations contributing a further ${\approx}2\%$ improvement (${\approx}6\%$ in $S_8$). Both measurements are consistent with our baseline full-shape result at the ${\approx}1\sigma$ level.}

{Table~\ref{tab:main2} additionally reports two robustness tests of our results. First, marginalizing over the magnification bias parameters $s_\mu$ with Gaussian priors of width $0.1$ centered on the externally measured values, rather than fixing them as in our baseline treatment, leaves the photometric posteriors unchanged, with all cosmological parameters shifting by less than $0.05\sigma$. Second, repeating the baseline analysis without applying the fiber-collision correction to the bispectrum, \textit{i.e.}\ retaining configurations with close pairs and/or triplets, leads to shifts of at most $0.12\sigma$ (in $\omega_{\rm cdm}$), with all other cosmological parameters shifting by less than $0.1\sigma$; since this comparison brackets the total impact of fiber collisions on the bispectrum, any residual systematic after our $\theta$-cut mitigation is negligible.}

Whilst stronger improvements may be wrought if one pushes the one-loop bispectrum analysis to smaller scales, we have previously argued that the one-loop EFT bispectrum calculation is not reliable beyond $\kmax=0.16~\hMpc$ (at least 
for the LRG sample). Nevertheless, it is interesting to see how much the nominal 
constraints could improve if the fit is extended to smaller scales. The results of an analysis using $k_{\rm max}^{B_\ell}=0.20~\hMpc$
are shown in Table~\ref{tab:main2} and Fig.~\ref{fig:bisp}. 
We caution that the inferred cosmological parameter constraints are subject to considerable theory-driven systematic error, thus our results are purely illustrative. 

When pushing to $k_{\rm max}^{B_\ell}=0.20~\hMpc$, we find that the errorbars on $\Omega_m$, $H_0$ and $\sigma_8$ improve by $10\%$, $5\%$, and $6\%$, respectively.
However, these improvements are accompanied by large shifts in the inferred mean values, with $\Delta\Omega_m$, $\Delta H_0$, and $\Delta \sigma_8$ equal to 
$-0.8\sigma$, $0.6\sigma$,  and $-0.8\sigma$, respectively. 
These shifts are much larger than those expected from statistical fluctuations given the small improvements in the error-bars.
We note that the (theory-systematic-induced) downwards shift of $\Omega_m$ is in
the direction favoring dynamical dark energy (which is sourced primarily by the preference of DESI for lower $\Omega_m$ than \textit{Planck}).
As such, we expect to find more significant biases if we repeat the aggressive bispectrum analyses within the $w_0w_a$CDM model. We conclude that the bispectrum analysis for $k_{\rm max}^{B_\ell}>0.16~\hMpc$ is affected by two-loop effects, consistent with the 
analysis of \cite{Bakx:2025pop} using high-fidelity simulations.

\begin{figure*}[!t]
\includegraphics[width=0.99\columnwidth]{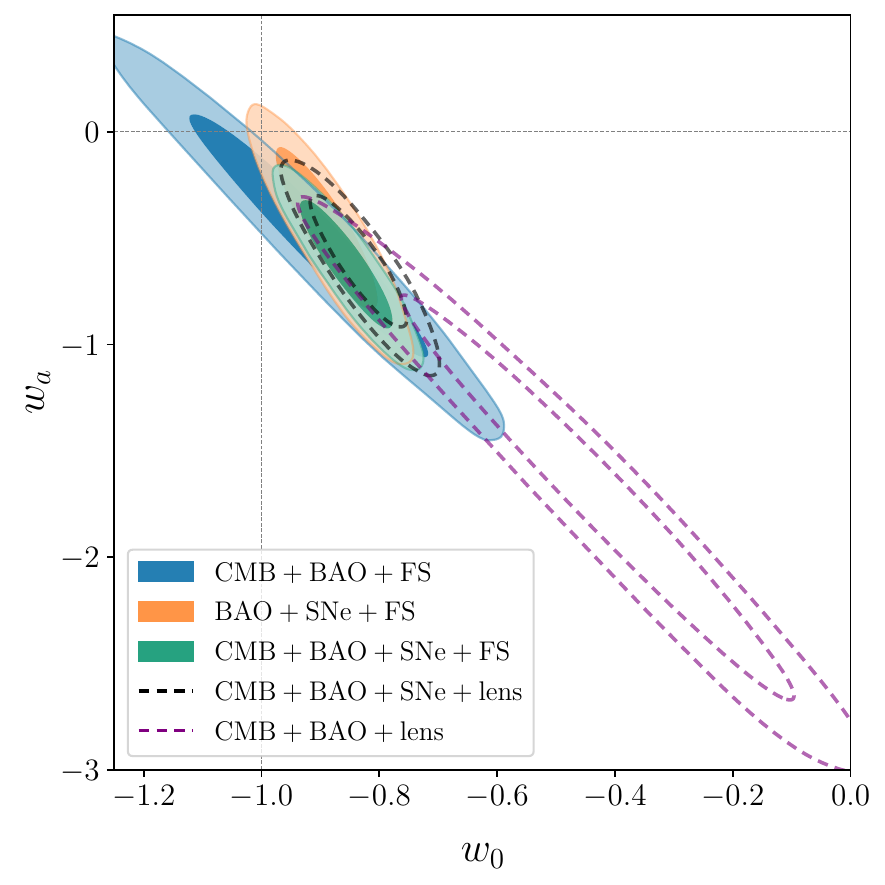}
\includegraphics[width=0.99\columnwidth]{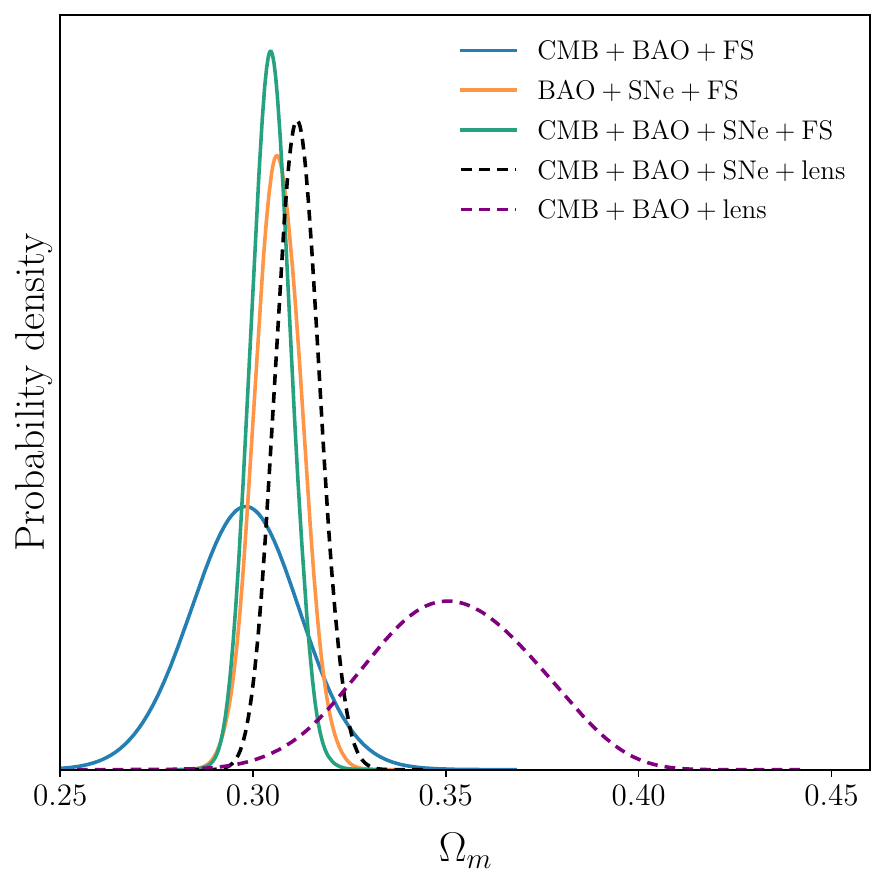}
\caption{\textbf{$w_0w_a$CDM parameter constraints.}
\textit{Left Panel:}
One- and two-dimensional posterior distributions of the dynamical dark energy parameters for various combinations of datasets, as indicated by the legend. Here FS refers to the combined DESI DR1 full-shape dataset, comprising three-dimensional galaxy power spectra and one-loop bispectra, photometric galaxy correlations, and cross-correlations of CMB lensing with photometric and spectroscopic samples. The results shown in orange use CMB priors on $\omega_b$ and $n_s$, and those in black represent the standard combination from \citep{DESI:2025zgx}. 
Black short-dashed lines mark the cosmological constant values of the dark energy equation of state parameters $(w_0,w_a)=(-1,0)$.
The EFT-based full-shape likelihood improves the dark energy figure of merit (given by the inverse area of the posterior in the $w_0-w_a$ plane) by $15\%$.
\textit{Right Panel:} $\Omega_m$ inference 
from the same data combinations.
}
\label{fig:w0wa}
\end{figure*}

\begin{figure}[!t]
\includegraphics[width=0.99\columnwidth]{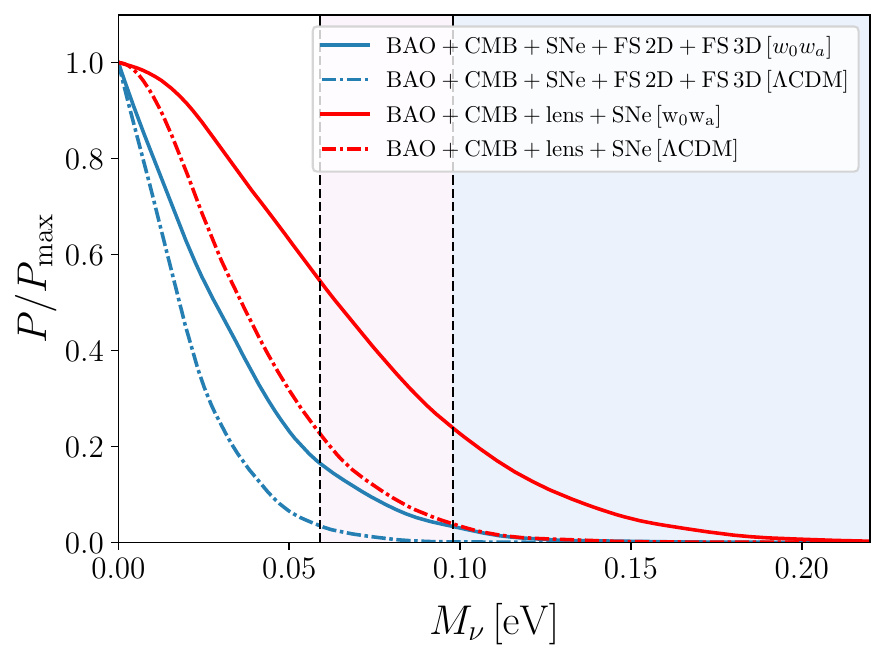}
\caption{\textbf{Neutrino mass constraints:} We show results including various combinations of DESI datasets in conjunction with the CMB likelihood and that of the Pantheon+ supernovae sample. Solid lines show constraints assuming the $w_0w_a$CDM background, while the dashed lines represent the limits obtained under the $\Lambda$CDM model. Marginalized constraints are given in Table~\ref{tab:mnu}. Importantly, we find $\approx 25\%$ tighter constraints when combining the geometric BAO, CMB, and SNe probes (red) with the full-shape clustering statistics considered in this work. 
% using either 
% the CMB lensing cross- or  auto correlations.
% using either the tree-level (blue) or one-loop (orange) bispectrum likelihood. Regardless of the bispectrum model, 
The combined neutrino mass constraints are in mild tension with the inverted hierarchy (rightmost vertical line) in both $\Lambda$CDM and $w_0w_a$CDM cosmological models.}
\label{fig:mnu}
\end{figure}

\begin{table*}[!htb]
    \centering
    %\resizebox{\linewidth}{!}{
    \begin{tabular}{lcccccc}
    \hline 
    \toprule
    Model/Dataset 
    & $M_\nu\,{\rm [eV]}$ 
    & $\Omega_m$ 
    & $H_0$ 
    & $\sigma_8$ 
    & $w_0$ %or $10^3\Omk$
    & $w_a$
    \\
        \hline 
    \midrule
    %\enspace
                %$\bm{o}\bm{\Lambda}$\textbf{CDM}+$\bm{M_\nu}$
    $\bm{w_0w_a\mathrm{CDM}}$  &  &  &  &  &  & \\
  $\cmb+{\rm lens}+\bao$   & $-$  
& $\enspace 0.351_{-0.022}^{+0.022}\enspace$ 
& $\enspace 63.86_{-2.10}^{+1.70}\enspace$ 
& $\enspace 0.784_{-0.017}^{+0.015}\enspace$
& $\enspace -0.43_{-0.21}^{+0.22}\enspace$ 
& $\enspace -1.71_{-0.63}^{+0.61}\enspace$ 
%final
    \\
   $\mathrm{FS~2D}+\mathrm{FS~3D}+\bao+\mathrm{CMB}$  
   & $-$ 
& $0.298^{+0.013}_{-0.013}$ 
& $69.1^{+1.3}_{-1.6} $ 
& $0.825^{+0.013}_{-0.013}$ %final-v2
& $-0.92^{+0.14}_{-0.14} $ 
& $-0.49^{+0.39}_{-0.39} $ 
    \\
   $\mathrm{FS~2D}+\mathrm{FS~3D}+\bao+\mathrm{SNe}$  
   & $-$ 
& $0.3066_{-0.0061}^{+0.0061}$ 
& $68.60_{-0.69}^{+0.69}$ 
& $0.831_{-0.017}^{+0.017}$ %final-v2
& $-0.886_{-0.057}^{+0.057} $ 
& $-0.47_{-0.23}^{+0.26}$ 
    \\
    $\cmb+\bao+{\rm SNe}+{\rm lens}$  
   & $-$ 
& $0.3108_{-0.0057}^{+0.0056}$ 
& $67.64_{-0.59}^{+0.59}$ 
& $0.8117_{-0.0085}^{+0.0085}$ %final
& $-0.840_{-0.055}^{+0.054}$ 
& $-0.60_{-0.19}^{+0.22}$ 
    \\
   $\cmb+\bao+{\rm SNe}+{\rm lens}+\mathrm{FS~2D}+\mathrm{FS~3D}$  
   &  $-$ 
&  $0.3041_{-0.0053}^{+0.0053}$ 
&  $68.35_{-0.56}^{+0.56}$ 
&  $0.8180_{-0.0075}^{+0.0075}$  
&  $-0.859_{-0.052}^{+0.052}$ 
&  $-0.60_{-0.19}^{+0.19}$ 
    \\
\color{blue}$\cmb+\bao+{\rm SNe}+\mathrm{FS~2D}+\mathrm{FS~3D}$  
& $-$ 
& \color{blue}$0.3046_{-0.0053}^{+0.0053}$ 
& \color{blue}$68.33_{-0.55}^{+0.55}$ 
& \color{blue}$0.8191_{-0.0090}^{+0.0090}$
& \color{blue}$-0.854_{-0.052}^{+0.052}$ 
& \color{blue}$-0.63_{-0.19}^{+0.21}$ 
    \\
    \bottomrule    \hline 
    $\bm{\Lambda}$\textbf{CDM+}$\bm{M_\nu}$ &  &  &  &  &  & \\
        % w/o FS
    $\cmb+\mathrm{lens}+\bao$ 
    & $<0.069$ 
& $0.2999_{-0.0038}^{+0.0038}$ 
& $68.56_{-0.30}^{+0.30}$ 
& $0.8178_{-0.0058}^{+0.0067}$ %final
& $-$
& $-$
    \\
       $\cmb+\mathrm{lens}+\bao+{\rm SNe}$  
   & $<0.077$ 
& $0.3009_{-0.0039}^{+0.0038}$ 
& $68.50_{-0.31}^{+0.31}$ 
& $0.8172_{-0.0061}^{+0.0072}$ %final
& $-$
& $-$
    \\
    $\cmb+\bao+{\rm SNe}+{\rm lens}+\mathrm{FS~2D}+\mathrm{FS~3D}$  
    &  $<0.049$ 
&  $0.2973_{-0.0033}^{+0.0033}$ 
&  $68.79_{-0.27}^{+0.27}$ 
&  $0.8223_{-0.0062}^{+0.0062}$ 
&  $-$ 
&  $-$  
\\
   \color{blue}$\cmb+\bao+{\rm SNe}+\mathrm{FS~2D}+\mathrm{FS~3D}$  
    & \color{blue}$<0.049$ 
& \color{blue}$0.2973_{-0.0034}^{+0.0034}$ 
& \color{blue}$68.79_{-0.27}^{+0.27}$ 
& \color{blue}$0.8202_{-0.0073}^{+0.0073}$ 
& $-$ 
& $-$ %final4
\\
    \midrule
        \hline 
    $\bm{w_0w_a}$\textbf{CDM+$\bm{M_\nu}$} &  &  &  &  &  & \\
    $\cmb+{\rm lens}+\bao$
    & $<0.166$ 
& $0.351_{-0.022}^{+0.023}$ 
& $63.86_{-2.19}^{+1.67}$ 
& $0.783_{-0.019}^{+0.019}$ 
& $-0.43_{-0.21}^{+0.24}$ 
& $-1.72_{-0.69}^{+0.66}$ %final
    \\
   $\cmb+{\rm lens}+\bao+{\rm SNe}$  
   & $<0.123$ 
& $0.3103_{-0.0059}^{+0.0056}$ 
& $67.68_{-0.61}^{+0.6}$ 
& $0.8136_{-0.0093}^{+0.01}$ %final
& $-0.847_{-0.057}^{+0.054}$ 
& $-0.55_{-0.2}^{+0.24}$ 
    \\
   $\cmb+\bao+{\rm SNe}+{\rm lens}+\mathrm{FS~2D}+\mathrm{FS~3D}$  
    & $<0.078$ 
&  $0.3034_{-0.0053}^{+0.0053}$ 
&  $68.37_{-0.57}^{+0.57}$ 
&  $0.8228_{-0.0083}^{+0.0083}$ 
&  $-0.870_{-0.052}^{+0.052}$ 
&  $-0.52_{-0.18}^{+0.20}$ %final-v3
    \\
       \color{blue}$\cmb+\bao+{\rm SNe}+\mathrm{FS~2D}+\mathrm{FS~3D}$   
    & \color{blue}$<0.077$ 
& \color{blue}$0.3038_{-0.0053}^{+0.0053}$ 
& \color{blue}$68.38_{-0.56}^{+0.56}$ 
& \color{blue}$0.8246_{-0.0097}^{+0.0097}$ 
&\color{blue} $-0.864_{-0.052}^{+0.052}$ 
& \color{blue}$-0.56_{-0.19}^{+0.21}$ %final-v2
    \\
    \midrule
        \hline 
    \end{tabular}
    %}
    \caption{\textbf{Constraints on $\Lambda$CDM extensions}: Marginalized constraints on cosmological parameters for various combinations of datasets and with different assumptions on the background cosmological model. Here, ``FS~2D'' refers to the complete combination of angular full-shape two-point functions discussed above (\textit{i.e.}\, photometric auto-spectra and both spectroscopic and photometric lensing cross-correlations), and
    FS~3D indicates the combination of the three-dimensional power spectra and bispectra. The rows shown in blue represent the baseline analyses of this work.  {The rows including both the CMB lensing auto-spectra (``lens'') and the full-shape data employ the full covariance of all lensing observables, as described in Section~\ref{sec:data}.} We impose the physical prior $M_\nu\geq 0$ in all analyses. We provide the mean and 68\% confidence intervals for all parameters, except $M_\nu$ for which we quote 95\% upper limits.\label{tab:mnu}
    }
\end{table*}

\subsection{Dynamical Dark Energy}
\noindent
Next, we present results on the dark energy equation-of-state parameters (fixing the neutrino mass to the minimal value of
$0.06$~eV as in the previous section). 
To efficiently probe the broad parameter space associated with the $w_0w_a$CDM model, we will perform only joint analyses in this section, combining our DESI likelihood with external data from type-Ia supernovae (SNe) and/or the CMB. Our main results are presented
in Fig.~\ref{fig:w0wa}, with the corresponding parameter constraints listed in the bottom panel of Table~\ref{tab:mnu}.
In this section, we will refer to the 
DESI DR1 photometric dataset
and the spectroscopic cross-correlations with CMB lensing as
``FS 2D'', while the 
spectroscopic two- and three-point
correlations will be called ``FS 3D''. Their combination (which is the fiducial set-up) will be simply referred to as 
``FS''.

First, we consider the constraints from DESI and the SNe in the absence of the primary CMB likelihoods. It was previously shown that the full-shape data from BOSS in combination with SNe and BAO (from eBOSS or DESI DR1) has enough statistical
power to determine the dark energy equation of state parameters
even in the absence of CMB data~\cite{Chudaykin:2020ghx,Chen:2024vuf}.
These works, however, found no evidence for dynamical dark energy
in the low-redshift data. Here we repeat this analysis using DESI full-shape data for the first time.

As discussed in \citep{DESI:2025zgx}, the combination of SNe and BAO from DESI DR2 (without full-shape) implies a weak preference for the dynamical dark energy (DDE) model over the cosmological constant; based on the difference in the $\chi^2$ statistic for the best-fit $\wa$ model relative to the best-fit $\ld$ model with ($w_0=-1$, $w_a=0$), 
the frequentist significance for this data combination is $1.7\sigma$ given the two additional free parameters (which increases to $2.9\sigma$ when adding the CMB and lensing likelihoods). As shown in Fig.~\ref{fig:w0wa}, the joint analysis of DR1 FS with DR2 BAO and SNe yield a 
a marginal posterior that includes the $\Lambda$CDM point within $95\%$. 
In contrast, the frequentist analysis yields a mild preference for DDE at {$2.2\sigma$} from the same data combination.
Whilst this is statistically not very significant, it is an interesting indication
that the preference for DDE is present even in the absence of the CMB primaries; moreover, the evidence increases by {$0.5\sigma$} upon adding the full-shape data.

Next, we consider exchanging the SNe likelihood with that of the CMB primaries. 
When adding full-shape information to the CMB and BAO likelihood (and dropping the lensing auto-spectrum, as before), we find a clear shift in the $w_0-w_a$ plane towards the $\Lambda$CDM values, with the posterior area in $w_0-w_a$ plane 
shrinking by $8\%$.
This improvement is somewhat smaller than that obtained in \papertwo (see Fig.~2), which did not include lensing cross-correlations, one-loop bispectra, or photometric samples, but did include $C_\ell^{\kappa\kappa}$.

Our frequentist analysis, however, shows that the preference for the $w_0w_a$CDM 
model is roughly the same for all three likelihoods discussed above. 
Specifically, we find a $2.7\sigma$, $2.8\sigma$ and $2.9\sigma$ significance
in the BAO+CMB+FS (from this work), BAO+CMB+lens+$P_\ell$
+$B^{\rm tree}_0$ (matching \papertwo), and BAO+CMB+lens (as in \citep{DESI:2025zgx}) analyses, respectively.
This implies that the additional constraining power of the full-shape data in the context of the $w_0w_a$CDM model is not very significant 
in the absence of SNe data. 
Nevertheless,  even though the 
frequentist preference does not improve 
significantly,  
the nominal figure-of-merit in the $w_0-w_a$ plane improves by 
almost $50\%$ upon adding the full-shape data.

Finally, we consider the joint analysis of CMB, SNe, BAO, and DESI FS data. From Fig.~\ref{fig:w0wa}, it is clear that adding full-shape information to the BAO, CMB, and SNe datasets does not lead to radical new insights into the nature of dark energy; however, the dark energy figure-of-merit does improve by {$15\%$} compared to the 
baseline CMB+BAO+SNe combination when including the full-shape data (FS~2D and FS~3D), {rising to $19\%$ when the CMB lensing auto-correlations are additionally included with their full covariance}. Based on \papertwo, these improvements
are primarily driven by the three-dimensional statistics.

The shifts in the $w_0$ and $w_a$ posteriors are related to the change in the $\Omega_m$ constraints. Adding FS information decreases the central value of $\Omega_m$ by $1.3\sigma$ compared to the joint CMB+BAO+lens result, as illustrated in the right panel of Fig.~\ref{fig:w0wa}. This shift, albeit somewhat less significant, is consistent with that observed when one adds the SNe dataset to the joint CMB+BAO result. In addition, in the presence of the SNe, the inferred value of $\Omega_m$ is $\approx 5\%$ more precise as a result of adding the FS data. This demonstrates the utility of the full-shape LSS measurements which help constrain the extended expansion history in the $\wa$ scenario.

\subsection{Massive Neutrinos in $\Lambda$CDM and Beyond}
\noindent

Lastly, we discuss the implications of our full-shape EFT likelihood for the neutrino mass sum. For this purpose, we will consider massive neutrinos cosmologies with both fixed ($\Lambda$CDM) and time-varying ($w_0w_a$CDM) dark energy sectors. Our main results for different analysis variations are displayed in Table~\ref{tab:mnu} and Fig~\ref{fig:mnu}.

Starting with the $\Lambda$CDM+$M_\nu$ model, we observe that the CMB+BAO dataset already places strong constraints on $M_\nu$ that disfavor the inverted  hierarchy at the $3\sigma$ level. This bound somewhat loosens upon including the SNe Pantheon+ data (with $M_\nu<0.077$ eV at 95\% CL). {Upon combining with the full-shape datasets introduced in this work, the constraint tightens by $36\%$, leading to the baseline result $M_\nu<0.049$ eV at 95\% CL. Further adding the CMB lensing auto-spectrum, with the full covariance of Section~\ref{sec:data}, leaves this bound unchanged.}

Our measurement is {$36\%$} stronger than the official constraint from DESI DR1 (three-dimensional) full-shape-plus-BAO-plus-CMB: $M_\nu<0.077~$eV~\cite{DESI:2024hhd}. Moreover, it is {$23\%$} tighter than the bound using DR2 BAO and the CMB: $M_\nu<0.064~$eV~\cite{DESI:2025ejh}. The extra improvement in our analysis is mainly due to our custom power spectrum and bispectrum 
likelihoods. Indeed, \papertwo already found $M_\nu<0.059$~eV in an analysis with the DESI DR1 full-shape data combined with the CMB and DR2 BAO, which is $8\%$ stronger than the official DESI CMB+BAO bound.  
{Within our limits, the inverted hierarchy floor is excluded at the ${\approx}3.5\sigma$ level, estimated by fitting the posterior with a Gaussian model truncated by the 
physical prior $M_\nu>0$.
Notably, even the normal hierarchy floor, $\mathrm{min}(M_\nu)=0.058$ eV, is disfavored at ${\approx}97.5\%$ CL (${\approx}2.3\sigma$). We conclude that within $\Lambda$CDM there is clear and strong evidence in favor of the normal hierarchy. 
At the same time, the fact that the bound starts to disfavor the minimal physical mass allowed by oscillation experiments is a result of the $\Lambda$CDM tension discussed in Section~\ref{sec:intro}, further motivating the $w_0w_a$CDM analysis below.}

{As an additional test, we consider supplementing our CMB+BAO+FS likelihood with the CMB lensing auto-spectrum ($C_\ell^{\kappa\kappa}$), employing the full covariance of all lensing observables described in Section~\ref{sec:data}. Since this leads to a considerable improvement in the bounds on the growth of structure, it should also lead to improved neutrino mass constraints \textit{if} the neutrino information content is dominated by the clustering signal. In practice, our neutrino mass bounds are essentially unchanged in both background models ($M_\nu<0.049$ versus $0.049$ eV in $\Lambda$CDM, and $0.078$ versus $0.077$ eV in $w_0w_a$CDM; cf.\ Table~\ref{tab:mnu}), which implies that most of our constraining power arises from three-dimensional correlations and the measurements of the shape and geometry parameters such as $\Omega_m$ and $H_0$.}

Upon promoting $\Lambda$CDM to $w_0w_a$CDM, the constraints on neutrino masses relax considerably. In particular, geometry-based neutrino mass probes\footnote{Strictly speaking, it is impossible to separate large-scale structure effects due to gravitational lensing from the CMB likelihood, so the demarcation into geometry- and clustering-based probes used here is somewhat imprecise.} (CMB, BAO, and SNe) are consistent with both the normal and inverted hierarchy 
within $2\sigma$. {However, the constraints tighten by $37\%$ when
adding the full-shape DESI datasets to the BAO and CMB datasets, yielding the 95\% CL upper limit $M_\nu < 0.077$ eV for our baseline combination, which is essentially unchanged ($M_\nu < 0.078$ eV) upon further adding the CMB lensing auto-spectrum. These limits exclude the inverted hierarchy floor, $M_\nu \geq 0.098$ eV~(eq.~\eqref{eq:floor}), at $98.3\%$ CL (${\approx}2.4\sigma$), \textit{even within the extended cosmological model}. {This exclusion level is computed directly from the Markov chains, as the posterior probability of exceeding the inverted-hierarchy floor, $p(M_\nu > 0.098~{\rm eV}) = 0.017$, which we translate into an equivalent (two-tailed) Gaussian significance.} 
Notably, this statement is possible only when including the full-shape datasets. Furthermore, the fact that our result is essentially unchanged by the addition of the CMB lensing auto-correlations suggests that the bounds are not majorly impacted by the $A_L$ anomaly~\cite{Green:2024xbb}.}
%% NOTE: exact chain computation: p(M_nu > 0.098 eV) = 0.01657 +- 0.00018
%% (Neff ~ 5.1e5), i.e. 98.34% CL = 2.40 sigma (two-tailed convention).

{
It is worth clarifying the origin of the $\sim 37\%$ improvement in the
$w_0w_a$CDM neutrino mass bound upon adding the full-shape data. Since the
error bars on $\sigma_8$, $w_0$, and $w_a$ are nearly unchanged by this
addition, the improvement is driven primarily by the full-shape measurement
of the geometric parameters, most notably $\Omega_m$, rather than by an
independent tightening of the growth-of-structure amplitude. In
$\Lambda$CDM, an analogous $\Omega_m$-driven tightening should be
interpreted with care: the mild tension between the CMB- and BAO-preferred
values of $\Omega_m$ can only be resolved through the neutrino sector,
pulling the inferred mass to anomalously low values via the CMB geometric
degeneracy \cite{DESI:2025zgx,Loverde:2024nfi}. In $w_0w_a$CDM, by contrast, this
tension is absorbed by the additional freedom in the dark energy sector,
so the residual $\Omega_m$ information supplied by the full-shape data
propagates into $M_\nu$ through the standard geometric degeneracy rather
than through a cross-dataset tension. For this reason, we regard the
$w_0w_a$CDM bound, despite its more modest significance, as the more
reliable arena for neutrino mass inference.}

{It is instructive to connect these results to the discussion of effective negative neutrino masses in Section~\ref{sec:intro}. 
We carry out an approximate analysis
based on our MCMC chains, leaving
a more careful exploration
of the effective (unrestricted)
neutrino mass $M_\nu$ 
in DESI full-shape data 
to future work. 
To this end, we model our marginalized $M_\nu$ posteriors as Gaussians truncated at the physical boundary $M_\nu = 0$. This model provides an excellent description of the chains: in the $w_0w_a$CDM case, it reproduces both the measured posterior mean and the exact tail probability $p(M_\nu > 0.098~{\rm eV})$ to within a few percent. We find the underlying, unrestricted Gaussians peak 
at effective masses $M_\nu^{\rm eff} \approx (-0.06 \pm 0.04)$ eV ($\Lambda$CDM) and $M_\nu^{\rm eff} \approx (-0.06 \pm 0.06)$ eV ($w_0w_a$CDM). 
Within $\Lambda$CDM, 
our datasets continue to exhibit the mild preference for negative effective neutrino masses discussed in Section~\ref{sec:intro}~\cite{Craig:2024tky,Loverde:2024nfi,Green:2024xbb,Elbers:2024sha,Graham:2025fdt,DESI:2025ejh}, where the pull towards $M_\nu^{\rm eff}<0$ is ${\approx}1.5\sigma$.
However, it reduces to 
${\approx}1\sigma$ in $w_0w_a$CDM, since the geometric
part of the ``negative'' neutrino mass tension gets absorbed by the dark energy sector. In this unrestricted Gaussian measure, the inverted-hierarchy floor lies $4.0\sigma$ ($2.8\sigma$) away from the peak of the $\Lambda$CDM ($w_0w_a$CDM) likelihood.}

{We note that a similar conclusion holds when the background is generalized
in the orthogonal direction of spatial curvature: our previous analysis of
the $\Lambda$CDM$+\Omega_k+M_\nu$ model \citep{desi2} found the inverted
hierarchy to be disfavored at the ${\sim}2\sigma$ level. The preference for
the normal ordering thus persists along both standard directions of
background freedom: 
the dark energy equation of state and spatial
curvature, which supports its interpretation as background-independent.}

\section{Discussion and Conclusions}
\label{sec:disc}
\noindent
In this work we have presented the most complete analysis of the public DESI galaxy 
clustering data to date, incorporating the three-dimensional galaxy power spectrum and bispectrum
multipoles, the projected angular two-point correlations of DESI photometric galaxies
and the cross-correlations of both photometric and spectroscopic galaxies with CMB lensing maps from \textit{Planck} and ACT. Notably, ours is the first analysis of DESI data to feature a consistent one-loop model for both the galaxy power spectrum and bispectrum, utilizing the systematic EFT prescription based on power counting rules. 
On the analysis side, we carefully account for a wide variety of systematic effects including the survey geometry, fiber collisions, integral constraints, and binning in all relevant statistics. We also vary the one-loop EFT  predictions as we scan over cosmology using the \textsc{cobra} technique -- this allows us to achieve an unprecedented subpercent accuracy in the one-loop EFT computations at minimal computational cost. 

Our work provides some of the strongest constraints on both the late-time matter abundance and the growth of structure. Specifically, in the $\ld$ model we find a 2\% limit on the mass fluctuation amplitude $\sigma_8 = 0.838\pm 0.017$, which is consistent and 
competitive with the $\ld$ bounds derived from the primary CMB data, e.g., $\sigma_8=0.811\pm 0.006$ from \textit{Planck} 2018~\cite{Planck:2018nkj} or $\sigma_8=0.8113\pm 0.005$ from \textit{Planck}+ACT \cite{LouisACT25}.
Our bound compares favorably to previous constraints from large-scale galaxy clustering, though remains somewhat shy of the CMB lensing - unWISE galaxy correlation result $\sigma_8=0.814\pm 0.010$~\cite{Farren:2024rla}. 
Other notable literature constraints include the lensing amplitude measurements from $3\times 2$-point analyses (including cosmic shear, galaxy clustering, and galaxy-galaxy lensing) using the DESI DR1 full-shape dataset combined with weak lensing maps from KiDS, DES, and HSC~\cite{Semenaite:2025ohg}. The strongest of these bounds is $S_8=0.791\pm 0.016$ (DESI$\times$DES), which is in reasonable agreement with our result of $0.834\pm0.018$.
Similarly, \cite{Porredon2025:DR1cross} perform a $3\times 2$-point analysis using projected data alone, with the strongest constraint $S_8 = 0.760_{-0.018}^{+0.020}$ when combining with KiDS weak lensing, which is somewhat lower than our result. Additionally, \cite{Lange2025:DR1_smallscale} adopt a simulation-based model for small-scale projected galaxy clustering and galaxy-galaxy lensing, finding $S_8 = 0.793 \pm 0.017$ when combining DESI DR1 with HSC or $S_8 = 0.793 \pm 0.023$ with DES and KiDS, which is again in fair agreement with our posterior.
Our result is also consistent
within $2.1\sigma$ with the most recent DES-Y6 measurement
$S_8=0.789\pm 0.012$~\cite{DES:2026fyc}.
Finally, we note that our growth of structure measurement is $50\%$ stronger than the official result from the the DESI DR1 full-shape clustering analyses: $\sigma_8=0.842\pm 0.034$~\cite{DESI:2024jis}, and matches the precision of the joint analysis of two point functions in two- and three-dimensions: $\sigma_8=0.803\pm0.017$~\cite{Maus2025:joint_3d_lensing_dr1}.

Our baseline matter density constraint of $\Omega_m=0.2974\pm 0.0050$ is also highly competitive, improving upon the collaboration results for three- and two-plus-three-dimensional clustering ($\Omega_m=0.2962\pm 0.0095$ and $\Omega_m=0.3037\pm 0.0069$) by $47\%$ and $28\%$ respectively \citep{DESI:2024jis,Maus2025:joint_3d_lensing_dr1}.
Importantly, our measurement
is already stronger than the CMB limit $\Omega_m=0.3150\pm 0.0072$, though interestingly lies somewhat low.
Notably, including the one-loop bispectrum sharpens the $\Omega_m$ constraint by $12\%$ compared to the result obtained with the tree-level bispectrum likelihood.

Our fiducial Hubble constant measurement: $H_0=(69.08\pm 0.37)~\mathrm{km}\,\mathrm{s}^{-1}\mathrm{Mpc}^{-1}$ can be compared to the collaboration DESI DR2 BAO+BBN result of $H_0=(68.51\pm 0.58)~\mathrm{km}\,\mathrm{s}^{-1}\mathrm{Mpc}^{-1}$ \cite{DESI:2025gwf} and the DR1 full-shape+BAO bound $H_0=(68.56\pm 0.75)~\mathrm{km}\,\mathrm{s}^{-1}\mathrm{Mpc}^{-1}$. These represent improvements by $36\%$ and $51\%$ respectively, which are mainly driven by 
a combination of 
our new 3D full-shape likelihood with the DR2 BAO.

In combination with CMB, BAO, and SNe data, our analysis yields some of the strongest neutrino mass constraints to date.
Specifically, we find $M_\nu<0.049$ eV in the $\Lambda$CDM background
and $M_\nu<0.077$ eV in the $w_0w_a$CDM background, both of which are significantly tighter than the bounds from CMB, BAO and SNe alone (though see \citep{Chebat:2025kes} for tight $M_\nu$ constraints from a frequentist analysis). Importantly, this implies a (largely) background-independent preference for the normal neutrino mass hierarchy over the inverted hierarchy.
Our $M_\nu$ constraints differ considerably from combining \textit{Planck} with the previous generation of spectroscopic surveys (BOSS; \cite{Ivanov:2019hqk,Philcox:2020vvt}).
Historically, neutrino mass constraints have been dominated by
the geometric expansion history information probed by the BAO; in this work, we find a $\simeq 37\%$ reduction in $\sigma(M_\nu)$ when we add the full-shape data to the geometric BAO and CMB probes.
{Within our analysis, this marks a tipping point in the development of the
large-scale structure full-shape technique: for the first time, full-shape
data sharpen the neutrino mass constraint substantially beyond what the
geometric BAO and CMB probes deliver, as has long been anticipated for
high-precision large-scale structure data~\cite{Chudaykin:2019ock,Sailer:2021yzm}.}

There are multiple ways to extend our analysis further. First, one may consider generalizing our treatment to other cosmological models, such as those featuring ultralight axions, light relics,
dark-matter-baryon interactions, and modified gravity, along the lines of previous EFT-based full-shape analyses
\citep{Ivanov:2019pdj,DAmico:2019fhj,Chen:2021wdi,Ibanez:2024uua,DAmico:2022osl,DAmico:2022gki,Ivanov:2024hgq,Ivanov:2021fbu,Ivanov:2021zmi,Chudaykin:2022nru,Ivanov:2021kcd,Ivanov:2023qzb,Cabass:2024wob,Zhang:2021yna,Chen:2022jzq,Cabass:2022ymb,Cabass:2022epm,Cabass:2024wob,Chen:2021wdi,Philcox:2021kcw,Wadekar:2020hax,Colas:2019ret,Ivanov:2019hqk,Ivanov:2020ril,Rogers:2023ezo,Chen:2024vuf,Spaar:2023his,He:2023oke,He:2023dbn,Xu:2021rwg,Chudaykin:2022nru,Chudaykin:2020ghx,Lu:2025gki,Lu:2025sjg,DAmico:2020tty,McDonough:2023qcu,Toomey:2025yuy,Silva:2025twg,Chen:2024vuf,Maus2025:joint_3d_lensing_dr1,DAmico:2025zui}. From a data-driven perspective,
it will be interesting to apply simulation-based priors in 
our analysis~\cite{Sullivan:2021sof,Ivanov:2024hgq,Ivanov:2024xgb,Chen:2025jnr,Ivanov:2024dgv,Akitsu:2024lyt,Ivanov:2025qie,DESI:2025wzd,kokron_stoch_heft}, though we leave this to future work. Finally, we could include more statistics, such as galaxy trispectra, galaxy clusters, galaxy lensing, and beyond -- there is always more room in the kitchen sink.

\vskip 8pt
\textit{Acknowledgments:} 
{\small
\begingroup
\hypersetup{hidelinks}
\noindent 
\\
We thank 
Laura Herold, 
Marc Kamionkowski, 
and Martin White 
for helpful discussions. AC acknowledges funding from the Swiss National Science Foundation. OHEP's contributions drew inspiration from the British \href{https://www.flickr.com/photos/198816819@N07/55027025538/}{willdlife}.
JMS acknowledges that support for this work was provided by The Brinson Foundation through a Brinson Prize. Support for this work was provided by NASA through
the NASA Hubble Fellowship grant HST-HF2-51572.001 awarded by the Space Telescope Science Institute, which is operated by the Association of Universities for Research in Astronomy, Inc., for NASA, under contract NAS5-26555. MM is funded by the DOE.
\endgroup

\vskip 2pt
This research used data obtained with the Dark Energy Spectroscopic Instrument (DESI). DESI construction and operations is managed by the Lawrence Berkeley National Laboratory. This material is based upon work supported by the U.S. Department of Energy, Office of Science, Office of High-Energy Physics, under Contract No. DE–AC02–05CH11231, and by the National Energy Research Scientific Computing Center, a DOE Office of Science User Facility under the same contract. Additional support for DESI was provided by the U.S. National Science Foundation (NSF), Division of Astronomical Sciences under Contract No. AST-0950945 to the NSF’s National Optical-Infrared Astronomy Research Laboratory; the Science and Technology Facilities Council of the United Kingdom; the Gordon and Betty Moore Foundation; the Heising-Simons Foundation; the French Alternative Energies and Atomic Energy Commission (CEA); the National Council of Humanities, Science and Technology of Mexico (CONAHCYT); the Ministry of Science and Innovation of Spain (MICINN), and by the DESI Member Institutions: \url{www.desi.lbl.gov/collaborating-institutions}. The DESI collaboration is honored to be permitted to conduct scientific research on I’oligam Du’ag (Kitt Peak), a mountain with particular significance to the Tohono O’odham Nation. Any opinions, findings, and conclusions or recommendations expressed in this material are those of the author(s) and do not necessarily reflect the views of the U.S. National Science Foundation, the U.S. Department of Energy, or any of the listed funding agencies.
}

\newpage

\pagebreak
\widetext
\appendix

% \begin{center}
% \textbf{\large Appendix}
% \end{center}
% \makeatletter
% \setcounter{equation}{0}
% \setcounter{figure}{0}
% \setcounter{table}{0}
% \setcounter{page}{1}
% \makeatletter
% \renewcommand{\theequation}{S\arabic{equation}}
% \renewcommand{\thefigure}{S\arabic{figure}}
% \renewcommand{\bibnumfmt}[1]{[S#1]}
% \renewcommand{\citenumfont}[1]{S#1}

\section{One-loop EFT for the power spectrum and bispectrum}
\label{sec:one-loopEFT}

\noindent In this appendix, we explicitly define the one-loop power spectrum and bispectrum model discussed in Section~\ref{sec:theory}. Further details can be found in \cite{Bakx:2025pop} and a number of previous works. The determinstic contributions to the one-loop bispectrum can be obtained from the standard perturbative series for the Fourier-space galaxy density $\delta_{g}^{{\rm (s)}}$ in terms of the linear density field $\delta_1(\k)$: \beq\label{eq:rsd_Zn}
    \delta^{~{\rm (s)}}_{g,(n)}(\k) = 
    \sum_{n}\int_{\vq_1\ldots\vq_n}(2\pi)^3\delta_D^{(3)}(\k-\vq_{1...n})Z_n(\vq_1,\cdots,\vq_n)\delta_1(\vq_1)\cdots\delta_1(\vq_n)\,,
\eeq
at $n$-th order, adopting the notation
$\vq_{i\cdots j}\equiv \vq_i+\cdots+\vq_j$ \cite{Scoccimarro:1995if,Scoccimarro:1996jy,Scoccimarro:1996se}. Up to fourth-order, the redshift-space kernels read
\beq\label{eq: RSD-kernels}
    Z_1(\vq_1) &=& K_1 + f\mu_1^2,\\\nonumber
    Z_2(\vq_1,\vq_2) &=& K_2(\vq_1,\vq_2) + f\mu_{12}^2G_2(\vq_1,\vq_2)+\frac{f\mu_{12}q_{12}}{2}K_1\left[\frac{\mu_1}{q_1}+\frac{\mu_2}{q_2}\right]+\frac{(f\mu_{12}q_{12})^2}{2}\frac{\mu_1}{q_1}\frac{\mu_2}{q_2} , \\\nonumber
    Z_3(\vq_1,\vq_2,\vq_3) &=& K_3(\vq_1,\vq_2,\vq_3)+f\mu_{123}^2G_3(\vq_1,\vq_2,\vq_3)\\\nonumber
    &&\,+\,(f\mu_{123}q_{123})\left[\frac{\mu_{12}}{q_{12}}K_1G_2(\vq_1,\vq_2)+\frac{\mu_3}{q_3}K_2(\vq_1,\vq_2)\right]\\\nonumber
    &&\,+\,\frac{(f\mu_{123}q_{123})^2}{2}\left[2\frac{\mu_{12}}{q_{12}}\frac{\mu_3}{q_3}G_2(\vq_1,\vq_2)+\frac{\mu_1}{q_1}\frac{\mu_2}{q_2}K_1\right]+\frac{(f\mu_{123}q_{123})^3}{6}\frac{\mu_1}{q_1}\frac{\mu_2}{q_2}\frac{\mu_3}{q_3} , 
        \eeq
    \beq\label{eq: RSD-kernels-2}
    Z_4(\vq_1,\vq_2,\vq_3,\vq_4) &=&  K_4(\vq_1,\vq_2,\vq_3,\vq_4)+f\mu_{1234}^2G_4(\vq_1,\vq_2,\vq_3,\vq_4)\\\nonumber
    &&\,+\,(f\mu_{1234}q_{1234})\left[\frac{\mu_{123}}{q_{123}}K_1G_3(\vq_1,\vq_2,\vq_3)+\frac{\mu_4}{q_4}K_3(\vq_1,\vq_2,\vq_3)\right.\\\nonumber
    &&\qquad\qquad\qquad\qquad+\,\left.\frac{\mu_{12}}{q_{12}}G_2(\vq_1,\vq_2)K_2(\vq_3,\vq_4)\right]\\\nonumber
    &&\,+\,\frac{(f\mu_{1234}q_{1234})^2}{2}\left[2\frac{\mu_{123}}{q_{123}}\frac{\mu_4}{q_4}G_3(\vq_1,\vq_2,\vq_3)+\frac{\mu_{12}}{q_{12}}\frac{\mu_{34}}{q_{34}}G_2(\vq_1,\vq_2)G_2(\vq_3,\vq_4)\right.\\\nonumber
    &&\qquad\qquad\qquad\qquad\left.+\,2\frac{\mu_{12}}{q_{12}}\frac{\mu_3}{q_3}K_1G_2(\vq_1,\vq_2)+\frac{\mu_1}{q_1}\frac{\mu_2}{q_2}K_2(\vq_3,\vq_4)\right]\\\nonumber
    &&\,+\,\frac{(f\mu_{1234}q_{1234})^3}{6}\left[3\frac{\mu_{12}}{q_{12}}\frac{\mu_3}{q_3}\frac{\mu_4}{q_4}G_2(\vq_1,\vq_2)+\frac{\mu_1}{q_1}\frac{\mu_2}{q_2}\frac{\mu_3}{q_3}K_1\right]\\\nonumber
    &&\,+\,\frac{(f\mu_{1234}q_{1234})^4}{24}\frac{\mu_1}{q_1}\frac{\mu_2}{q_2}\frac{\mu_3}{q_3}\frac{\mu_4}{q_4}\,,
\eeq
where $\mu_{i\cdots j}\equiv \mu_{\vq_i+\cdots+\vq_j}$, and the real-space galaxy kernels are given by
\beq\label{eq: real-kernels}
    K_1(\vq_1) &=& {b_1} , \\\nonumber
    K_2(\vq_1,\vq_2) &=&\left\{{b_1}F_2(\vq_1,\vq_2)\right\}+\left\{\frac{{b_2}}{2}+{\gamma_2}\,\kappa(\vq_1,\vq_2)\right\} , \\\nonumber
    K_3(\vq_1,\vq_2,\vq_3)&=& \left\{{b_1}F_3(\vq_1,\vq_2,\vq_3)\right\} + \left\{{b_2}F_2(\vq_1,\vq_2)+2{\gamma_2}\,\kappa(\vq_1,\vq_{23})G_2(\vq_2,\vq_3)\right\}\\\nonumber
    &&\,+\,\left\{\frac{{b_3}}{6}+{\gamma_2^\times}\,\kappa(\vq_1,\vq_2)+{\gamma_3}\,L(\vq_1,\vq_2,\vq_3)+{\gamma_{21}}\kappa(\vq_1,\vq_{23})\kappa(\vq_2,\vq_3)\right\} , \\\nonumber
\eeq    

\beq\label{eq: real-kernels}
    K_4(\vq_1,\vq_2,\vq_3,\vq_4)&=&\left\{{b_1}F_4(\vq_1,\vq_2,\vq_3,\vq_4)\right\}\\\nonumber
    &&\,+\,\left\{\frac{{b_2}}{2}\left[F_2(\vq_1,\vq_2)F_2(\vq_3,\vq_4)+2F_3(\vq_1,\vq_2,\vq_3)\right]\right.\\\nonumber
    &&\qquad\left.+{\gamma_2}\left[\kappa(\vq_{12},\vq_{34})G_2(\vq_1,\vq_2)G_2(\vq_3,\vq_4)+2\kappa(\vq_{123},\vq_4)G_3(\vq_1,\vq_2,\vq_3)\right]\right\}\\\nonumber
    &&\,+\,\left\{\frac{{b_3}}{2}F_2(\vq_1,\vq_2)+{\gamma_2^\times}\left[2\kappa(\vq_{12},\vq_3)G_2(\vq_1,\vq_2)+\kappa(\vq_3,\vq_4)F_2(\vq_1,\vq_2)\right]\right.\\\nonumber
    &&\qquad+\,3{\gamma_3}\,L(\vq_1,\vq_2,\vq_{34})G_2(\vq_3,\vq_4)\\\nonumber
    &&\qquad\left.+\,{\gamma_{21}}\left[\kappa(\vq_{12},\vq_{34})\kappa(\vq_1,\vq_2)F_2(\vq_3,\vq_4)+2\kappa(\vq_{123},\vq_4)\kappa(\vq_{12},\vq_3)F_2(\vq_1,\vq_2)\right]\right\}\\\nonumber
    &&\,+\,\left\{{\gamma_{21}^\times}\,\kappa(\vq_1,\vq_{23})\kappa(\vq_2,\vq_3)+{\gamma_{211}}L(\vq_1,\vq_2,\vq_{34})\kappa(\vq_3,\vq_4)\right.\\\nonumber
    &&\qquad+\,{\gamma_{22}}\,\kappa(\vq_{12},\vq_{34})\kappa(\vq_1,\vq_2)\kappa(\vq_3,\vq_4)\\\nonumber
    &&\qquad+\,{\gamma_{31}}\left[\frac{1}{18}\kappa(\vq_1,\vq_{234})\left(\frac{15}{7}\kappa(\vq_{23},\vq_4)\kappa(\vq_2,\vq_3)-L(\vq_2,\vq_3,\vq_4)\right)\right.\\\nonumber
    &&\qquad\qquad\left.\left.+\frac{1}{14}\left(M(\vq_1,\vq_{23},\vq_4,\vq_{234})-M(\vq_1,\vq_{234},\vq_{23},\vq_4)\right)\kappa(\vq_2,\vq_3)\right]\right\},
\eeq
defining
\beq\label{eq: ang-def}
    \kappa(\vq_1,\vq_2) &\equiv& (\hq_1\cdot\hq_2)^2-1 , \\\nonumber
    L(\vq_1,\vq_2,\vq_3) &\equiv& 2(\hq_1\cdot\hq_2)(\hq_2\cdot\hq_3)(\hq_3\cdot\hq_1)-(\hq_1\cdot\hq_2)^2-(\hq_2\cdot\hq_3)^2-(\hq_3\cdot\hq_1)^2+1 , \\\nonumber
    M(\vq_1,\vq_2,\vq_3,\vq_4) &\equiv& (\hq_1\cdot\hq_2)(\hq_2\cdot\hq_3)(\hq_3\cdot\hq_4)(\hq_4\cdot\hq_1).
\eeq
This depends on the standard $F_n$ and $G_n$ density and velocity kernels of Eulerian perturbation theory, which we compute in the Einstein-de-Sitter approximation~\cite{Fry:1983cj,Scoccimarro:1995if,Bernardeau:2001qr}. 
The above basis uses tidal bias parameters, which are related to those appearing in our one-loop power spectrum model by
\be 
\begin{split} 
% & \kappa(\k_1,\k_2)\equiv \G(\k_1,\k_2) \,,\quad 
% \kappa(\k_{1},\k_{23})
% \kappa(\k_2,\k_3)=-\frac{4}{7}\Gamma_3(\k_1,\k_2,\k_3)
% \Gamma_3(\k_1,\k_2,\k_3)\equiv -\frac{4}{7}\left[\kappa(\k_{1},\k_{23})
% \kappa(\k_2,\k_3)\right]_{\rm sym.}
% (F_2(\k_2,\k_3)-G_2(\k_2,\k_3))
% \,, \\
&\gamma_2=b_{\G}\,,\quad \gamma_{21}=-\frac{4}{7}(b_{\G}+b_{\Gamma_3})\,.
\end{split}
\ee 

The deterministic part of the one-loop power spectrum is given by 
\be 
\begin{split}
P_{\rm det,gg}^{~{\rm (s)}}(\k)&=Z_1^2(\k)P_{\rm lin}(k)+P_{\rm 1-loop}(\k)=Z_1^2(\k)P_{\rm lin}(k)+P_{22}(\k)+2P_{13}(\k)\\
&\equiv Z_1^2(\k)P_{11}(k)+2\int_{\q} P_{11}(|\k-\q|)P_{11}(q)[Z_2(\q-\k,\k)]^2+6Z_1(\k)P_{11}(k)
\int_{\q} P_{11}(q)Z_3(\k,-\q,\q)\,,
\end{split}
\ee 
setting $P_{11}=P_{\rm lin}$ to match the notation of~\cite{Bakx:2025pop}.
Note that all contributions from cubic bias operators except for $\G$
and $\Gamma_3$ disappear in the one-loop power spectrum upon renormalization~\cite{Assassi:2014fva}. The resulting power spectrum depends on just four bias parameters at one-loop order: $\{b_1,b_2,b_{\G},b_{\Gamma_3}\}$, whilst the one-loop bispectrum depends also on
$\{b_3,\gamma_2^\times,
\gamma_3,\gamma_{21}^\times,\gamma_{211},\gamma_{22},\gamma_{31}\}$.

The one-loop power spectrum counterterm is given by
\be 
\label{eq:Z1ctrdef}
\begin{split}
    P^{~{\rm (s)}}_{\rm ctr,gg}(\k)=&2Z_1^{\rm ctr}(\k)Z_1(\k)P_{\rm lin}(k)\,,\quad Z_1^{\rm ctr}(\k) =  \frac{k^2}{k_{\rm NL}^2}\bigg(-b_{\nabla^2\delta}+\left(e_1 -\frac{1}{2}c_1 f\right)f\mu^2 -\frac{1}{2}c_2 f^2\mu^4\bigg)\,.
\end{split}
\ee 
where the kernel can be ``improved'' by including a higher order velocity dispersion counterterm 
\be 
\label{eq:k4}
Z_1^{\rm ctr}(\k)\to Z_1^{\rm ctr}(\k)-\frac{\tilde{c}}{2}
\frac{k^4}{k_{\rm NL}^4}
f^4\mu^4  Z_1(\k)\,,
\ee
which captures a dominant two-loop contribution. Finally, the stochastic contribution takes the form
\be 
\label{eq:stochP1}
P^{{\rm (s)}}_{\rm stoch,gg}(k,\mu) = \frac{1}{\bar n}\left(P_{\rm shot}+a_0\frac{k^2}{k^2_{\rm NL}} +a_2\mu^2 \frac{k^2}{k^2_{\rm NL}} \right)~\,.
\ee 

To form the real-space galaxy auto-spectrum used in the description of the angular correlation
functions, we can take the above expressions and set $f=0$ and $a_2=0$. This leads to the following expression for the deterministic real-space one-loop galaxy-matter cross-spectrum and matter-matter auto-spectrum:
\be 
\begin{split}
P^{(r)}_{\rm 1-loop,gm}(k)&=2\int_{\q} P_{11}(|\k-\q|)P_{11}(q)K_2(\q-\k,\k)F_2(\k-\q,-\k)+3P_{11}(k)
\int_{\q} P_{11}(q)[b_1F_3(\k,-\q,\q)+K_3(\k,-\q,\q)]\,,\\
P^{(r)}_{\rm 1-loop,mm}(k)&=2\int_{\q} P_{11}(|\k-\q|)P_{11}(q)[F_2(\q-\k,\k)]^2+6P_{11}(k)
\int_{\q} P_{11}(q)F_3(\k,-\q,\q)\,.
\end{split}
\ee

Next, we turn to the one-loop bispectrum contributions. The deterministic pieces are given by~\cite{Scoccimarro:1997st,Bernardeau:2001qr,Baldauf:2014qfa,Philcox:2022frc}:
\be 
B_{\rm det}^{\rm 1-loop~(s)}=B_{222}+
B_{321}^I+
B_{321}^{II}+B_{411}\,,
\ee 
where
\beq\label{eq: Bk-one-loop}
    B_{222}(\k_1,\k_2,\k_3) &=& 8\int_{\vq}Z_2(\k_1+\vq,-\vq)Z_2(\k_1+\vq,\k_2-\vq)Z_2(\k_2-\vq,\vq)
    % \\\nonumber
    % &&\qquad\times\,
    P_{11}(q)P_{11}(|\k_1+\vq|)
    P_{11}(|\k_2-\vq|), \\\nonumber
    B_{321}^I(\k_1,\k_2,\k_3) &=&  6\,Z_1(\k_1)P_{11}(k_1)\int_{\vq}Z_3(-\vq,\vq-\k_2,-\k_1)Z_2(\vq, \k_2-\vq)
    % \\\nonumber
    % &&\qquad\times\,
    P_{11}(q)P_{11}(|\k_2- \vq|)+\text{5 perm.}, \\\nonumber
    B_{321}^{II}(\k_1,\k_2,\k_3) &=& 6\,Z_2(\k_1,\k_2)Z_1(\k_2)
    P_{11}(k_1)
    P_{11}(k_2)\int_{\vq}Z_3(\k_1,\vq,-\vq)P_{11}(q)+\text{5 perm.}, \\\nonumber
    B_{411}(\k_1,\k_2,\k_3) &=& 12\,Z_1(\k_1)Z_1(\k_2)
    P_{11}(k_1)P_{11}(k_2)\int_{\vq}Z_4(\k_1,\k_2,\vq,-\vq)P_{11}(q)+\text{2 cyc.}
\eeq
This is renormalized by two counterterm contributions, 
\beq
& B^{\rm ctr~(s)~I}=2Z_1(\k_1)Z_1(\k_2)Z_2^{\rm ctr}(\k_1,\k_2)P_{11}(k_1)P_{11}(k_2) +\text{2 cyc.}\,,
\eeq 
defining
\be
\begin{split}
& k_{\rm NL}^2 F^{\rm ctr.~(s)}_2(\k_1,\k_2) = b_{\nabla^2\delta}f k_{3z}\left(\frac{k_1^2k_{2z}}{2k_2^2}+(2\leftrightarrow 1)\right)  +e_1 \left(f k_{3z}^2 F_2(\k_1,\k_2)+\frac{f^2k_{3z}^2k_{1z}k_{2z}}{2}\left(\frac{1}{k_1^2}+\frac{1}{k_2^2}\right) \right) \\
&+ e_5f k_{3z}^2\frac{(\k_1\cdot\k_2)(\k_1\cdot\k_3)(\k_2\cdot\k_3)}{k_1^2 k_2^2 k_3^2} 
- e_4fk_{3z}\left(\frac{k_{1z}(\k_1\cdot \k_2)}{2k_1^2}+(2\leftrightarrow 1)\right)
\\
&
+c_1\left(-\frac{ f^2}{2}k^2_{3z}F_2(\k_1,\k_2)+f^3\frac{k^3_{3z} k_{1z}}{4 k^2_{1}}+(2\leftrightarrow 1)\right)
+c_2\left(-\frac{f^2}{2}
\frac{k^4_{3z}}{k_3^2}F_2(\k_1,\k_2)+f^3 k_{3z}^3\frac{k_{1z}k^2_{2z}}{4k_1^2k_2^2}+(2\leftrightarrow 1)\right) \\
& -\frac{c_3f^2k^2_{3z}}{2} 
 -\frac{c_4f^2}{2}k_{3z}^2\frac{(\k_1\cdot\k_2)k_{1z}k_{2z}}{k_1^2k_2^2}
% -c_5f^2\left(k_{3z}^2\frac{k^2_{1z}}{4k_1^2}
% +(2\leftrightarrow 1)\right)
-\frac{c_6}{2}f^2k_{3z}^2\frac{(\k_1\cdot \k_2)^2}{k_1^2k_2^2}
+\frac{c_7}{2}f^2 \frac{k_{3z}^4}{k^2_3}\frac{2}{7}\G(\k_1,\k_2)~\,.
\end{split}
\ee 
and 
\be 
B^{\rm ctr~(s)~II}=2Z^{\rm ctr}_1(\k_1)Z_1(\k_2)Z_2(\k_1,\k_2)P_{11}(k_1)P_{11}(k_2)
+\text{5 perms.}\,,
\ee 
where $Z_1^{\rm ctr}$ is given in \eqref{eq:Z1ctrdef}\,\&\,\eqref{eq:k4}.

The pure bispectrum stochasticity is given by
\be 
B_{\rm stoch}(\k_1,\k_2,\k_3) = \frac{1}{\bar n^2}\left(
A_{\rm shot} + \frac{1}{k_{\rm NL}^2}(a_1(k_1^2+k_2^2+k_3^2)+a_6(k_{1z}^2+k_{2z}^2+k_{3z}^2))
\right)\,.
\ee 
whilst the mixed contributions include the tree-level piece of \eqref{eq:Bmixtree} plus the one-loop term
\be\label{eq:mixed1loopstoch_s}
\begin{split}
& B_{\rm mixed}^{\rm 1-loop~(s)}(\k_1,\k_2,\k_3)=\frac{1+P_{\rm shot}}{\bar n}(\mathcal{P}_{22}^{\text{~(s)}}(k_1,k_2) + \mathcal{P}^{I~\text{~(s)}}_{13}(k_1,k_2) + 
\mathcal{P}^{II~\text{~(s)}}_{13}(k_1,k_2) +\text{5 perms} )\,,\\
& \mathcal{P}^{\text{~(s)}}_{22}(k_1,k_2) = 2\int_{\q} \mathcal{Z}_2(\q,\k_1-\q,\k_2)Z_2(-\q,-\k_1+\q)P_{11}(q)P_{11}(|\k_1-\q|)\,,\\
& \mathcal{P}^{I~\text{~(s)}}_{13}(k_1,k_2) = 3Z_1(\k_1) P_{11}(\k_1)\int_{\q} \mathcal{Z}_3(\k_1,\q,-\q,\k_2)P_{11}(q) \,,\\
& \mathcal{P}^{II~\text{~(s)}}_{13}(k_1,k_2) = 3\mathcal{Z}_1(\k_1,\k_2) P_{11}(\k_1)\int_{\q} Z_3(\k_1,\q,-\q)P_{11}(q)\,.
\end{split}
\ee 
This uses the mixed stochastic kernels defined as 
\be 
\begin{split}
\mathcal{Z}_1(\k_1,\k_2) &  =d_1 + f\frac{k_{12z}k_{1z}}{k_1^2}\\
 \mathcal{Z}_2(\k_1,\k_2,\k_3) &=d_1 F_2(\k_1,\k_2)+f\frac{k_{123z}k_{12z}}{k_{12}^2}G_2(\k_1,\k_2)+\frac{fk_{123z}d_1}{2}\left(\frac{k_{1z}}{k_1^2}+\frac{k_{2z}}{k_2^2}\right)\\
 & ~~~~~ + \frac{f^2 k_{123z}^2 }{2}\frac{k_{1z}}{k_1^2}\frac{k_{2z}}{k_2^2}+\frac{d_2}{2}+d_{\G} \G(\k_1,\k_2) \,,
\end{split}
\ee 
with the un-symmetrized $\mathcal{Z}_3$ kernel
\be 
\begin{split}
& \mathcal{Z}_3(\k_1,\k_2,\k_3,\k_4) = d_1F_3(\k_1,\k_2,\k_3)
+f \frac{k_{1234z}k_{123z}}{k_{123}^2}G_3(\k_1,\k_2,\k_3)
+f k_{1234z}\frac{k_{1z}}{k_1^2}d_1 F_2(\k_2,\k_3)
\\
&+ f^2\frac{k_{1234z}^2k_{1z}}{k_1^2}G_2(\k_2,\k_3)\frac{k_{23z}}{k_{23}^2}+fk_{1234z}d_1\frac{k_{23z}}{k_{23}^2}G_2(k_2,\k_3)
+d_1\frac{f^2k_{1234z}^2}{2}
\frac{k_{1z}k_{2z}}{k_1^2k_2^2}\\
&+\frac{f^3k_{1234z}^3}{6}
\frac{k_{1z}}{k_1^2}
\frac{k_{2z}}{k_2^2}
\frac{k_{3z}}{k_3^2}+d_2F_2(\k_2,\k_3) +2 d_{\G}\G(\k_{23},\k_1)F_2(\k_2,\k_3) +d_2 \frac{fk_{1234z}k_{1z}}{2k_1^2}\\
& + d_{\G} fk_{1234z}
\G(\k_2,\k_3)\frac{k_{1z}}{k_1^2}+2d_{\Gamma_3}\G(\k_{23},\k_1)(F_2(\k_2,\k_3)-G_2(\k_2,\k_3))~\,.
\end{split}
\ee 
The linear contribution of \eqref{eq:Bmixtree} can be obtained by identifying~\cite{Ivanov:2021kcd}
\be 
b_1 B_{\rm shot}\equiv 2d_1 (1+P_{\rm shot})\,.
\ee 
In this work we subtract the Poisson contributions from all statistics (which involves the non-linear power spectrum), such that the final mixed term is given by 
\be\label{eq:mixed1loopstoch_s}
\begin{split}
B_{\rm mixed}^{\rm 1-loop~(s)}(\k_1,\k_2,\k_3)&=\frac{1+P_{\rm shot}}{\bar n}(\mathcal{P}_{22}^{\text{~(s)}}(k_1,k_2) + \mathcal{P}^{I~\text{~(s)}}_{13}(k_1,k_2) + 
\mathcal{P}^{II~\text{~(s)}}_{13}(k_1,k_2) +\text{5 perms} )\,,\\
&-\frac{1}{\bar n}\left(P_{\rm 1-loop}(k_1,\mu_1)+P_{\rm 1-loop}(k_2,\mu_2)+P_{\rm 1-loop}(k_3,\mu_3)\right)\,,
\end{split}
\ee 
where $P_{\rm 1-loop}$
is the one-loop EFT power spectrum.

Finally, we require mixed counterterms, which come in two forms. The first is given by 
\be 
B_{\rm mixed}^{\rm ctr~(s)~I}(\k_1,\k_2,\k_3)=\sum_{n=1}^{10} s_n Z^{(n)}_{k^2\bar n^{-1}P}(\k_2,\k_3) \frac{1}{\bar{n}}Z_1(\k_1)P_{11}(k_1)+\text{2 cyc.}\,,
\ee 
where $s_n=\{a_0,a_3,a_4,a_5,a_7,...,a_{12}\}$ (which notably does not include $a_1$, $a_2$ or $a_6$), and the kernels are defined as
\be 
\begin{split}
&  Z^{(1)}_{k^2\bar n^{-1}P}(\k_2,\k_3)=-\frac{k_1^2(k_2^2+k_3^2)-(k^2_2-k_3^2)^2
+2fk_{1z}(k_{3z}k_2^2+k_{2z}k_3^2)
}{2k_1^2k_{\rm NL}^2}\,, \\
& 
 Z^{(2)}_{k^2\bar n^{-1}P}(\k_2,\k_3)=-\frac{k_1^4 + (k_2^2-k_3^2)^2}{2k_1^2k_{\rm NL}^2}\,, \quad Z^{(3)}_{k^2\bar n^{-1}P}(\k_2,\k_3)=-\frac{k_1^2}{k_{\rm NL}^2}\,,\quad 
 Z^{(4)}_{k^2\bar n^{-1}P}(\k_2,\k_3)=-\frac{(k_2^2+k_3^2)}{k_{\rm NL}^2}\,,\\
 & Z^{(5)}_{k^2\bar n^{-1}P}(\k_2,\k_3)=f\frac{\left[(k_3^2-k_1^2-k_2^2)k_{2z}^2+(k_2^2-k_3^2-k_1^2)k^2_{3z}+2k_{3z}k_{2z}k_1^2-2f k^2_{1z}k_{2z}k_{3z}\right]}{2k_1^2k_{\rm NL}^2}\,,\\
 &  
 Z^{(6)}_{k^2\bar n^{-1}P}(\k_2,\k_3)=\frac{f^2}{2k_{\rm NL}^2}k^2_{1z}(1-f\mu_1^2)\,,\quad Z^{(7)}_{k^2\bar n^{-1}P}(\k_2,\k_3)=f\frac{k_{1z}^2}{k_{\rm NL}^2}\,,\\
 & Z^{(8)}_{k^2\bar n^{-1}P}(\k_2,\k_3)=f\frac{k_{2z}k_{3z}}{k_{\rm NL}^2}\,,\quad 
Z^{(9)}_{k^2\bar n^{-1}P}(\k_2,\k_3)=
\frac{f}{4k_1^2k_2^2k_3^2k_{\rm NL}^2}
\left(
(k_2^2-k_3^2-k_1^2)^2 k_2^2 k^2_{3z}
+(k_3^2-k_2^2-k_1^2)^2 k_3^2 k^2_{2z}
\right) \,,\\
& Z^{(10)}_{k^2\bar n^{-1}P}(\k_2,\k_3)=\frac{f}{k_{\rm NL}^2}\left(\frac{k_{1z}}{k_1^2}\left(
k_{1z}(k_1^2+k_2^2-k_3^2)
+2k_{2z}(k_2^2-k_3^2)
\right)\right)\,.
\end{split}
\ee 
The second type of mixed counterterms is given by 
\be 
\begin{split}
B_{\rm mixed}^{\rm ctr~(s)~II}(\k_1,\k_2,\k_3)
&=\frac{1}{\bar{n}}(b_1 B_{\rm shot}+f\mu_1^2P_{\rm shot})Z_1^{\rm ctr}(\k_1)P_{11}(k_1)+\text{2 cyc.}\,,
\end{split}
\ee 
where we subtracted the Poisson contribution as before. In principle, these terms are second-order in EFT parameters, which prevents them from being analytically marginalized. 
Since explicitly sampling $\vec{b}\equiv\{B_{\rm shot},P_{\rm shot},b_{\nabla^2\delta},e_1,c_1,c_2,\tilde{c}\}$ in the analysis would lead to a significant reduction in speed, we adopt a Fisher approximation
for the likelihood parts involving 
$B_{\rm mixed}^{\rm ctr~(s)~II}$, linearizing around fiducial
values
$\vec{b}^{\rm fid}\equiv\{B^{\rm fid}_{\rm shot},P^{\rm fid}_{\rm shot},b^{\rm fid}_{\nabla^2\delta},e^{\rm fid}_1,c^{\rm fid}_1,c^{\rm fid}_2,\tilde{c}^{\rm fid}\}$. This leads to the approximate expression 
\be \label{eq; mixed-approx}
\begin{split}
B_{\rm mixed}^{\rm ctr~(s)~II}\Big|_{\rm approx}
&=\frac{1}{\bar{n}}(b_1  B_{\rm shot}+f\mu_1^2 P_{\rm shot})Z_1^{\rm ctr}[b^{\rm fid}_{\nabla^2\delta},e^{\rm fid}_1,c^{\rm fid}_1,c^{\rm fid}_2,\tilde{c}^{\rm fid}](\k_1)P_{11}(k_1)\\
&+\frac{1}{\bar{n}}(b_1 B^{\rm fid}_{\rm shot}+f\mu_1^2P^{\rm fid}_{\rm shot})Z_1^{\rm ctr}[b_{\nabla^2\delta},e_1,c_1,c_2,\tilde{c}](\k_1)P_{11}(k_1)\\
&-\frac{1}{\bar{n}}(b_1 B^{\rm fid}_{\rm shot}+f\mu_1^2P^{\rm fid}_{\rm shot})Z_1^{\rm ctr}[b^{\rm fid}_{\nabla^2\delta},e^{\rm fid}_1,c^{\rm fid}_1,c^{\rm fid}_2,\tilde{c}^{\rm fid}](\k_1)P_{11}(k_1)+\text{2 cyc.}\,,
\end{split}
\ee 
which is accurate up to corrections of order $\mathcal{O}((\vec{b}-\vec{b}^{\rm fid})^2)$. Notably, setting the fiducial values equal to the prior means given in Table~\ref{tab:priors} leads to \eqref{eq; mixed-approx} vanishing identically -- to avoid this, we could perform an analysis without $B_{\rm mixed}^{\rm ctr~(s)~II}$ and use the corresponding best-fit values to update $\vec{b}^{\rm fid}$ (and optionally iterate until convergence). 
In practice, our parameter constraints do not have any appreciable sensitivity to $B_{\rm mixed}^{\rm ctr~(s)~II}$. 
As such, the leading-order approximation is acceptable, \textit{i.e.}\ we can set $\bf{b}^{\rm fid}=\bf{b}^{\rm prior}_{\rm mean}$.
This is consistent with the fact that $B_{\rm mixed}^{\rm ctr~(s)~II}$
is numerically quite small for the fiducial $k^{B_\ell}_{\rm max}=0.16~\hMpc$. Should this change with DESI DR2, however, the approach outlined above can be easily used to add a non-trivial $B^{\rm ctr~(s)~II}_{\rm mixed}$ contribution to the theory prediction with high accuracy.

\bibliography{refs.bib,short.bib}

\end{document}